\documentclass[12pt,aps,epsf,twocolumn,numberedappendix,appendixfloats]{emulateapj}

\usepackage[breaklinks,colorlinks,urlcolor=blue,citecolor=blue,linkcolor=blue]{hyperref}
\usepackage{savesym}
\savesymbol{tablenum}
\usepackage{siunitx}
\restoresymbol{SIX}{tablenum}
\usepackage{mathtools}
\usepackage{physics}
\usepackage{graphicx}% Include figure files
\usepackage[FIGTOPCAP]{subfigure}
\usepackage{rotating}
\usepackage{epstopdf}
\usepackage{dcolumn}% Align table columns on decimal point
\usepackage{bm}% bold math
\usepackage{enumitem} 
\usepackage{makecell}
\usepackage{dblfloatfix}    % To enable figures at the bottom of page

\def\msun{{\mathrm{M}_\odot}}

\def\beq{\begin{equation}}
\def\eeq{\end{equation}}
\newcommand{\eq}[1]{\begin{align}#1\end{align}}
\newcommand{\subeq}[1]{\begin{subequations}\begin{align}#1\end{align}\end{subequations}}
\def\barr{\begin{eqnarray}}
\def\earr{\end{eqnarray}}
\def\bali{\begin{align}}
\def\eali{\end{align}}
\def\bsub{\begin{subequations}}
\def\esub{\end{subequations}}

\def\mrem{m_{\rm rem}}

\def\mimbh{M_\mathrm{IMBH}}
\def\mgc{M_\mathrm{GC}}

\def\tdf{t_\mathrm{DF}}

\def\mmin{{m_\mathrm{min}}}
\def\mmax{{m_\mathrm{max}}}

\def\tgw{{t_\mathrm{GW}}}

\def\forb{f_\mathrm{orb}}

\def\mchirp{\mathcal{M}}

\begin{document}

\title{Binary intermediate-mass black hole mergers in globular clusters}

\author{Alexander Rasskazov\altaffilmark{1}, Giacomo Fragione\altaffilmark{2,3} and Bence Kocsis\altaffilmark{1}}
\affil{$^1$E\"otv\"os University, Institute of Physics, P\'azm\'any P. s. 1/A, Budapest, Hungary 1117}
\affil{$^2$Department of Physics \& Astronomy, Northwestern University, Evanston, IL 60208, USA}
\affil{$^3$Center for Interdisciplinary Exploration \& Research in Astrophysics (CIERA), Northwestern University, Evanston, IL 60208, USA}

\begin{abstract}
We consider the formation of binary intermediate black holes (BIMBH) in globular clusters (GC), which could happen either in situ or due to the mergers between clusters. We simulate the evolution of the BIMBH orbit (and its subsequent merger) due to stellar ejections. We also take into account the evaporation of GCs due to the tidal field of the host galaxy and two-body relaxation. Our results show that if at least $10^{-3}$ of all GCs become BIMBH hosts and the BIMBH masses are $\sim1\%$ of the GC mass, at least one of the inspiralling (or merging) BIMBHs will be detected by LISA during its 4-year mission lifetime. Most of the detected BIMBHs come 1) from heavy GCs ($\gtrsim\SI{3e5}{\msun}$), as lower-mass GCs end up being disrupted before their BIMBHs have time to merge, and 2) from  redshifts $1<z<3$, assuming that most of GCs form around $z\sim4$ and given that the merger timescale for most BIMBHs is $\sim1$ Gyr. If the BIMBH to GC mass ratio is lower ($\sim10^{-3}$) but the fraction of BIMBH hosts among GCs is higher ($\gtrsim10^{-2}$), some of their mergers will also be detected by LIGO, VIRGO, and KAGRA and the proposed Einstein Telescope.
\end{abstract}

\section{Introduction}

While there is solid observational evidence for the existence of the supermassive (SMBHs, $M\gtrsim10^5\msun$) and stellar-mass black holes {\citep[SBHs, $10\msun\lesssim M\lesssim\bm{50}\msun$,][]{bhmass1,bhmass2}}, intermediate-mass black holes (IMBHs, $100\msun\lesssim M\lesssim10^5\msun$) remain elusive. Assuming the $M-\sigma$ relation \citep{m-sigma} can be extended down to IMBH masses, such objects may be hosted by globular clusters (GC). \citet{imbhBaumgardt} claimed there is evidence of a $\sim\num{4e4}\msun$ IMBH in $\omega$ Cen\footnote{{ It is likely, however, that $\omega$ Cen is not a canonical GC but rather a stripped dwarf galaxy \citep{omegacen}.}}, based on the observed velocity dispersion profiles, while recent analyses showed no evidence for an IMBH \citep{baum2019}. { \citet{Abbate2019}, using the measured pulsars' accelerations in M62, have found a central excess of mass in the range $[1200,6000]\,\msun$ (which could be an IMBH or a system of stellar mass dark remnants).} However, there is a growing evidence that some X-ray sources \citep{xray1,xray2} and tidal disruption events \citep{imbhTDE,peng2019} can be explained by accreting IMBHs. In particular, a tidal disruption of a white dwarf could only be visible for IMBHs of mass $\lesssim 2 \times 10^5\msun$, since for larger masses the Schwarzschild radius would be larger than the tidal disruption radius of a typical white dwarf \citep{imbhWD2009,imbhWD2016}. {Furthermore, empirical correlations between the quasiperiodic oscillation (QPO) frequencies and black hole masses show tentative evidence for IMBHs \citep{wu2016}.} IMBHs can lurk in galactic nuclei as well. For example, a GC can spiral due dynamical friction into the galactic nucleus, carrying its IMBH close enough to the SMBH so that the two could form a binary \citep*{GCinspiral1,GCinspiral2}. In the center of the Milky Way, the existence of such an IMBH still has not been ruled out \citep{gual2010,DEGN}, and it could be the origin of some of the observed hypervelocity stars \citep{yut03,Rasskazov2019}. {If present, they may be responsible for disrupting stellar and black hole binaries in galactic centers \citep{Deme2019}.} However, there is still no direct compelling dynamical evidence for IMBHs in the innermost region of the Galactic center \citep[e.g.][]{sch2018}.
{Recently more than ten low luminosity active galactic nuclei (AGN) have been identified to harbor IMBHs \citep{Zolotukhin2018,Mezcua2018}. Recent studies of dynamical and accretion signatures alike point to a high fraction of low mass galaxies hosting IMBHs \citep[see][for a review]{Greene2019}.}

Several IMBH formation mechanisms have been suggested: the collapse of a supermassive star produced via runaway mergers of massive stars in the dense center of a GC  \citep{Portegies_Zwart_2002,freitag2006,giers2015}; repeated mergers of stellar-mass BHs \citep{miller2002}; direct collapse of gas or Pop III stars in the early Universe \citep{Madau_2001,Whalen_2012,Woods_2017,Tagawa2019}; fragmentation of SMBH accretion disks \citep{mckernan2012,mckernan2014}.

{In this paper we examine the possibility that}
IMBHs can form a binary IMBH (BIMBH). \citet{gurkan2006} found that in sufficiently dense and/or centrally concentrated clusters with primordial binary fractions higher than $\sim10\%$, two supermassive stars generally form instead of one. It is also possible for an IMBH to dynamically form a binary with a stellar-mass BH \citep{Mapelli2016,frag2018b,mocca}. Alternatively, BIMBH can be the product of a merger between two star clusters \citep{asf2006}, which is followed by the BIMBH sinking to the center of the resulting merger cluster due to dynamical friction. Whatever scenario is invoked to produce a BIMBH in a star cluster, its orbit starts shrinking soon after its formation due to stellar ejections, similarly to the well-known case of binary SMBHs in galactic nuclei \citep{Quinlan1996,Sesana2006,sesana2015,Rasskazov2017}. Eventually, the BIMBH semimajor axis becomes small enough for the gravitation wave (GW) emission to take over, which quickly leads to the merger of two IMBHs. 

A way to detect such mergers is via GWs emitted by them during inspiral and merger. Merging BIMBHs can be detected by future space-borne observatories such as LISA \citep{lisa} or the proposed ground-based GW observatory Einstein Telescope \citep{ET}. Mergers of sufficiently low-mass IMBHs ($\lesssim 10^3 \msun$), at frequencies $\gtrsim\SI{10}{Hz}$, can also be detected by Advanced LIGO, VIRGO, and KAGRA at design sensitivity. Using the non-detection of massive binaries in the first two LIGO observational runs, \citet{ligoIMBH} placed upper limits on merging BIMBHs. For instance, they derived an upper limit of $0.2$ \si{Gpc^{-3}yr^{-1}} on the merger rate of $100\msun+100\msun$ binaries. \citet{fregeau2006} estimated tens of inspirals/yr detected by LISA and $\sim 10$ mergers and ringdowns/yr detected by LIGO , assuming $10\%$ of all GCs form BIMBHs. Using similar assumptions, \citet{gair2011} predicted from a few to a few thousand detection events for ET. Modelling the waveform of GW emission and assuming $\sim 40$ times lower GC formation rate, \citet{santamaria} estimated $\sim 1$ event/yr for LIGO and $\sim20$ events/yr for ET.

In this paper, we calculate the rate of BIMBH detections by various present and upcoming GW observatories: LISA, LIGO and ET. We improve on the previous papers by including the dynamical evolution of BIMBHs due to stellar ejections, and also taking into account the evaporation of GCs caused by stellar relaxation and tidal stripping in their host galaxy potential. The central density and velocity dispersion values of GCs, necessary to model the stellar ejections, are taken from the observed GC catalogue \citep{harris}. We examine how future GW instruments will be able to constrain the values of two parameters -- the fraction of GCs hosting BIMBHs and the BIMBH mass/GC mass ratio.

This paper is organized as follows. In Section~\ref{section:dynamics}, we describe the dynamics of BIMBHs in the core of star clusters, which we couple with the GC evolution in the host galaxy across the cosmic time. In Section~\ref{section:gw} we calculate the masses, the rates and the number of IMBHs that can detected by LISA, LIGO and ET assuming circular orbits. In Section~\ref{sec:eccentricity} we extend our analysis to eccentric orbits and in Section~\ref{sec:kick+spin} we discuss the case of velocity and spin of merger remnants. Finally, we summarize our results and discuss the implications of our findings in Section~\ref{section:conclusions}.

\section{Joint evolution of binary IMBHs and GCs}
\label{section:dynamics}

We assume the BIMBH to be in the center of the GC, neglecting its Brownian motion, which could possibly increase the hardening rate \citep{Chatterjee2003,Bortolas2016}. However, sufficiently massive BIMBH are expected to have a small wandering radius compared to the size of the host cluster core \citep{Chatterjee2003}. After their formation, BIMBHs harden due to 3-body interactions with stars in the GC core and later due to GW emission
\subeq{\label{eq:dadt}
\dv{a}{t} &= -\frac{a}{t_h(a)} + \qty(\dv{a}{t})_{\rm GR},\\
t_h(a) &= \frac{\sigma}{HG\rho_c a},\label{eq:t_h(a)}\\
\qty(\dv{a}{t})_{\rm GR} &= - \frac{64}{5}\frac{q}{(1+q)^2}\frac{G^3\mimbh^3}{c^5a^3},\label{eq:dadtgr} 
}
where $\rho_c$ is the GC central density, $\sigma$ is its central velocity dispersion, $\mimbh$ and $q=M_{\rm IMBH,2}/M_{\rm IMBH,1}\leq1$ are the total mass and mass ratio of the IMBH binary, {and $H$ is a dimensionless constant that varies within 15 and 20 depending on BIMBH mass ratio and eccentricity \citep{Sesana2006,Rasskazov2019}.} 
For simplicity, we first assume all BIMBH to form at nearly zero eccentricity and with $a=a_h$, i.e. at the hard-binary separation semi-major axis (see Section~\ref{sec:eccentricity} for the eccentric case), 
\eq{\label{eq:a_h}
a_h = \frac{GM_{\rm IMBH,2}}{\sigma^2},
}
where $M_{\rm IMBH,2}$ is the mass of the smaller IMBH. The exact initial value of $a$ is not important as long as it is $\lesssim a_h$: the hardening timescale increases as the binary shrinks (Eq.~\ref{eq:t_h(a)}), and as a result the binary spends most of the time at the shortest $a$ where the stellar ejections still dominate the IMBH binary evolution.

\begin{figure}
	\subfigure{\includegraphics[width=0.48\textwidth]{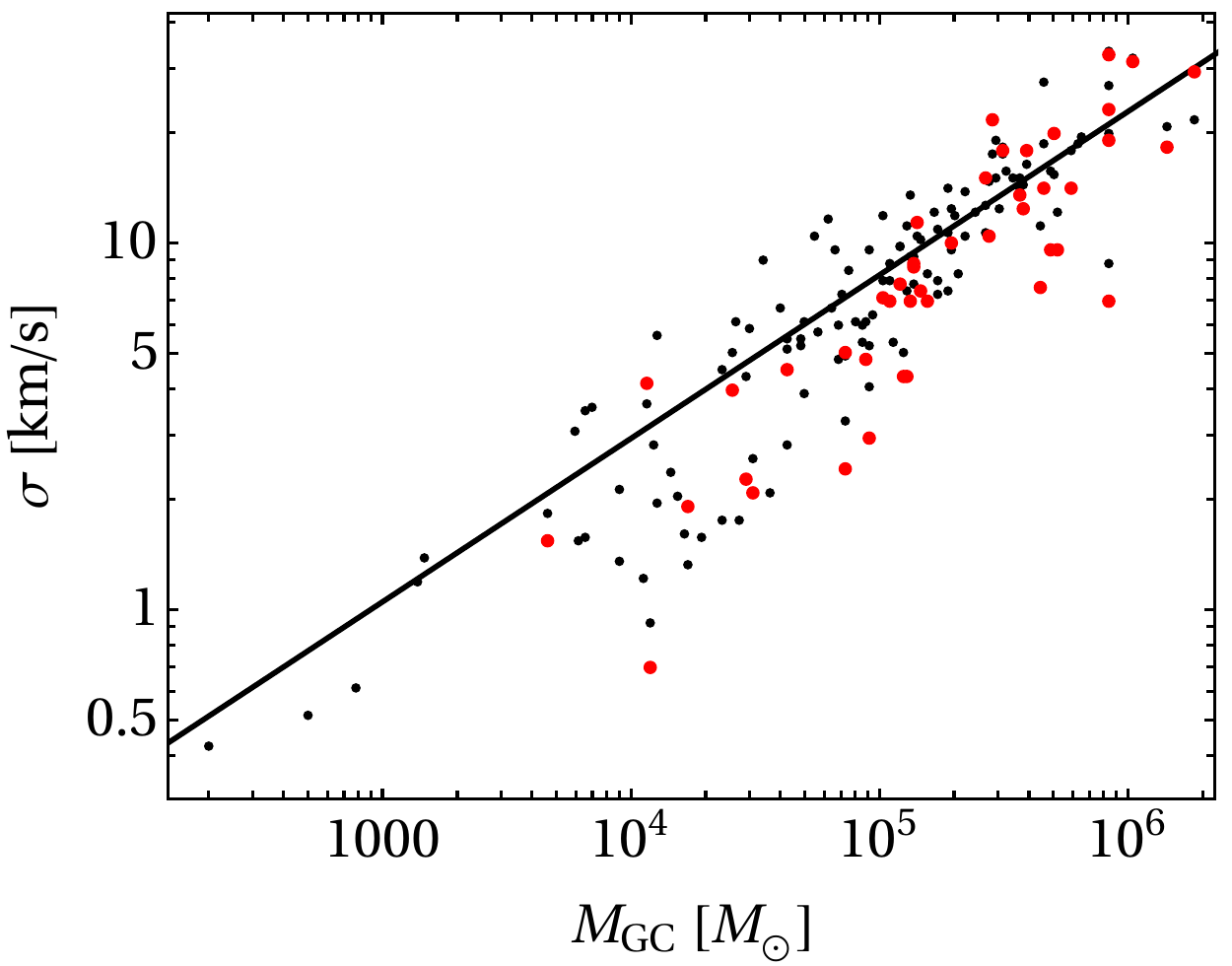}}
\caption{
Central velocity dispersion vs. the total GC mass and its best power-law fit. Black: $\sigma$ calculated according to \eqref{eq:sigma}, red: observed $\sigma$ from the Harris catalogue.
}
\label{fig:sigma}
\end{figure}

{ In our model, we neglect the recoil kicks imparted on the BIMBH during binary-single interactions, since they are typically insufficient to eject the whole BIMBH from the GC \citep[e.g.][]{ANTORASIO2016}. We also neglect the possible mergers with stellar mass BHs and the corresponding GW recoil kicks, which may be relevant for small IMBH masses \citep{frag2018b}.}

\begin{figure}
	\subfigure{\includegraphics[width=0.48\textwidth]{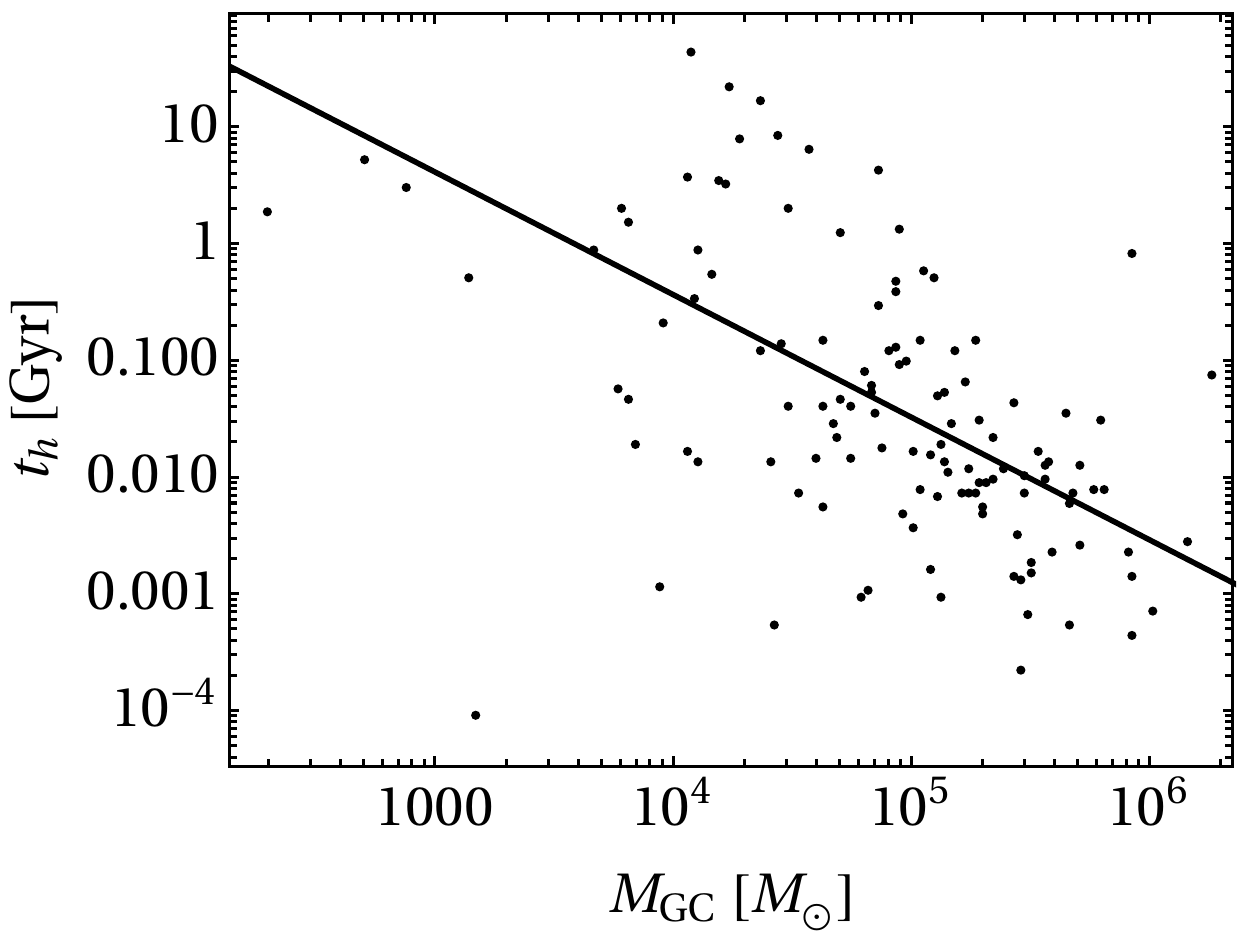}}
\caption{
Hardening timescale at $a=\SI{1}{mpc}$ as a function of the total GC mass. The solid line represents the best power-law fit to data.
}
\label{fig:th}
\end{figure}

We calculate $t_h$ for a GC of a given mass using the Harris catalogue of observed GC parameters \citep{harris}, where we have removed all GCs marked as core-collapsed. We convert the GC V-band luminosity into mass using mass-to-light ratio $1.7$ \citep{harris2017}. We calculate the central density assuming the density distribution introduced in \citet{GCdensity},
\eq{
\rho(r) = \frac{\rho_c}{(1+r^2/r_c^2)(1+r^2/r_h^2)},
\label{eqn:rhor}
}
where $r_c$ is the core radius and $r_h$ is the half-mass radius ($r_c\ll r_h$). Rewriting Eq.~\eqref{eqn:rhor} in terms of the total cluster mass $\mgc$
\eq{
\rho_c = \frac{\mgc(r_h+r_c)}{2\pi^2r_c^2r_h^2},
}
while the central velocity dispersion can straightforwardly be computed as
\eq{\label{eq:sigma}
\sigma \approx \sqrt\frac{6G\mgc(\pi^2/8-1)}{\pi r_h}
} 
The observed values of $\sigma$ are available only for a small fraction of all GCs. Therefore, we use Eq.~\eqref{eq:sigma} to estimate the cluster central velocity dispersion. As shown in Fig.~\ref{fig:sigma}, our estimated values are in good agreement with the Harris catalogue. Moreover, we find that there is a correlation between $\sigma$ and $\mgc$ (Fig. \ref{fig:sigma}),
\eq{
\sigma &= \SI{8.18}{km\ s}^{-1} \qty(\frac{\mgc}{10^5\msun})^{0.446},
}
and, using Eq.~\ref{eq:t_h(a)}, a weaker one between $t_h$ and $\mgc$ (Fig.~\ref{fig:th})\footnote{ To avoid confusion note that $t_h$ should not be confused with the GC half-mass relaxation time. Throughout this paper $t_h$ refers to the hardening timescale Eq.~(\ref{eq:t_h(a)}).} 
\eq{\label{eq:t_h(mgc)}
t_h = \SI{32.6}{Myr} \qty(\frac{\mgc}{10^5\msun})^{-1.05} \qty(\frac{a}{\SI{1}{mpc}})^{-1}.
}

{ It is not clear how much of the dispersion in Fig.~\ref{fig:th} is intrinsic and how much is due to measurement errors\footnote{ Harris catalogue does not have uncretainties of both GC luminosities and characteristic radii.}. Therefore, we decide to neglect the intrinsic dispersion in the scaling relation between $t_h$ and $\mgc$. Moreover, we use the present-day $t_h$--$\mgc$ relation, ignoring its possible evolution throughout the GC lifetime, and assume that the hardening timescale depends only on the GC mass.}

We use the above expression for $\sigma$ to calculate $a_h$ at the beginning of the BIMBH evolution (Eq.~\ref{eq:a_h}), and the one for $t_h$ to calculate the value of $t_h(a)$ parameter in Eq.~\eqref{eq:dadt}, during the BIMBH evolution. Simultaneously with the BIMBH evolution, we also track the evolution of the host $\mgc$. During the GC lifetime, its mass decreases due to stellar evolution, as well as the dynamical ejection of stars via two-body relaxation and stripping by a host galaxy's tidal field \citep{GOT}
\eq{\label{eq:dmgcdt}
\dv{\mgc}{t} = -\mgc \qty(\frac{1}{t_{\rm ev}}+\frac{1}{t_{\rm iso}}+\frac{1}{t_{\rm tid}}).
}

Following \citep{GOT}, we calculate the GC mass loss via stellar evolution assuming all stars were born coeval following a \citet{KroupaIMF} initial stellar mass function, in the range $0.1\msun$--$100\msun$
\beq\label{eq:KroupaIMF} 
f_\mathrm{IMF} (m) \propto \begin{dcases} 
2\left(\frac{m}{\msun}\right)^{-1.3}\qc m<0.5\msun\,,\\
\left(\frac{m}{\msun}\right)^{-2.3}\qc m>0.5\msun.
\end{dcases}
\eeq
We adopt a main-sequence lifetime 
\begin{equation}\label{eq:tMS}
t_{\rm MS} = 10^{10} \qty(\frac{m}{\msun})^{-2.5} \si{yr}
\end{equation}
and an initial-to-final mass function for stellar remnants \citep{DEGN}\footnote{We are neglecting the details of stellar evolution models, which may depend on the stellar metallicity, winds, and rotation \citep{hur00}.}
\eq{
\mrem &= \begin{dcases}
0.109m+0.394\msun, & m<8\msun \\
1.4\msun, & 8\msun<m<30\msun \\
0.1m, & 30\msun<m
\end{dcases}.
}
All the mass lost via stellar winds and supernova explosions is assumed to leave the cluster. In particular, the stars with masses between $m(t+\dd{t})$ and $m(t)$ end their main-sequence life during the time span $t$ and $t+\dd{t}$, where $m(t)$ is the mass of the heaviest stars that are still on the main sequence at time $t$, {the inverse function of Eq.~\eqref{eq:tMS}}. Thus, the relative change in the GC mass
\eq{
\frac{\dd{\mgc}}{\mgc} = \frac{f_\mathrm{IMF}(m(t)) \qty[m(t)-\mrem(m(t))] |\dv*{m(t)}{t}| \dd{t} }{\int_\mmin^{m(t)} f_\mathrm{IMF}m\dd{m} + \int_{m(t)}^\mmax f_\mathrm{IMF}\mrem(m)\dd{m}},
}
which gives a characteristic timescale for the stellar evolution-induced mass loss
\eq{
t_{\rm ev} &= \frac{\mgc}{\qty|\dv*{\mgc}{t}|} \nonumber\\
&= \frac{\int_\mmin^{m(t)} f_\mathrm{IMF}m\dd{m} + \int_{m(t)}^\mmax f_\mathrm{IMF}\mrem(m)\dd{m}}{f_\mathrm{IMF}(m(t)) \qty[m(t)-\mrem(m(t))] |\dv*{m(t)}{t}|}
}
In the absence of the other mass loss mechanisms, it leads to the GC losing up to $\sim 40\%$ of it mass in $\SI{10}{Gyr}$.

The dynamical mass loss is modeled using the prescription for $t_{\rm iso}$ from \citet{GOT}
\eq{\label{eq:tiso}
t_{\rm iso} &\approx \SI{17}{Gyr} \frac{\mgc}{\num{2e5}\msun},
}
and for $t_{\rm tid}$ from \citet{millisecondPulsars}
\subeq{
t_{\rm tid} &\approx \SI{10}{Gyr} \qty(\frac{\mgc}{\num{2e5}\msun})^{2/3} P(r),\\
P(r) &= 100 \qty(\frac{r}{\SI{1}{kpc}}) \qty(\frac{V_c(r)}{\SI{1}{km {s}^{-1}}})^{-1}.
}
In the previous equation, $r$ is the GC distance from the center of its host galaxy, while $V_c(r)$ and $P(r)$ are the circular velocity of the galaxy and the rotation period, respectively.

We include the effect of dynamical friction on GC orbit \citep{BinneyTremaine}
\subeq{
\dv{r^2}{t} &= -\frac{r^2}{\tdf(r,\mgc)},\\
\tdf &= \SI{0.45}{Gyr} \qty(\frac{r}{\SI{1}{kpc}})^2 \frac{V_c(r)}{\SI{1}{km/s}} \qty(\frac{\mgc}{10^5\msun})^{-1} f_\epsilon,
}
where $f_\epsilon$ is the correction for eccentricity of cluster orbits, for which we assume $f_\epsilon=0.5$, following \citet{GOT}.

The initial positions of GCs are chosen to map the host galaxy stellar density distribution. In our calculations, we assume the host galaxies to be a Milky Way-like galaxy, with a spherical Sersic profile of total mass $M_\ast=\num{5e10}\msun$, effective radius $r_e = \SI{4}{kpc}$ and concentration index $n_s=2.2$ \citep{GOT}. We also include a dark matter halo with a Navarro-Frenk-White profile \citep{NFW} having mass $M_h=10^{12}\msun$ and scale radius $r_s=\SI{20}{kpc}$.

\begin{figure}
	\subfigure{\includegraphics[width=0.48\textwidth]{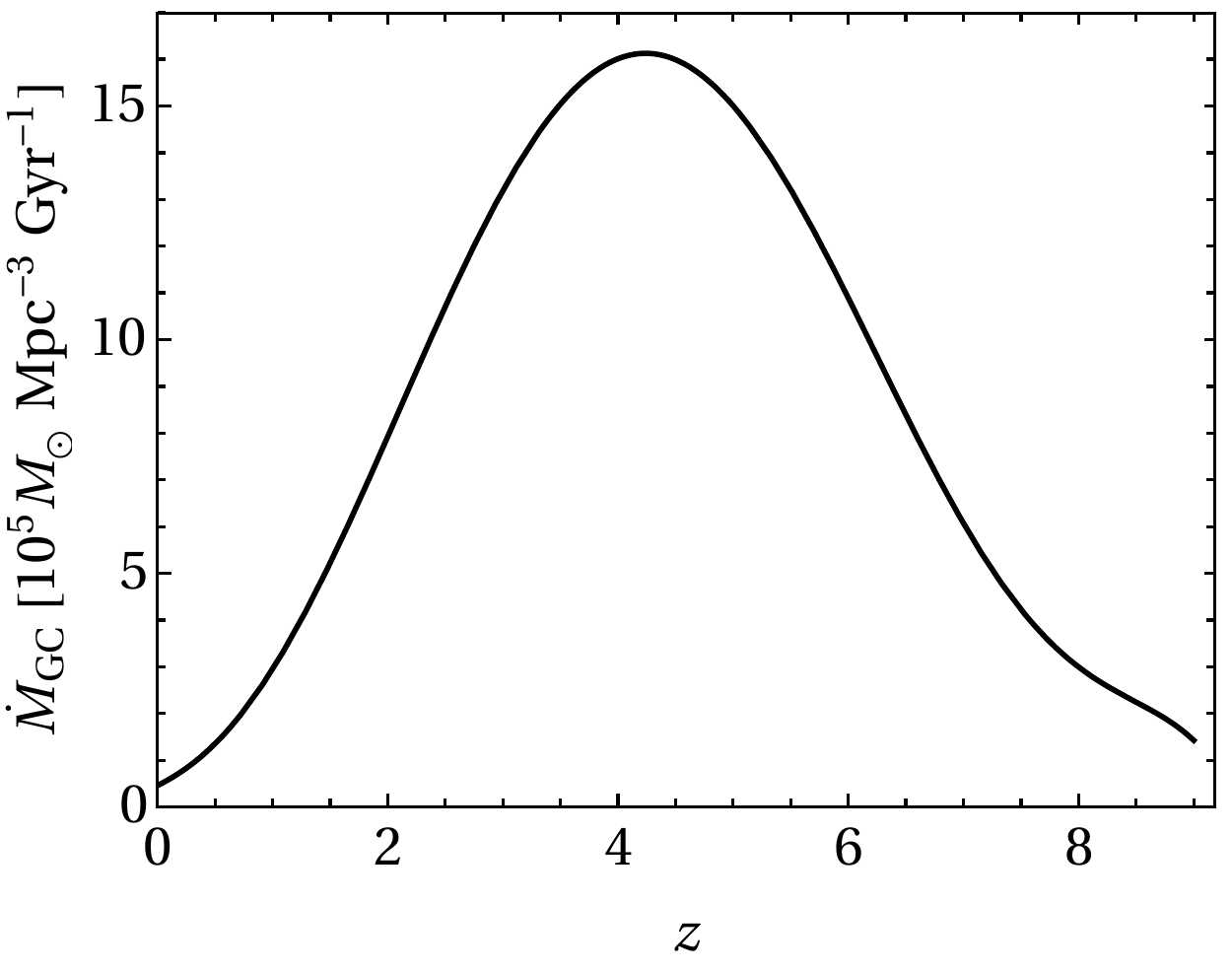}}
\caption{
GC formation rate as a function of redshift, taken from \citet{ElBadry} and \citet{Rodriguez}.
}
\label{fig:formationRate}
\end{figure}

Finally, we adopt the GC formation rate described in \citet{Rodriguez} and \citet{ElBadry}, illustrated in their Fig.~\ref{fig:formationRate}. The GC masses are sampled from a power-law distribution, as observed for young massive clusters in nearby galaxies \citep{gcmass1,gcmass2,gcmass3,gcmass4,gcmass5,gcmass6}
\eq{\label{eq:dndm}
\dv{N}{\mgc} \propto \mgc^{-\beta}\qc \mmin<\mgc<\mmax
}
with $\mmin=10^4\msun$, $\mmax=10^7\msun$, $\beta=2$. { It is worth noting that GCs might actually form with heavier initial masses due to multiple stellar populations \citep{gratton2012}}.

\section{Demographics of merging binary intermediate-mass black holes}

The inclusion of GC disruption has a significant effect on the demographics of merging BIMBH. For most of the low-mass GCs the hardening timescales are long enough and GC disruption timescales are short enough, that the GC ends up getting disrupted before the BIMBH has time to coalesce (see Figure~\ref{fig:th} and Eq.~\ref{eq:tiso}). This is mainly due to the isolated 2-body relaxation term $t_{\rm iso}$ in Eq.~\eqref{eq:dmgcdt},which dominates GC evaporation at low $\mgc$.

More generally, there are two categories of GCs in our model: 
\begin{enumerate}
\item Heavy GCs, where the BIMBH merges before the cluster is disrupted. For every such GC we record the merger time. For any fixed GC mass and IMBH masses, that time depends only on the initial radius of GC's orbit in the galaxy $r_0$ (as it affects the strength of tidal stripping). In what follows, we use the merger times averaged over all randomly chosen values of $r_0$.
\item Low-mass clusters, which are disrupted before their BIMBHs can merge. It is possible that after the GC disruption the IMBHs are close enough to merge due to GW emission, but those cases are rare and do not significantly affect the merger rate.
\end{enumerate}

\begin{figure}
	\subfigure{\includegraphics[width=0.48\textwidth]{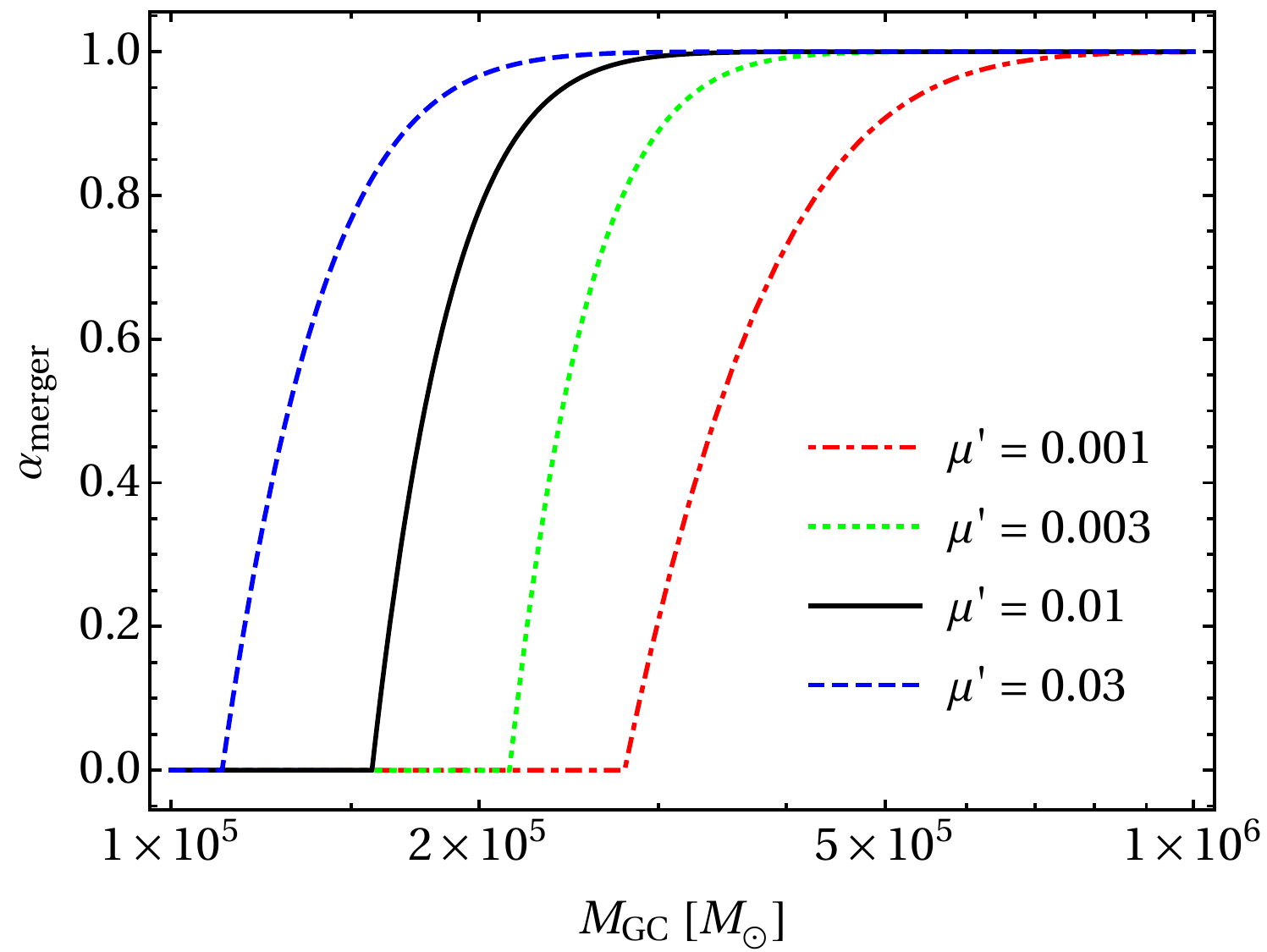}}
\caption{
The fraction of GCs where the BIMBH has time to merge before the GC is disrupted. 
Dashed blue, solid black, dotted green and dot-dashed red lines are for $\mu' = [4q/(1+q)^2]^{1/3}\mimbh/\mgc = 0.03,\,0.01,\,0.003,\,0.001$, respectively.
}
\label{fig:mergerFraction}
\end{figure}

Fig.~\ref{fig:mergerFraction} shows the fraction of GCs where the BIMBH has time to merge before the GC is disrupted, for $q=1$ and various values of the ratio between BIMBH mass and GC mass $\mu = \mimbh/\mgc$. All the BIMBH mergers come from GCs heavier than $\sim \SI{2e5}{\msun}$; 
the exact number depends on $\mu$ as heavier IMBHs shrink faster due to GW emission and the hardening timescale does not depend on $\mu$ for hard binaries (Eq.~\ref{eq:dadt}).

\begin{figure}
	\includegraphics[width=0.48\textwidth]{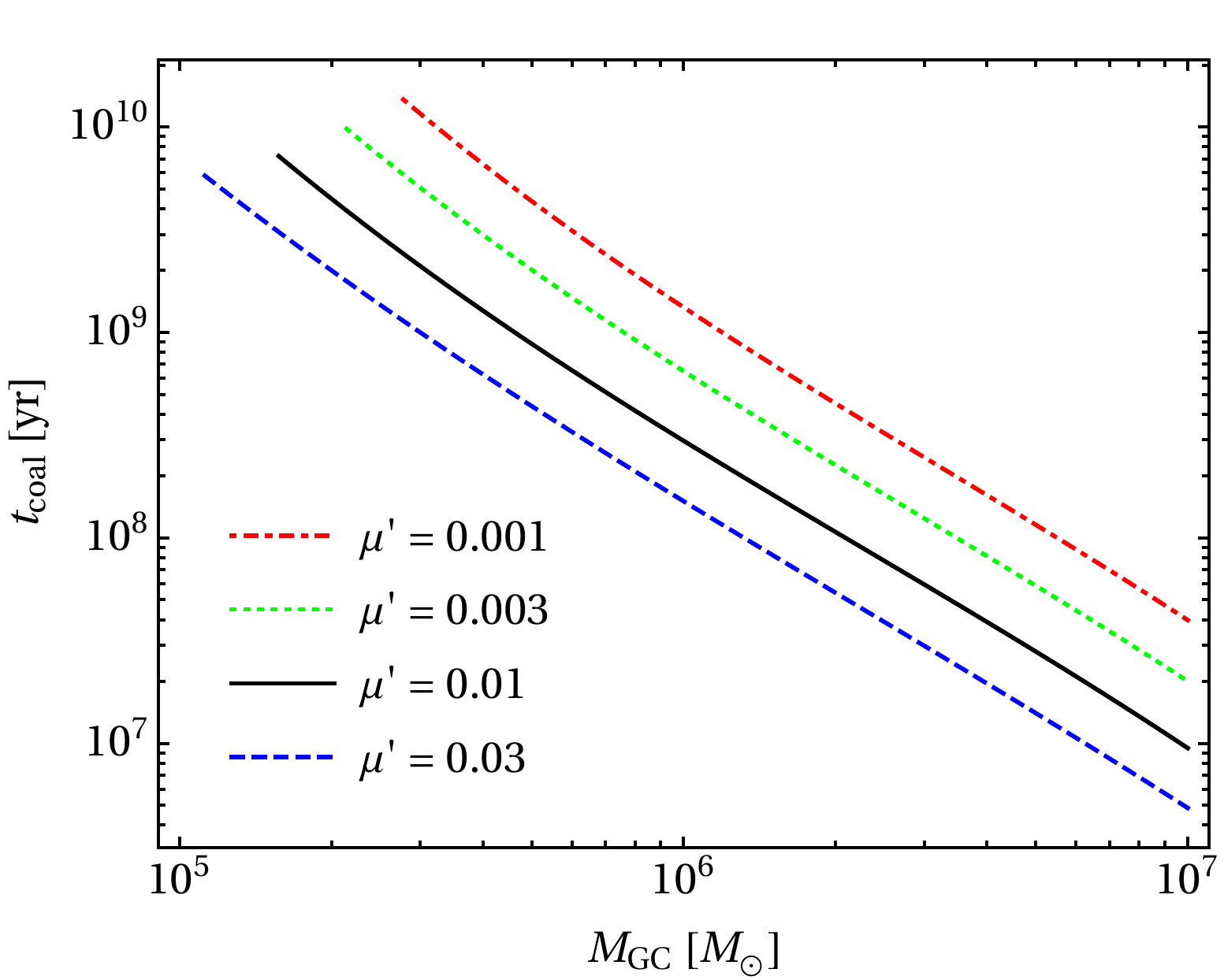}
\caption{
BIMBH coalescence timescales (from binary formation at $a=a_h$ to coalescence). Line styles are as in Fig.~\ref{fig:mergerFraction}.
}
\label{fig:tcoal}
\end{figure}

Fig.~\ref{fig:tcoal} reports the time from BIMBH formation to coalescence $t_{\mathrm{coal}}$ for various initial GC and BIMBH masses. As discussed above, $t_{\mathrm{coal}}$ decreases for heavier GCs at fixed $\mu$ and for heavier IMBHs. The cutoff at low GC masses corresponds to coalescence time always being longer than the GC disruption time. 
The figure shows that coalescence always occurs in less than a Hubble time. 

{ However, that conclusion relies on the assumption that the loss-cone depletion is insignificant and therefore the BIMBHs are not affected by the ``final parsec problem'' which was shown to stall the hardening of supermassive BHs in spherical galaxies \citep{Begelman1980,Vasiliev2015,Bortolas2016}. We consider this assumption to be reasonable 
as 2-body relaxation time in GC centers is rather short and therefore the loss-cone repopulates quickly.}

All these results assume $q=1$ for all BIMBHs. 
A lower $q$ would decrease the GW timescale $\dv*{a}{t}$ in Eq.~\eqref{eq:dadt}. However, setting $q<1$ would be equivalent to reducing $\mu$ at $q=1$. Indeed, Eq.~\eqref{eq:dadtgr} can be rewritten as 
\eq{\label{eq:da/dt}
\qty(\dv{a}{t})_{\rm GR} &= - \frac{64G^3\mgc^3}{5c^5a^3}\mu^3\eta,
}
where $\eta=q/(1+q)^2$. At the same time, neither $\mu$ nor $\eta$ enter the stellar ejection term (Eq.~\ref{eq:t_h(a)}); there is a dependence of $H$ on $q$, but it is rather weak for hard binaries \citep{Sesana2006}. Also $a_h$ depends on $\mu$ and $q$, however, as was mentioned above, its value is not too important. 
Therefore, all the results in this section remain valid for any $q$ (the same for all binaries) if we replace $\mu$ with $\mu'=\mu\,(4\eta)^{1/3}$. 

\section{Gravitational Wave signal}
\label{section:gw}

In this Section, we estimate the number of merging BIMBHs that can be observed by present and upcoming instruments as a function of their parameters including mass, mass ratio, and singal to noise ratio.

LISA is only sensitive to GW frequencies higher than $f_{\rm min}\sim 10^{-5}$ Hz \citep{robson} where all IMBHs are in gravitational radiation-dominated regime already.
To determine which of the BIMBHs are detectable, we first select the binaries that radiate at $f_{\rm min}\gtrsim 10^{-5}$ Hz. Given their formation rate and merger timescales, the number of BIMBHs formed in the GCs with initial masses $\qty[\mgc,\, \mgc+\Delta\mgc]$ and observed at the redshift interval $\qty[z,\, z+\Delta z]$ which radiate at frequencies $f>f_{\rm min}$ is
\eq{\label{eq:observed}
&\frac{\dd^2{N_\mathrm{obs}}}{\dd{z}\dd{\mgc}} \Delta z \Delta\mgc = 
\alpha_{\rm IMBH} \, \alpha_{\rm merger}(\mgc) \nonumber\\
&\quad\times \frac{1}{\langle\mgc\rangle} \frac{\partial^2{\mgc}}{\partial{t}\partial{V}}\Bigg|_{t=t(z)-t_\mathrm{coal}(\mgc)} \dv{V}{z}\Delta z  \nonumber\\
&\quad\times \dv{N}{\mgc}\Delta\mgc \, t_{\rm GW}(f_{\rm min}(1+z),\,\mimbh,\,q).
}
Here we have assumed a fixed ratio between BIMBH total mass $\mimbh$ and its (initial) host GC mass $\mgc$. The variables are as follows:
\begin{itemize}
\item $\alpha_{\rm IMBH}$ is the fraction of GCs forming BIMBHs,
\item $\alpha_{\rm merger}(\mgc)$ is the fraction of GCs of mass $\mgc$ where the BIMBH reaches coalescence before Hubble time, 
\item $\dv{N}{\mgc}$ is the initial GC mass distribution (Eq.~\ref{eq:dndm}), normalized so that $\int_\mmin^\mmax \dv{N}{\mgc} \dd{\mgc} = 1$,
\item $\langle\mgc\rangle = \int_\mmin^\mmax \dv{N}{\mgc} \mgc \dd{\mgc}$ is the mean initial GC mass,
\item $\frac{\partial^2{\mgc}}{\partial{t}\partial{V}}$ is the total mass of GCs forming per unit time per unit comoving volume (shown in Fig.~\ref{fig:formationRate}),
\item $t(z)$ is the time since the Big Bang,
\item $t_{\rm coal}$ is the time from BIMBH formation (which is assumed to happen at $a\approx a_h$) to coalescence,
\item $t_{\rm GW}(f,\,\mgc)$ is the time (in the source frame) required for GWs to bring to coalescence an BIMBH radiating at the observed GW frequency $f$ with total mass $\mimbh$ and mass ratio $q$. We are assuming all BIMBHs to be circular so that the radiation frequency is twice the orbital frequency  and  
\eq{
\tgw = \frac{5}{256} \frac{c^5}{\eta(G\mimbh)^{5/3}(\pi f)^{8/3}}.
}
\end{itemize}

Note that $\frac{1}{\langle\mgc\rangle} \frac{\dd{\mgc}}{\dd{t}\dd{V}}$ is the total {\it number} of GCs forming per unit time per unit comoving volume and $\dv{N}{\mgc}\Delta\mgc$ is the fraction of them that have masses between $\mgc$ and $\mgc+\Delta\mgc$. To account for non-negligible coalescence time, we calculate this formation rate at the redshift at which the BIMBH was formed, $z_{\rm form}$, rather than at the redshift at which it is observed, $z$. To do that, we calculate the cosmic time $t(z)$, corresponding to redshift $z$
\footnote{{ We have used $H=\SI{69.6}{km/s/Mpc}$, $\Omega_M=1-\Omega_\Lambda=0.29$ as the cosmological parameters \citep{cosmology}.}}
; then $t=t(z)-t_{\rm coal}$ is the cosmic time corresponding to $z_{\rm form}$. 
If $t<0$, then GCs of mass $\mgc$ do not produce observable BIMBHs at redshift $z$, and we exclude those particular $\{z,\mgc\}$ bins from our model.

To estimate the number of merging BIMBH observable by LISA, we use the analytic approximation for the LISA sensitivity from \citet{robson} assuming 4 year observation time.
% \begin{eqnarray}
% S_n(f) &=& \frac{10}{3L^2} \qty(P_{\rm OMS}(f) + \frac{4P_{\rm acc}(f)}{(2\pi f)^4}) \qty(1+\frac{6}{10}\qty(\frac{f}{f_\ast})^2) \nonumber\\
% &+& S_c(f),
% \end{eqnarray}
% where
% \subeq{
% L &= \SI{2.5}{Gm},\\
% f_\ast &= \frac{c}{2\pi L} = \SI{19.09}{mHz},\\
% P_{\rm OMS} &= (\SI{1.5e-11}{m})^2\,\mathrm{Hz}^{-1},\\
% P_{\rm acc} &= (\SI{3e-15}{m})^2 \qty(1+\qty(\frac{\SI{0.4}{mHz}}{f})^2)\,\mathrm{Hz}^{-1}.
% }
% Here $S_c$ is the galactic confusion noise from unresolved binaries which is well fit by the function
% \eq{
% S_c(f) &= Af^{-7/3}e^{-f^\alpha+\beta f\sin(kf)}[1+\tanh(\gamma(f_k-f))]\,\mathrm{Hz}^{-1},
% }
% where $A= \num{9e-45}$, $\alpha= 0.138$, $\beta= -221$, $k= 521$, $\gamma= 1680$, $f_k = 0.00113$, for a 4 year observation time.

Since all BIMBHs are in the GW-dominated regime in the LISA frequency band, the GW emission frequency of the merging binary evolves during the observation period as follows:
\eq{
\dv{f}{t} = \frac{96}{5} \frac{(G\mchirp)^{5/3}\pi^{8/3}f^{11/3}}{c^5}. \label{eq:dfdt}
}
This gives us the probability distribution of observed GW frequency in the LISA band: 
\eq{
\dv{N}{f}=\dv{N}{t}\qty(\dv{f}{t})^{-1}\propto f^{-11/3} \qc f>f_{\rm min}.\label{eq:dndf}
}
Eq.~\eqref{eq:dfdt} describes the emitted GW frequency, but since it differs from the observed one only by a factor of $1+z$, the same distribution applies to both of them. 

\begin{figure}
\includegraphics[width=0.48\textwidth]{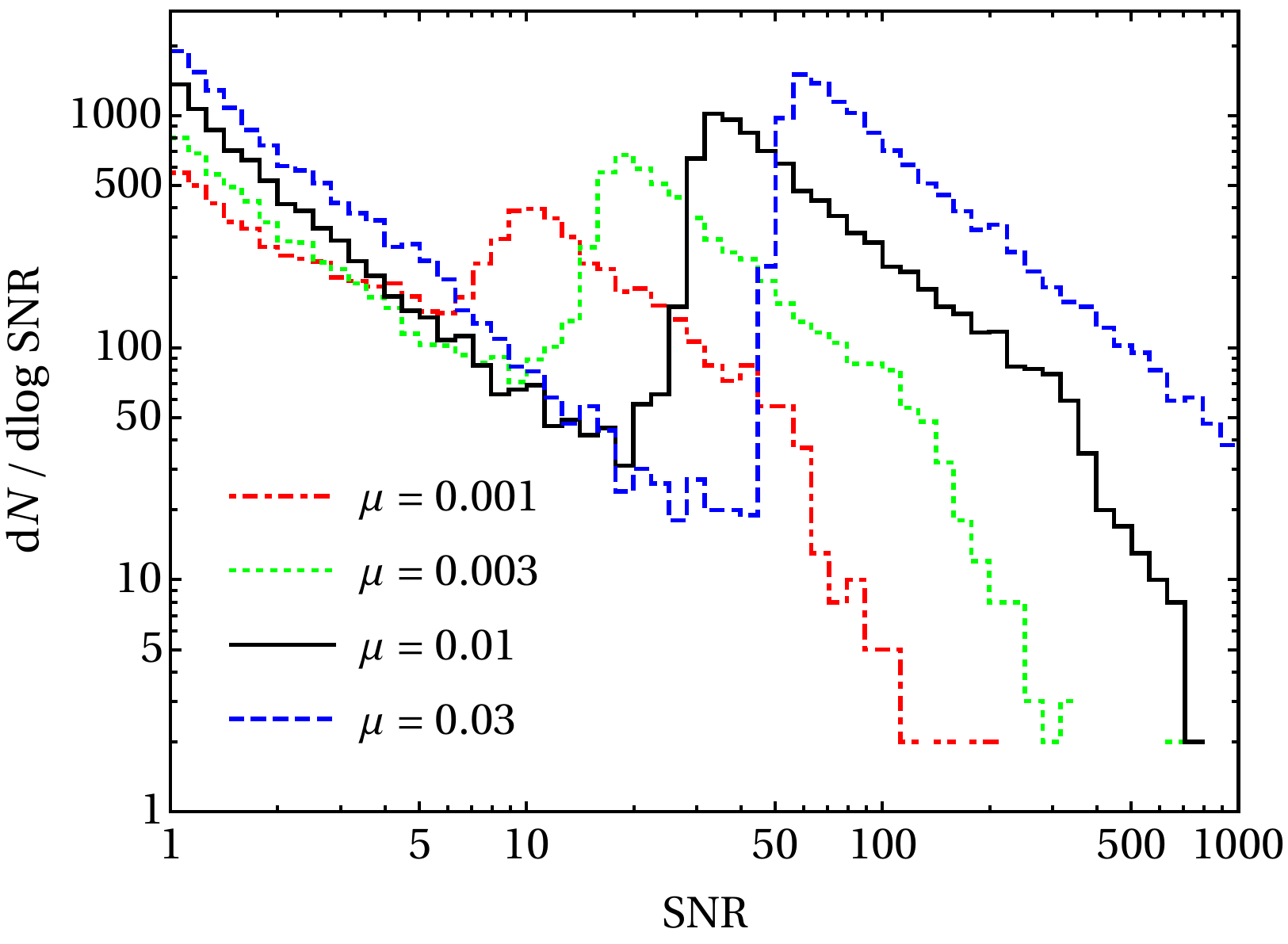}
\includegraphics[width=0.48\textwidth]{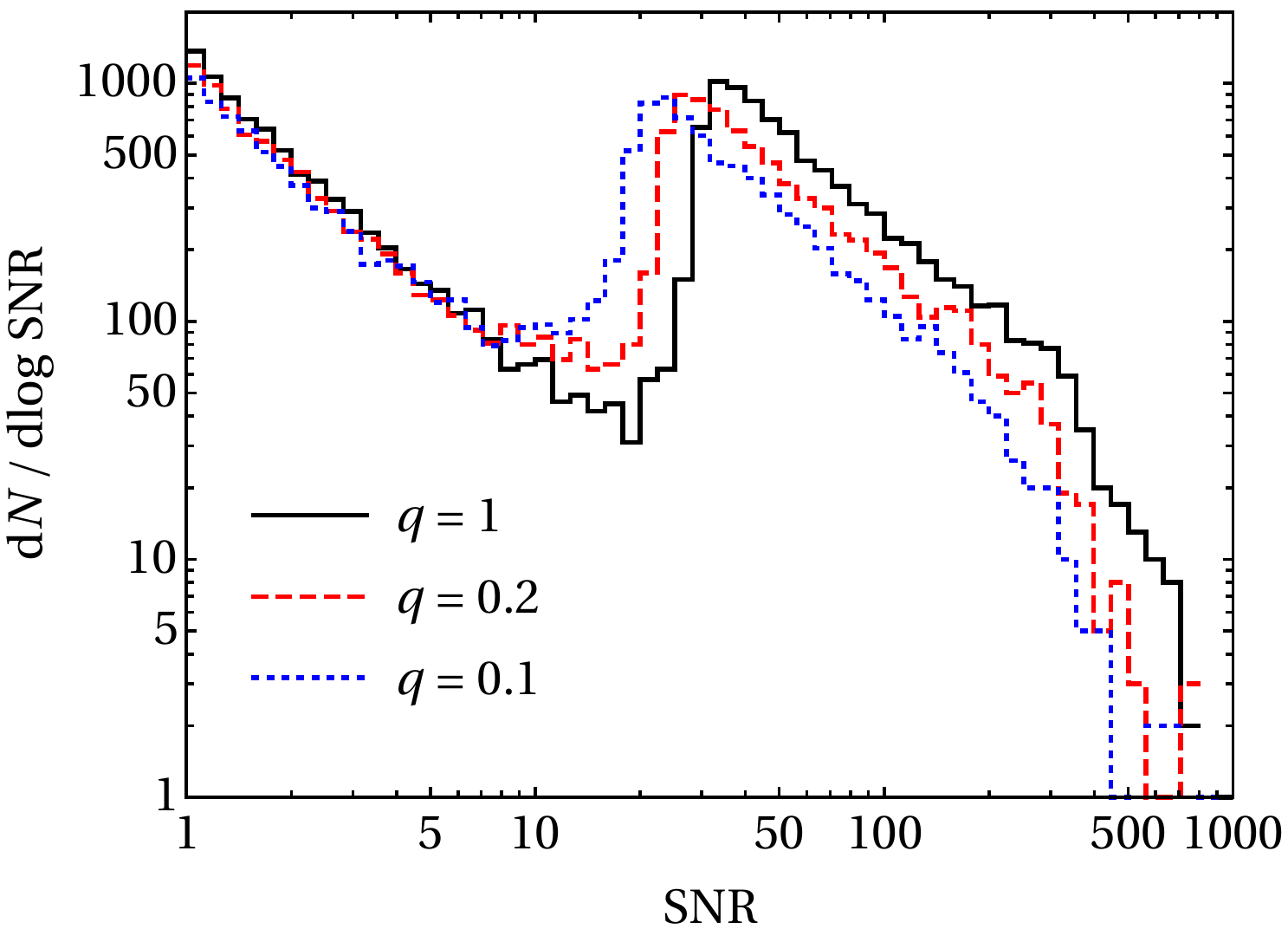}
\caption{The distribution of observed BIMBHs' SNR. Top: dashed blue, solid black, dotted green and dot-dashed red lines are for $\mimbh/\mgc = 0.03,\,0.01,\,0.003,\,0.001$, respectively, for equal binary component masses. Bottom: solid black, dashed red and dotted blue lines are for { $\bm{\mu=0.01}$ and} $q=1,\,0.2,\,0.1$, respectively.}
\label{fig:snr}
\end{figure}

After we have determined the number of observable (i.e. emitting in the LISA band) BIMBHs in a $\{z,\mgc\}$ bin from Eq.~\eqref{eq:observed}, we randomly assign a GW frequency to each of them, randomly selecting from the distribution \eqref{eq:dndf}. The frequency is increasing as the LISA during the LISA observation (Eq.~\ref{eq:dfdt}), which may lead to the binary merging before the LISA mission is over if the initial frequency is sufficiently close to merger. Assuming $\Delta f$ is the frequency change during the mission lifetime, we calculate the signal-to-noise ratio (SNR) $\kappa$ to determine whether that BIMBH would actually be detectable for an isotropic source direction and orientation \citep{robson}
\begin{equation}\label{eq:snr}
\kappa = 
\sqrt{\frac{16}{5}\int_f^{f+\Delta f} \frac{A^2(f)}{S_n(f)} \dd{f}},
\end{equation}
where $A(f)$ is the strain amplitude for the face-on orientation 
\begin{align}\label{eq:A}
A(f) =& \sqrt{\frac{5}{24}} \frac{(G\mchirp_z/c^3)^{5/6}f_0^{-7/6}}{\pi^{2/3}D_L/c}\times \nonumber\\
&\times
\begin{dcases}
(f/f_0)^{-7/6}\qc f<f_0\\
(f/f_0)^{-2/3}\qc f_0\leq f<f_1\\
w\mathcal{L}(f,f_1,f_2)\qc f_1\leq f<f_3.
\end{dcases}
\end{align}
In the previous equation,
\begin{subequations}
\begin{align}
f_k&\equiv \frac{a_k\eta^2+b_k\eta+c_k}{\pi(GM_z/c^3)},\\
\mathcal{L} &\equiv \qty(\frac{1}{2\pi}) \frac{f_2}{(f-f_1)^2+f_2^2/4},\\
w &\equiv \frac{\pi f_2}{2} \qty(\frac{f_0}{f_1})^{2/3}.
\end{align}
\end{subequations}
We take the values of $\{f_k,a_k,b_k,c_k\}$ from Table 2 in \citet{robson}. Whenever $\Delta f< 0.1f$, we use the first-order approximation
\eq{\label{eq:kappa_point}
\kappa &= \frac{h(f)}{\sqrt{S_n(f)}},
}
where
\eq{
h(f) &= \frac{8(G\mchirp_z/c^3)^{5/3}(\pi f)^{2/3}}{5^{1/2}D_L/c}T^{1/2}, 
}
$T$ is the mission lifetime, $\mchirp_z = M_z\eta^{3/5} = \mimbh(1+z)\eta^{3/5}$ is the BIMBH chirp mass in the detector frame and $D_L$ is its luminosity distance. 

The total number of BIMBHs observed by LISA with SNR $>8$ (standard detection threshold) can be well approximated as
\eq{
N_\mathrm{LISA} \approx 870 \,\alpha_{\rm IMBH} \qty(\frac{\mu}{0.01})^{0.39}.
}
Therefore, as long as BIMBHs form in $\gtrsim10^{-3}$ of all GCs, we should expect LISA to detect at least one of them, during its 4 year mission lifetime.

We show the distribution of the SNR's of merging BIMBHs in Fig.~\ref{fig:snr}. That distribution is bimodal: SNR $\gtrsim20\,(\mu/0.01)^{0.53}$ correspond to merging BIMBHs, while smaller SNR correspond to the BIMBHs observed at lower frequencies which do not merge during the LISA observation. The latter have predominantly low SNR due to observable BIMBHs being concentrated at low frequencies (Eq.~\ref{eq:dndf}) and the steep increase of noise $S_n(f)$ at low $f$.
At both low and high values of SNR ($\kappa$) the distribution is $\dv*{N}{\log\kappa}\propto\kappa^{-1.3}$. We note this is remarkably different from the expected $dN/d\ln \kappa \propto \kappa^{-3}$ distribution that we would observe for sources at low redshifts if their distribution is homogeneous in space 
(as observed for the LIGO stellar mass BH binaries in the O1 and O2 observing runs) 
\footnote{Assuming all BH binaries have the same parameters and are only observed at low redshifts, SNR depends only on their distance from Earth $r$. Then $\kappa\propto1/r$ and $\dv*{N}{\kappa}\propto\dv*{N}{r}\dv*{r}{\kappa}\propto r^4\propto\kappa^{-4}$.}.

\begin{figure}
\includegraphics[width=0.48\textwidth]{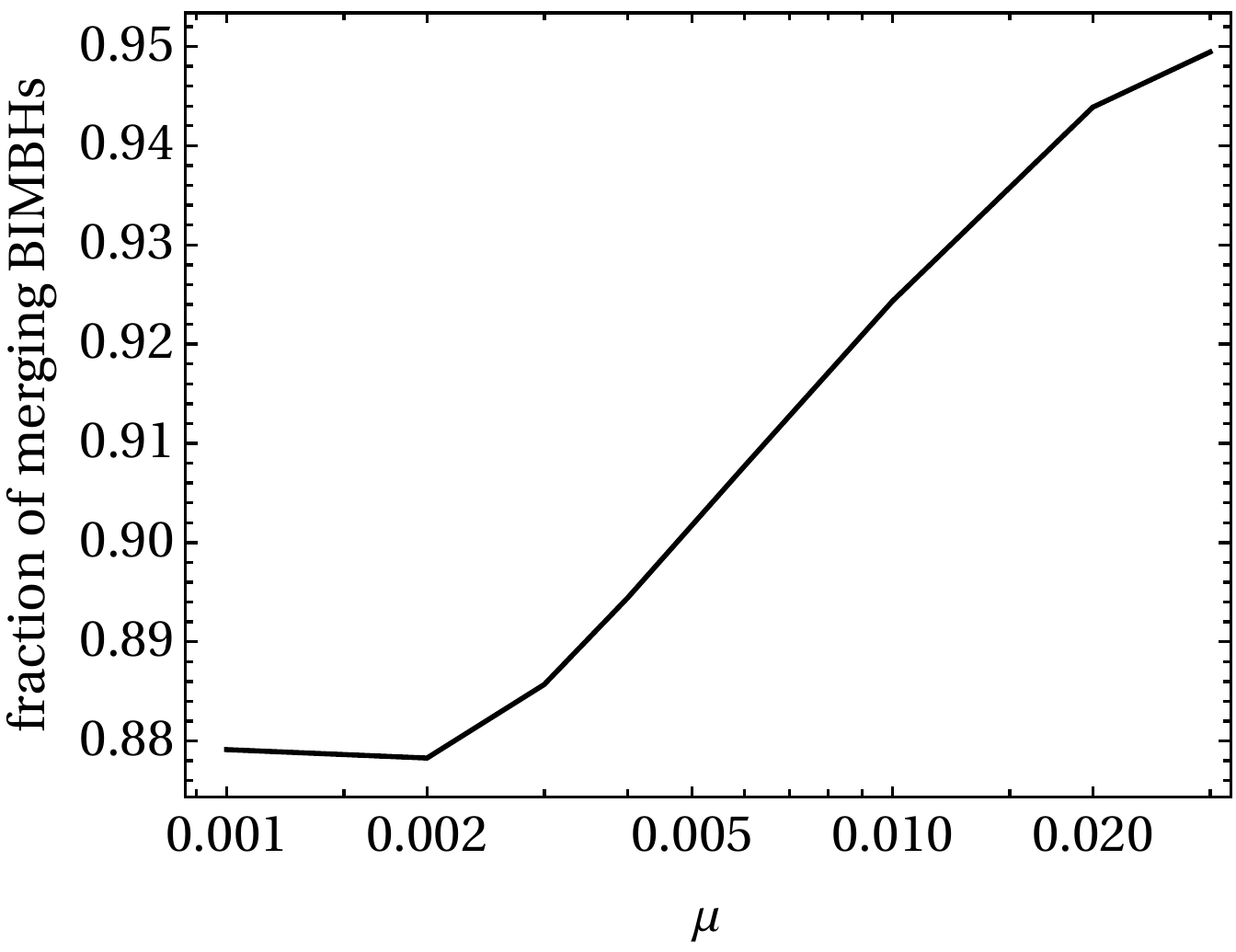}
\includegraphics[width=0.48\textwidth]{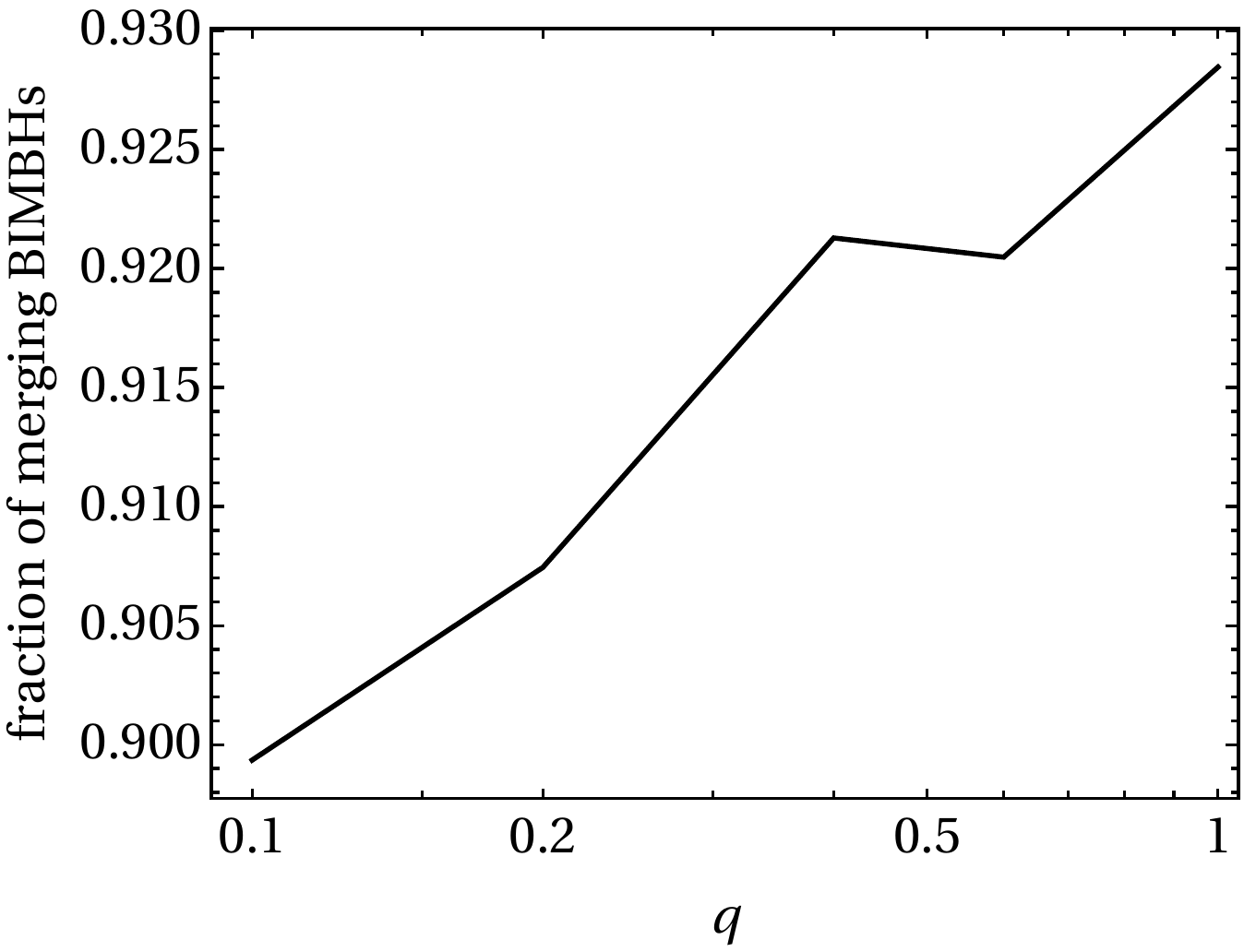}
\caption{The fraction of detected BIMBHs that are observed to merge for $q=1$ and various $\mu$ (top) or $\mu=0.01$ and various $q$ (bottom).}
\label{fig:fraction-of-merging-bimbhs}
\end{figure}

{As shown in Fig.~\ref{fig:fraction-of-merging-bimbhs}, most of the detected BIMBHs are observed to merge, and their fraction increases with $\mu$ and $q$ due to stronger GW emission.}

\begin{figure}
\includegraphics[width=0.48\textwidth]{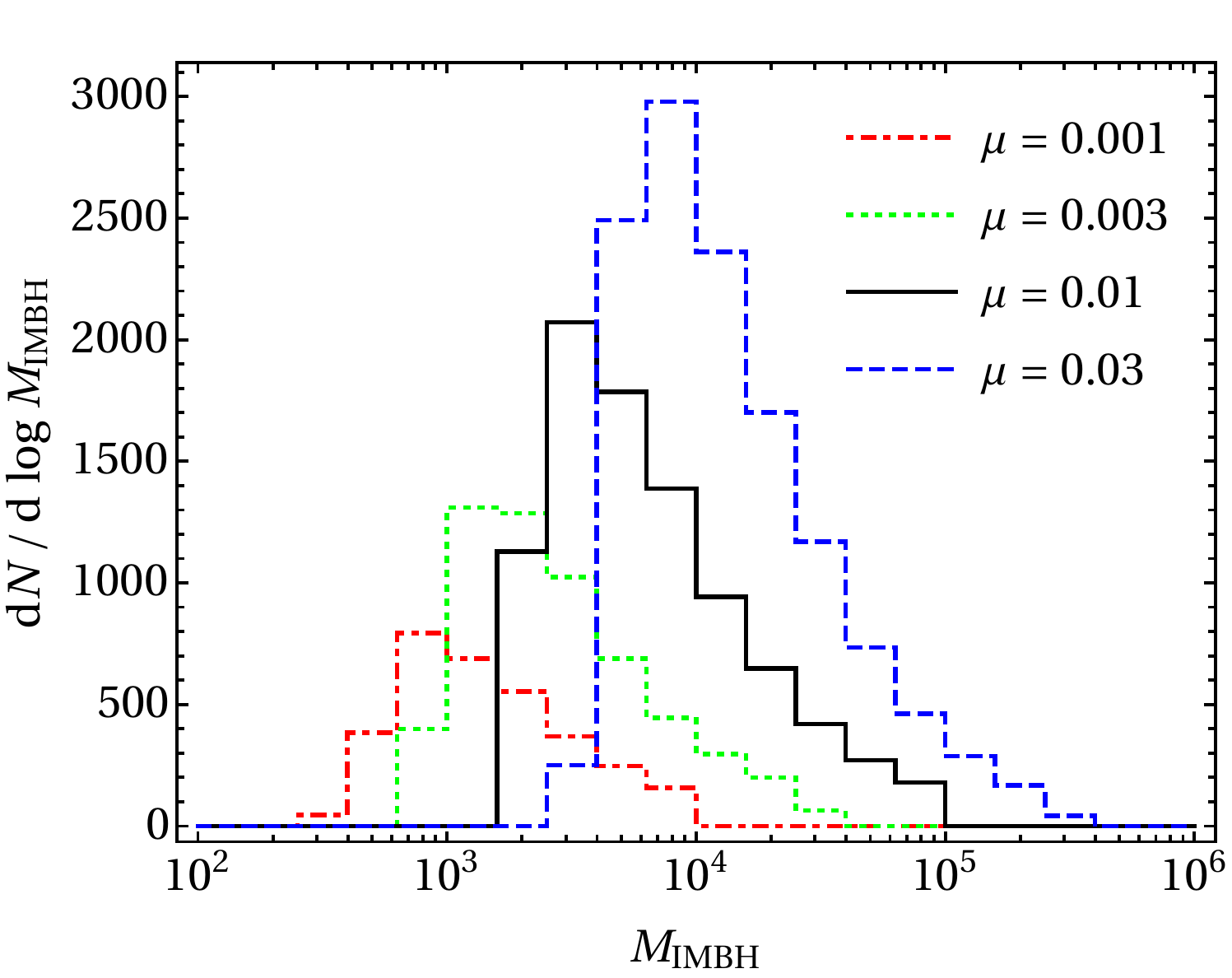}
\includegraphics[width=0.48\textwidth]{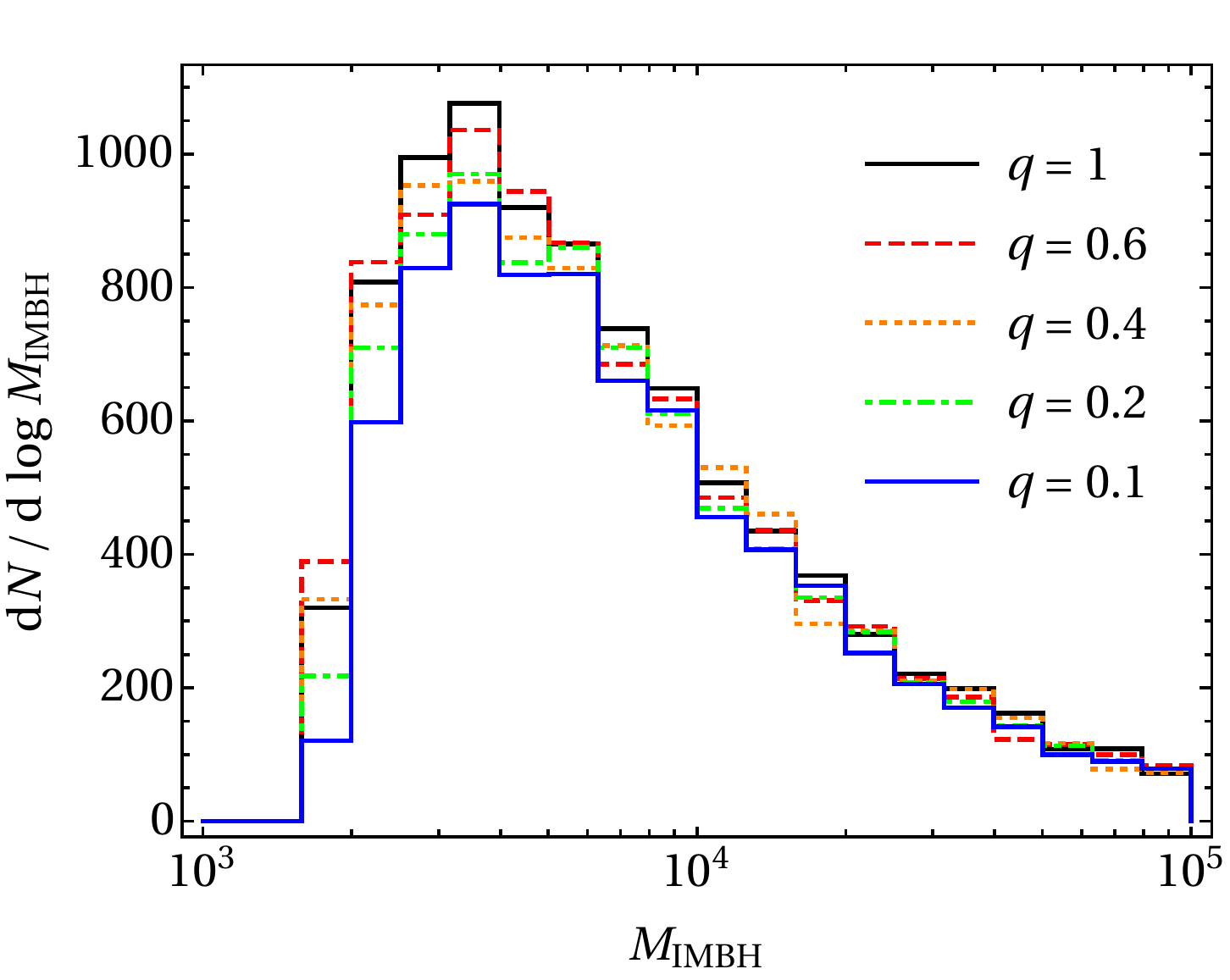}
\caption{The distribution of observed BIMBHs' total masses for $q=1$ with various values of $\mimbh/\mgc$ (top) and $\mimbh/\mgc=0.01$ with various values of $q$ (bottom). Top: line styles as in Fig.~\ref{fig:snr} (top). Bottom: solid black, dashed red, dotted orange, dot-dashed green and solid blue lines correspond to $q=1,\,0.6,\,0.4,\,0.2,\,0.1$, respectively.}
\label{fig:m_imbh}
\end{figure}

The distribution of detected BIMBH masses is shown in Fig.~\ref{fig:m_imbh} (top). Higher $\mimbh$ masses for higher values of $\mu$ are not only due to a higher IMBH/GC mass ratio, but also due to the fact that BIMBHs in lower-mass GCs become massive enough to merge before the GC is disrupted. 

We have also made some simulations for $\mu=0.01$ and various values of $0.1\leq q\leq 1$ and found that the number of detected BIMBHs and their mass distribution are almost independent of $q$ at constant $\mu$ ($\lesssim 15\%$ difference between $q=0.1$ and $q=1$; Fig.~\ref{fig:m_imbh}, bottom). The decrease of $q$ decreases the number of detectable BIMBHs since the GW amplitude $A(f)\propto\eta^{1/2}$, but increases at the same time the number of events in a given frequency range due to increased GW timescale $\tgw\propto\eta^{-1}$. {That can be seen in the dependence of the fraction of merging BIMBHs (Fig.~\ref{fig:fraction-of-merging-bimbhs}, bottom) and the SNR distribution (Fig.~\ref{fig:snr}, bottom) on $q$.}

\begin{figure}
\includegraphics[width=0.49\textwidth]{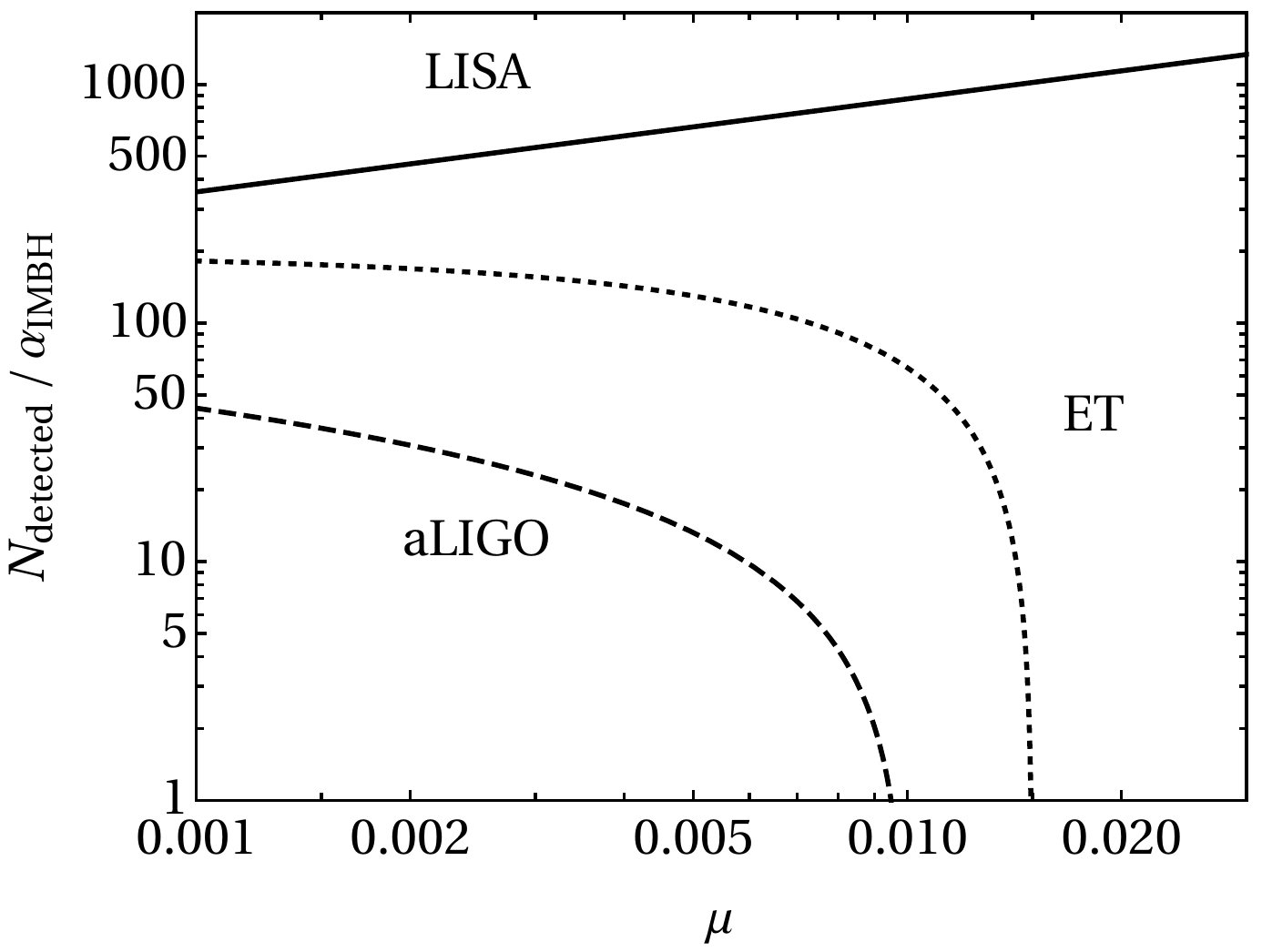}
\caption{
The number of detections by LISA, aLIGO and ET during the 4 year LISA mission time as a function of $\mimbh/\mgc$. It is assumed that all GCs produce a BIMBH ($\alpha_{\rm IMBH} = 1$); if $\alpha_{\rm IMBH} < 1$, the numbers should be rescaled $\propto\alpha$. Line styles as in Fig.~\ref{fig:snr} (top).
}
\label{fig:detectionRate}
\end{figure}

Low-mass BIMBHs
can be detected by Advanced LIGO \citep{ligo} or the proposed Einstein Telescope \citep{ET}. To find such sources and calculate their SNR, we use Eq.~\eqref{eq:snr} with the sensitivity curves taken from \citet{ligoSensitivity} and \citet{ETSensitivity} for LIGO and ET, respectively.
Assuming the same SNR $>8$ detection threshold, the number of detections during the same $4$ year time span are nicely fitted by
\eq{
N_\mathrm{LIGO} &\approx 44 \,\alpha_{\rm IMBH} \log_{10}\frac{0.01}{\mu},
}
for LIGO, and
\eq{
N_\mathrm{ET} &\approx 130 \,\alpha_{\rm IMBH} \qty(1.5-\frac{\mu}{0.01}),
}
for ET, respectively. Fig.~\ref{fig:detectionRate} illustrates the dependence of those numbers (as well as the LISA rate) on $\mu$.

These results agree (within a factor of few) with \citet{santamaria}, who assumed $\alpha_{\rm IMBH}=0.1$ and $\mu=0.002$, and estimated $N_\mathrm{LIGO}=1-10$, $N_\mathrm{ET}=60-100$. However, our ET rate is much lower compared to \citet{gair2011}, who predicted a rate of $\sim1000$ events/yr, for two reasons. First, they used a GC formation rate which is argued in \citet{santamaria} to be unrealistically high; second, they use a different, somewhat older, model of ET noise curve (ET-C rather than ET-D).

% \begin{figure}[h!]
% \includegraphics[width=0.49\textwidth]{mergerRate.pdf}
% \includegraphics[width=0.49\textwidth]{totalMergerRate-log.pdf}
% \caption{
% BIMBH merger rate per unit volume (top) and total merger rate within redshift $z$ (bottom). 
% The rates are normalized per $\alpha_{\rm IMBH}$, as in Fig.~\ref{fig:detectionRate}.
% Line styles as in Fig.~\ref{fig:snr} (top).
% }
% \label{fig:mergerRate}
% \end{figure}

LIGO and ET would be very unlikely to detect any events if $\mu\gtrsim0.01$ or  $\mu\gtrsim0.015$, respectively, which correspond to $\mimbh\gtrsim1000\msun$ and $\mimbh\gtrsim1500\msun$.
The majority of the LIGO detections ($\sim 65\%$ for $\mu=0.001$ and $\gtrsim 90\%$ for $\mu=0.003$) and especially ET detections ($\sim 65\%$ for $\mu=0.001$ and $\gtrsim 96\%$ for $\mu\geq0.002$) will be detected by LISA as well (before they merge, when they are still radiating in the LISA band). As in the case of LISA rates, those depend weakly on $q$.

Finally, we can estimate the total rate of IMBH mergers per unit cosmic volume
\begin{align}
\frac{\dd{N_{\rm merger}}}{\dd{V}\dd{t}} =& 
\alpha_{\rm IMBH} \int \dd{\mgc} \alpha_{\rm merger}(\mgc)  \nonumber\\
&\times \frac{1}{\langle\mgc\rangle} \frac{\dd{\mgc}}
{\dd{t}\dd{V}}\Bigg|_{t=t(z)-t_\mathrm{coal}(\mgc)} 
\nonumber\\
&\times\dv{N}{\mgc}\Delta\mgc 
\end{align}
In Fig.~\ref{fig:mergerRate} (top panel), we report the total merger rate. It peaks at $z\sim 2$ and, as expected, increases with $\mu$; {it's almost identical to the distribution of detected BIMBHs' redshifts.}
Fig.~\ref{fig:mergerRate} (bottom) shows the total merger rate contributed by sources within redshift $z$. We find that even in the most optimistic case $\alpha_{\rm IMBH}$ we would not detect any sources closer than $z=0.3$. 
Nearly all sources ($(250-1000)\alpha_{\rm IMBH}$ events/yr) are at $z<4$. BIBMBHs at $z<1$ contribute $(20-40)\alpha_{\rm IMBH}$ events/yr, the ones with $z<2$ -- $(90-330)\alpha_{\rm IMBH}$ events/yr.

{
\section{The case of nonzero eccentricity}
\label{sec:eccentricity}

\begin{figure}
\includegraphics[width=0.49\textwidth]{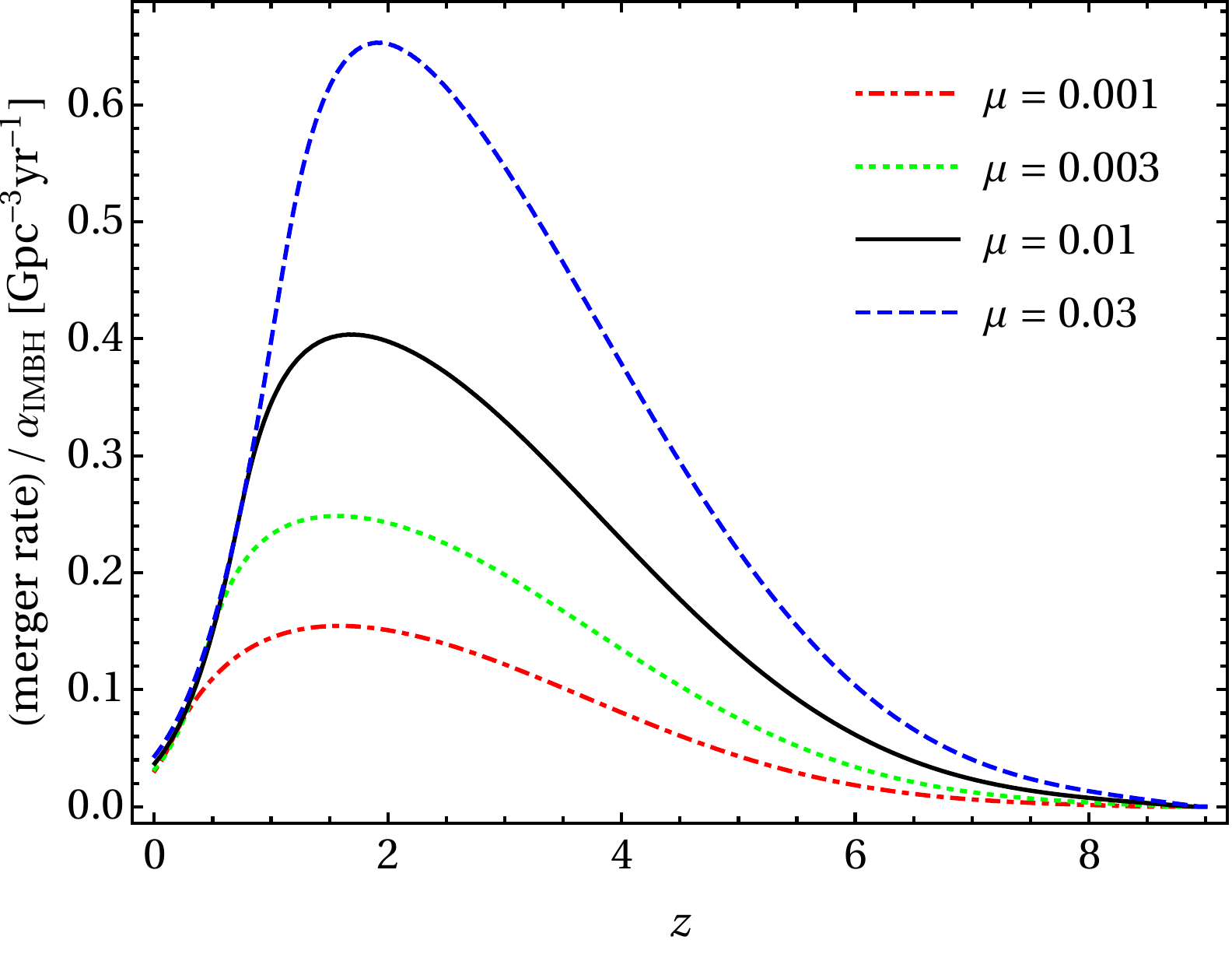}
\includegraphics[width=0.49\textwidth]{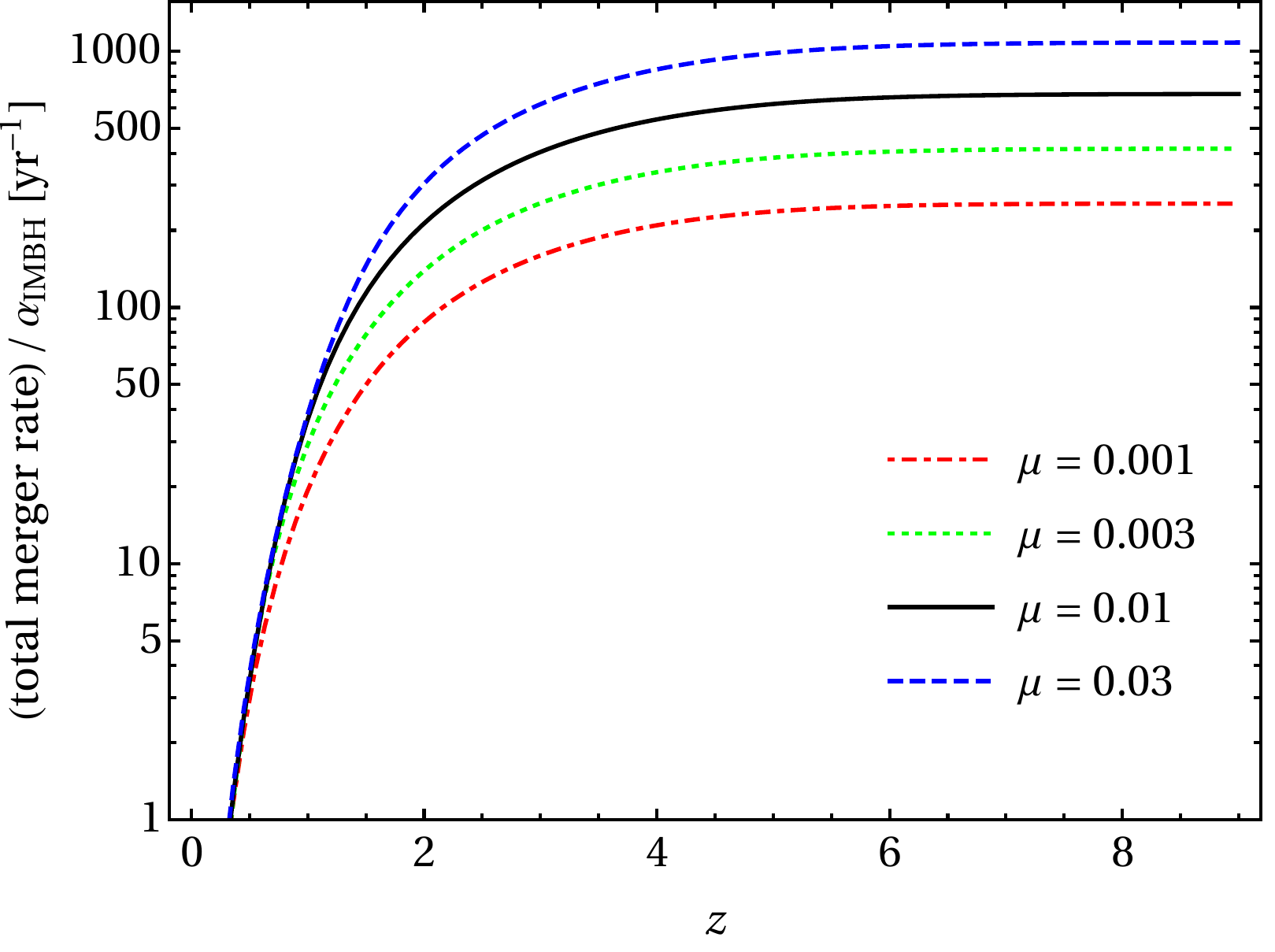}
\caption{
BIMBH merger rate per unit volume (top) and total merger rate within redshift $z$ (bottom). 
The rates are normalized per $\alpha_{\rm IMBH}$, as in Fig.~\ref{fig:detectionRate}.
Line styles as in Fig.~\ref{fig:snr} (top).
}
\label{fig:mergerRate}
\end{figure}

Let us now examine the case where the initial BIMBH eccentricity is nonzero. Eccentricity can significantly affect both BIMBH evolution and GW emission. Stellar ejections by a massive binary with nonzero eccentricity will further increase its eccentricity, provided the binary is approximately equal-mass and the stellar environment is nonrotating \citep{Quinlan1996,Sesana2006,RasskazovMerritt,Rasskazov2019,bonetti}. As a result, it is possible for the eccentricity to reach extreme values such as $\sim 0.99$, even with a moderate initial eccentricity at $a=a_h$. 

GR effects, namely GW losses, on the other hand, tend to decrease eccentricity and speed up the hardening \citep{Peters64}:
% \begin{subequations}
\bsub
\label{eq:dadt-dedt}
\begin{align}
\qty(\dv{a}{t})_{\rm GR} &=
\frac{64G^3\mimbh^3\eta}{5c^5a^3}\, \frac{1+(73/24)e^2+(37/96)e^4}{(1-e^2)^{7/2}},\\
\qty(\dv{e}{t})_{\rm GR} &= 
\frac{304G^3\mimbh^3\eta}{15c^5a^4}\, e\frac{1+(121/304)e^2}{(1-e^2)^{5/2}}.
\end{align}
\esub
% \end{subequations}
% \esub
When GW emission dominates the shrinking of the binary this leads to \citep{Peters64}
\begin{equation}\label{eq:a(e)}
    a(e) = a_0 \left(\frac{e}{e_0}\right)^{12/19}\frac{1-e_0^2}{1-e^2}\left(\frac{1+\frac{121}{304}e^2}{1+\frac{121}{304}e_0^2}\right)^{870/2299}
\end{equation}
(where $a_0$ and $e_0$ are the semimajor axis and eccentricity at the beginning of GW-dominated regime)
and the orbital frequency decreases accordingly as
\begin{equation}\label{eq:forb}
    \forb =\frac{1}{2\pi} \left(\frac{G M}{a(e)^3}\right)^{1/2}\,.
\end{equation}

Because of the dramatic increase of the GW-driven hardening rate at high eccentricities, the BIMBH can reach the GW-dominated stage much earlier compared to the circular case, after which the eccentricity and semimajor axis decrease as {in Eqs.~(\ref{eq:dadt-dedt}).} 
%Eq.~(\ref{eq:a(e)}). 
For $a\ll a_0 (1-e_0)$, the eccentricity decreases rapidly as $e \approx e_0 \qty(\frac{a}{a_0(1-e_0^2)})^{19/12}$.

In addition, eccentricity also affects the GW spectrum of the binary. Eccentric binaries radiate GWs not only at $f = 2\forb$, but at all integer harmonics of the orbital frequency with GW power distributed between harmonics in the following way \citep{PetersMathews}:
\subeq{
P_n &\propto g(n,e),\\
f_n &= \frac{n\forb}{1+z} \qc n = 1,2,3,\dots,\\
g(n,e) &= \frac{n^4}{32} \Bigg[  \bigg\{J_{n-2}(ne) -2e J_{n-1}(ne) + \frac{2}{n} J_{n}(ne) \nonumber\\
&+ 2eJ_{n+1}(ne)-J_{n+2}(ne)\bigg\}^2+\left(1-e^2\right)\nonumber\\
&\times\bigg\{ J_{n-2}(ne) -2J_{n}(ne)+J_{n+2}(ne)\bigg\}^2 \nonumber\\
&-+ \frac{4}{3n^2}J^2_{n}(ne)\Bigg].
}
The higher the eccentricity for fixed semimajor axis, the higher is the peak harmonic. \citet{Wen2003} derived an analytical approximation for the peak harmonic
\eq{\label{eq:n_peak}
n_\mathrm{peak} = 2\frac{(1+e)^{1.1954}}{(1-e^2)^{1.5}}\ .
}
Combining with Eqs.~\eqref{eq:a(e)}, \eqref{eq:forb} and \eqref{eq:n_peak} shows that in the GW-driven evolutionary stage the peak frequency evolves as
\begin{align}
    f_\mathrm{peak} =& f_\mathrm{peak,0}     
        \left(\frac{e}{e_0}\right)^{-18/19}
    \nonumber\\&\times
    \left(\frac{1+e}{1+e_0}\right)^{1.1954}
    \left(\frac{1+\frac{121}{304}e^2}{1+\frac{121}{304}e_0^2}\right)^{-1305/2299}\,.\nonumber
\end{align}
Thus, the eccentricity becomes much less than unity for frequencies that satisfy $f_{\rm peak}\gg f_{\rm peak,0}$.
Here $f_{\rm peak,0}$ is the peak GW frequency when GWs start to dominate the evolution of the binary over dynamical hardening, which we approximate with $|(\mathrm{d}a/\mathrm{d}t)_\mathrm{GR}| = a/t_h(a)$:
\bsub
\begin{align}
a_0 &= \left(
\frac{64}{5}\frac{G^3}{c^{5}} \mimbh^3\eta F(e)\times\mathrm{32.6\,Myr}\frac{10^5\msun}{M_{\rm GC}}\mathrm{mpc}
\right)^{1/5}\\
&= \SI{5.1e-6}{pc} \left(\frac{\mimbh}{10^3\msun}\right)^{2/5} \qty(\frac{\mu}{10^{-2}})^{1/5} \nonumber\\
&\times \qty(\frac{\eta}{1/4})^{1/5} \frac{\qty(1+(73/24)e_0^2+(37/96)e_0^4)^{1/5}}{\qty(1-e_0^2)^{0.7}},\\
f_{\rm peak,0} &= \SI{1.8e-6}{Hz} \left(\frac{\mimbh}{10^3\msun}\right)^{-0.1} \qty(\frac{\mu}{10^{-2}})^{-0.3} \nonumber\\
&\times \qty(\frac{\eta}{1/4})^{-0.3} \qty(1+(73/24)e_0^2+(37/96)e_0^4)^{-0.3} \nonumber\\
&\times \qty(1-e_0^2)^{-0.45} (1+e_0)^{1.1954}.
\end{align}
\esub
{Here we have used Eq. \eqref{eq:t_h(mgc)} to replace the $\sigma/HG\rho$ factor with the $t_h-\mgc$ scaling relation used in our model. The value of $e_0$ in our simulations (with initial eccentricity 0.6) is between 0.95 and 0.99 depending on $\mu$ and $\mgc$.} {Thus, when GWs start driving the binary, the initial peak frequency is between $\SI{7e-6}{Hz}$ and $\SI{1.6e-5}{Hz}$, and $e\propto f_{\rm peak,0}/f_{\rm peak}$ approximately at higher peak frequencies. Thus, typically $e\leq 0.1$ for $f_{\rm peak}\geq10^4\,\si{Hz}$
in the LISA frequency band.}

SNR is calculated in the following way \citep{BarackCutler,EnokiNagashima}:
\subeq{\label{eq:snr_eccentric}
\kappa^2 &= \sum_{n=1}^\infty \int_{f_n}^{f_n+\Delta f_n} \frac{h_{c,n}^2}{f_n S_n(f_n)} \frac{\dd{f_n}}{f_n},\\
h_{c,n} &= \frac{1}{\pi d_L} \sqrt{2\dv{E}{f_n}},\\
\dv{E}{f_n} &= \frac{(1+z)(2\pi)^{2/3}G^{2/3}\mchirp}{3n[\forb/(1+z)]^{1/3}},
}
where we follow \citet{OLeary2009} and truncate the spectrum at the maximum harmonic
\eq{
n_\mathrm{max} = 5\frac{(1+e)^{1/2}}{(1-e)^{3/2}}.
}
and $f_n = n \forb/(1+z)$ and $\Delta f_n=n \Delta \forb/(1+z)$ are the observed GW frequency of the $n^{\rm th}$ orbital harmonic and its change during the observation time. 
Here we have replaced the $1/5$ factor from \citet{BarackCutler} and \citet{EnokiNagashima} with the general sky and polarization averaged signal response function for LISA/aLIGO/ET.

In practice, we use one of two approximations to evaluate Eq.~\eqref{eq:snr_eccentric}. {Whenever the eccentricity is significantly nonzero (we use the criterion $e>0.1$), the BIMBH is at low enough orbital frequencies that the changes in its orbital parameters are negligible.}
In particular, $\Delta f_n \ll f_n$ so that we may assume that the integrand is nearly constant and
\subeq{
\frac{\Delta f_n}{f_n} &\approx \frac{96}{5}\qty(\frac{G\mchirp}{c^3})^{5/3} (2\pi\forb)^{8/3} F(e) \frac{T}{1+z} \nonumber\\
&\approx \num{1.6e-4} \,\left(\frac{\mchirp_z}{10^3\msun}\right)^{5/3} \left(\frac{f_{\mathrm{peak}}}{0.1\mathrm{mHz}}\right)^{8/3} \frac{T}{4\,\mathrm{yr}}\nonumber\\
&\times \frac{(1-e^2)^{1/2}(1+(73/24)e^2+(37/96)e^4)}{(1+e)^{3.1877}},\\
\kappa &\approx \sum_{n=1}^\infty\frac{16(1+z)(G\mchirp)^{5/3}(2\pi\forb)^{2/3}\sqrt{Tg(n,e)}}{\sqrt{5S_n}c^4D_Ln} \label{eq:snr_eccentric_point}
}
where $\mchirp=M\eta^{3/5}$ is the chirp mass of the BIMBH, where $M=m_1+m_2$, $\eta=m_1m_2/M^2$ for BIMBH component masses $m_1$ and $m_2$. {We can see that high eccentricity doesn't increase the relative frequency change compared to the circular case given the same peak frequency.} Thus when the eccentricity is low enough ($e<10^{-3}$) we use the circular approximation, i.e. that $n=2$ is the only nonzero term (with $g(2,0)=1$). In that case, Eq.~\eqref{eq:snr_eccentric_point} is equivalent to Eq.~\eqref{eq:kappa_point} or Eq.~\eqref{eq:snr} if $f_n\leq0.1$ or $f_n\geq0.1$, respectively. 

The way we sample the initial orbital frequency is also different from the circular case. For every GC mass bin we 
{solve the time evolution equations, recording the orbital parameters at every timestep:
\subeq{
\dv{a}{t} &= -\frac{a}{t_h(a)} + \qty(\dv{a}{t})_{\rm GR},\\
\dv{e}{t} &= \frac{K(a/a_h,e)}{t_h(a)} + \qty(\dv{e}{t})_{\rm GR},
}
where $K$ is calculated in the same way as in \citet{Rasskazov2019}. Then the random values of $a$ and $e$ are taken at a time uniformly distributed between the moment the BIMBH enters the LISA band and the moment of merger. 
}
The former is determined as the moment when the peak harmonic (Eq.~\ref{eq:n_peak}) reaches the frequency $10^{-4}\,\si{Hz}$. Here we have increased the minimum LISA frequency compared to $10^{-5}\,\si{Hz}$ for the circular case: the number of BIMBHs radiating in the frequencies above $10^{-5}\,\si{Hz}$ is too high to calculate all their signals in a reasonable time (because of high eccentricities BIMBHs enter that frequency range at orbital frequencies far below $10^{-5}\,\si{Hz}$ and, consequently, spend much more time there compared to the circular case). This is a reasonable approximation as the LISA sensitivity below $f\sim 10^{-4}\,{\rm Hz}$ deteriorates sharply with decreasing frequency and it is very low anyway -- as a result, only a tiny fraction of sources are detected at those frequencies.\footnote{We note that the stochastic background of small SNR sources may possibly be significant at low frequencies, but this is beyond the scope of this paper.} 

Due to high eccentricities, many BIMBHs start radiating in the LISA band when they are still not quite in the GW-dominated regime yet. This decreases the residence time that the binary spends at a particular frequency and thereby decreases the number of sources at a given frequency by multiplying the right-hand side of Eq.~\eqref{eq:snr_eccentric_point} by the factor $(\dv*{a}{t})_\mathrm{GR} / ((\dv*{a}{t})_\mathrm{GR} + (\dv*{a}{t})_\ast)$ -- the fraction of energy lost to GWs (not to ejected stars).

In our eccentric models, to simplify the calculations, we have fixed the initial distance from the galactic center for all the GCs at $r=\SI{10}{kpc}$. It doesn't have a significant effect on the GC mass evolution except for the smallest GCs where the BIMBH doesn't merge anyway. We assume the initial eccentricity $e_0=0.6$ and mass ratio $q=1$.

As mentioned above, one of the effects of nonzero eccentricity is a significant reduction in the merger timescale. It is illustrated in Fig.~\ref{fig:tcoal_ecc}. Comparing with Fig.~\ref{fig:tcoal}, we see that the initial eccentricity $e_0=0.6$ can decrease the merger timescales by one or two orders of magnitude. However, we find that $>99\%$ of all the detected BIMBHs (at any $\mu$) are at the point where the eccentricity was reduced below $0.05$. So, the initial eccentricity of $\sim0.6$ mostly affects the merger time but not the GW spectrum.

\begin{figure}
\includegraphics[width=0.48\textwidth]{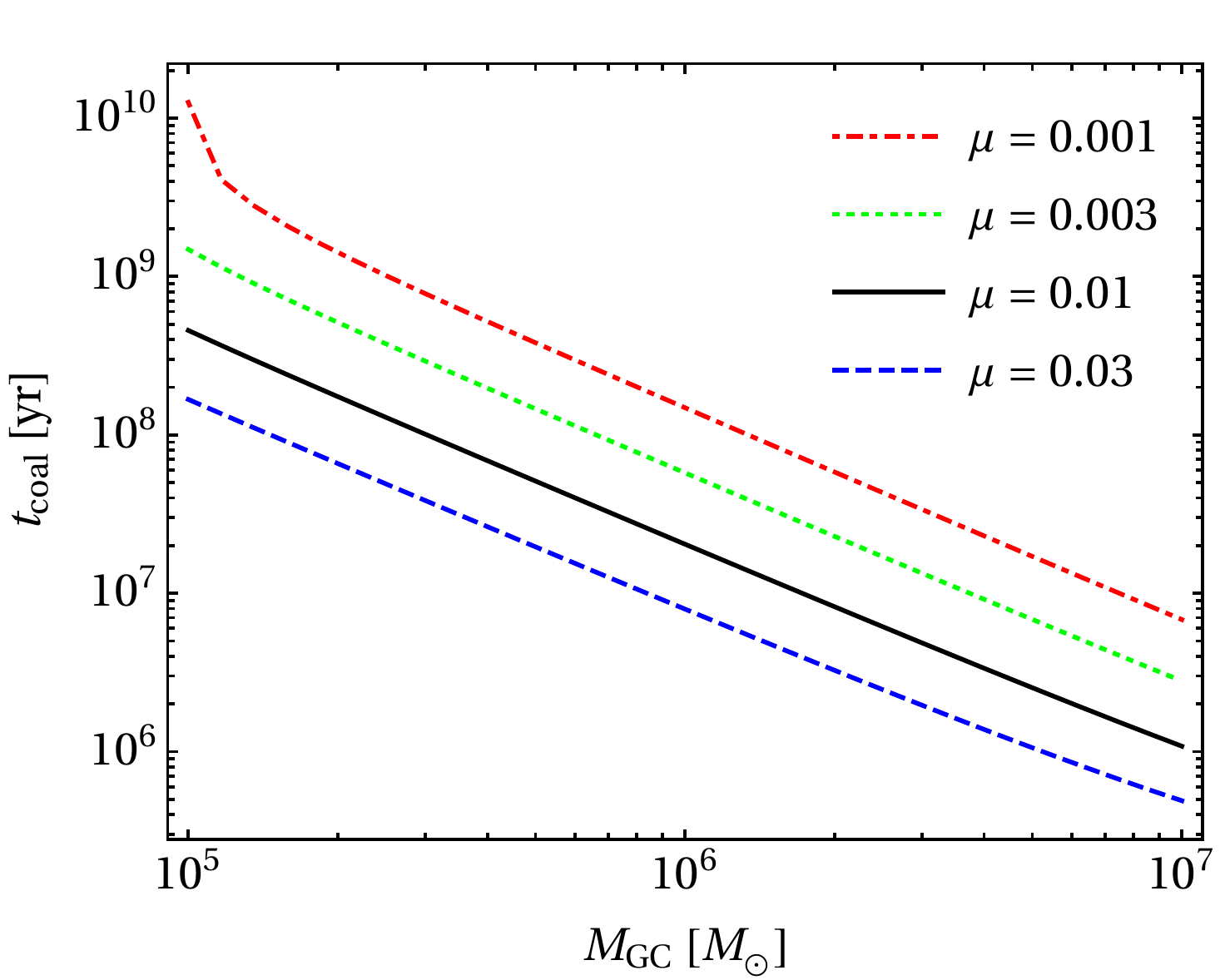}
\caption{BIMBH coalescence timescales given initial eccentricity 0.6. Line styles are as in Fig.~\ref{fig:snr} (top panel).}
\label{fig:tcoal_ecc}
\end{figure}

Because of lower merger times, more of the low-mass GCs contribute to the merger rate, with even GCs of mass $\sim 10^5\,\msun$ producing a merger in less than a Hubble time. Fig.~\ref{fig:increase} shows the ratio of $N_{\rm detected}$ with $e_0=0.6$ to $N_{\rm detected}$ with $e_0=0$. aLIGO and ET detection rates are increased by a factor of $\sim 6$--$10$. The increase in LISA detections is less pronounced: $\sim3$ times for high $\mu$ almost no increase for low $\mu$. On one hand, the decrease in merger timescale means that more low-mass IMBHs merge, but at the same time long merger timescales at $e_0=0$ bring BIMBHs (most of which are born at $z\sim 4$, Fig.~\ref{fig:formationRate}) to much lower redshifts at the moment of their merger, making them easier to detect. These two effects result in the shape of all three curves in Fig.~\ref{fig:increase} with the peak at some average $\mu$. The superposition of these two effects is also illustrated in Fig.~\ref{fig:comparison}. In the case of heavy IMBHs ($\mu=0.01$) the decrease in merger time is much more important, while for low-mass IMBHs ($\mu=0.001$) those two effects more or less balance each other. 
%The SNR distribution is similar to that for zero eccentricity (Fig.~\ref{fig:snr-ecc}). 

\begin{figure}
\includegraphics[width=0.48\textwidth]{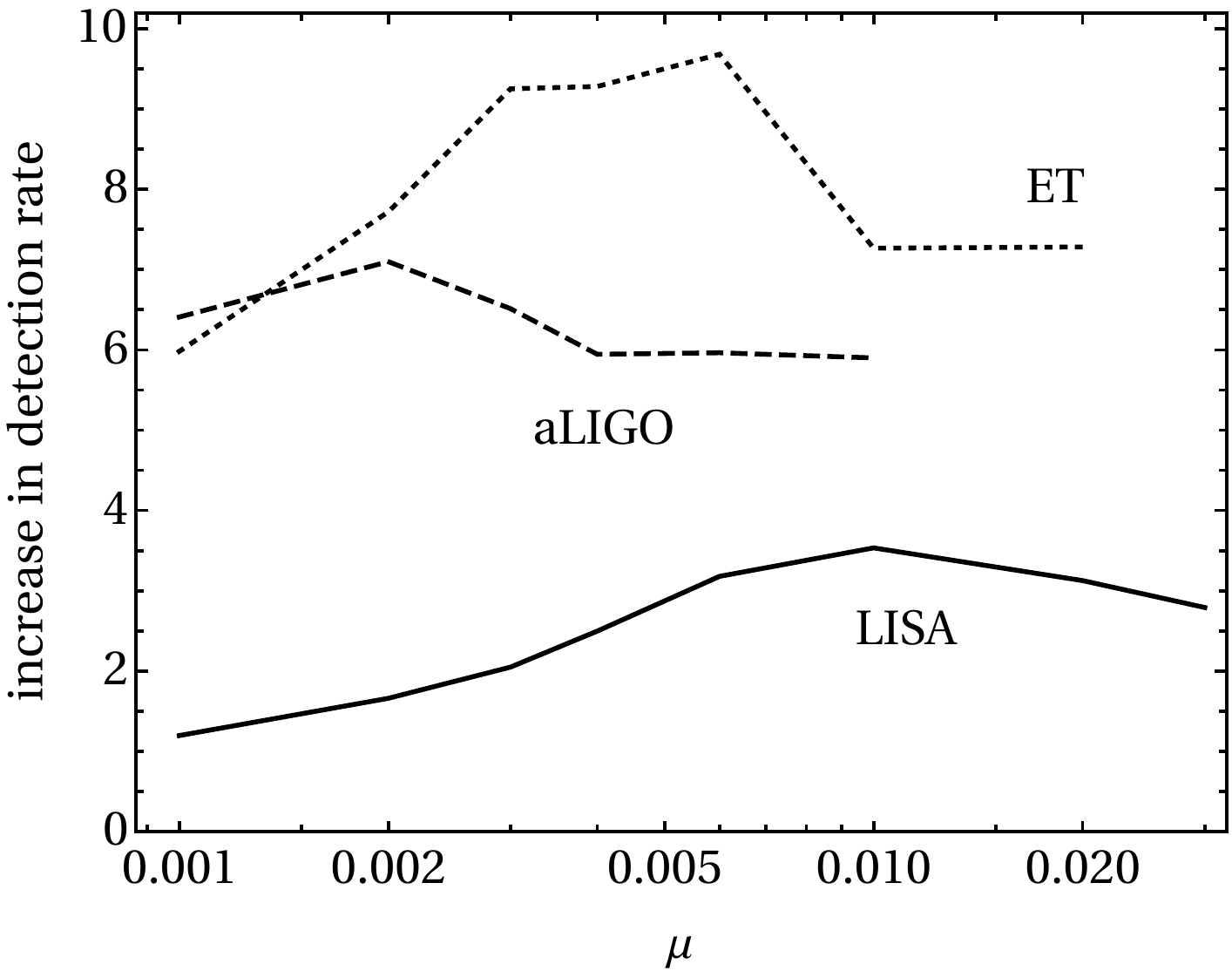}
\caption{Relative increase in LISA/aLIGO/ET detection rates with $e=0.6$ compared to the circular case.}
\label{fig:increase}
\end{figure}

% \begin{figure}
% \includegraphics[width=0.48\textwidth]{comparison-ecc-1.pdf}
% \includegraphics[width=0.48\textwidth]{comparison-ecc-2.pdf}
% \caption{The distribution of observed BIMBHs' total masses for $q=1$ and $e_0=0$ (solid) or $e_0=0.6$ (dashed). Top: $\mu=0.01$, bottom: $\mu=0.001$.}
% \label{fig:comparison}
% \end{figure}

% \begin{figure}
% \includegraphics[width=0.48\textwidth]{snr-ecc.pdf}
% \caption{As Fig.~\ref{fig:snr} but for $e_0=0.6$.}
% \label{fig:snr-ecc}
% \end{figure}
}

\section{Kick velocity and spin of merger remnants}
\label{sec:kick+spin}

Due to the anisotropic emission of GWs at merger, the merger remnant is imparted a recoil kick \citep{lou12}, which can eject the IMBH merger remnant from the host GC. The recoil kick depends on the asymmetric mass ratio $\eta$ and on the magnitude of the IMBH reduced spins, $|\mathbf{{\chi_1}}|$ and $|\mathbf{{\chi_2}}|$.
We model the recoil kick as \citep{lou10}
\begin{equation}
\textbf{v}_{\mathrm{kick}}=v_m \hat{e}_{\perp,1}+v_{\perp}(\cos \xi \hat{e}_{\perp,1}+\sin \xi \hat{e}_{\perp,2})+v_{\parallel} \hat{e}_{\parallel}\ ,
\label{eqn:vkick}
\end{equation}
where
\begin{eqnarray}
v_m&=&A\eta^2\sqrt{1-4\eta}(1+B\eta)\\
v_{\perp}&=&\frac{H\eta^2}{1+q}(\chi_{2,\parallel}-q\chi_{1,\parallel})\\
v_{\parallel}&=&\frac{16\eta^2}{1+q}[V_{1,1}+V_A \tilde{S}_{\parallel}+V_B \tilde{S}^2_{\parallel}+V_C \tilde{S}_{\parallel}^3]\times \nonumber\\
&\times & |\mathbf{\chi}_{2,\perp}-q\mathbf{\chi}_{1,\perp}| \cos(\phi_{\Delta}-\phi_{1})\ .
\end{eqnarray}

\begin{figure}
\includegraphics[width=0.48\textwidth]{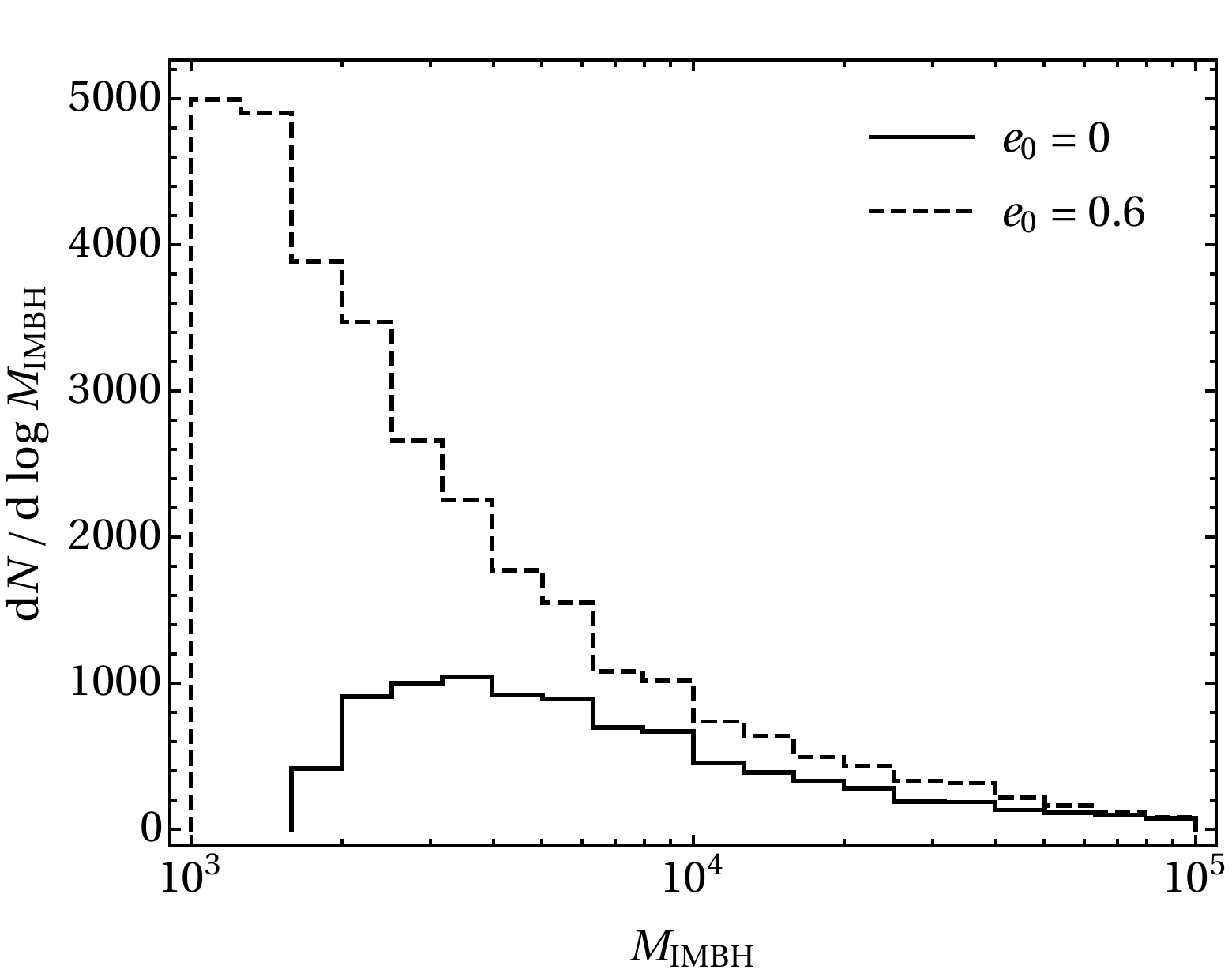}
\includegraphics[width=0.48\textwidth]{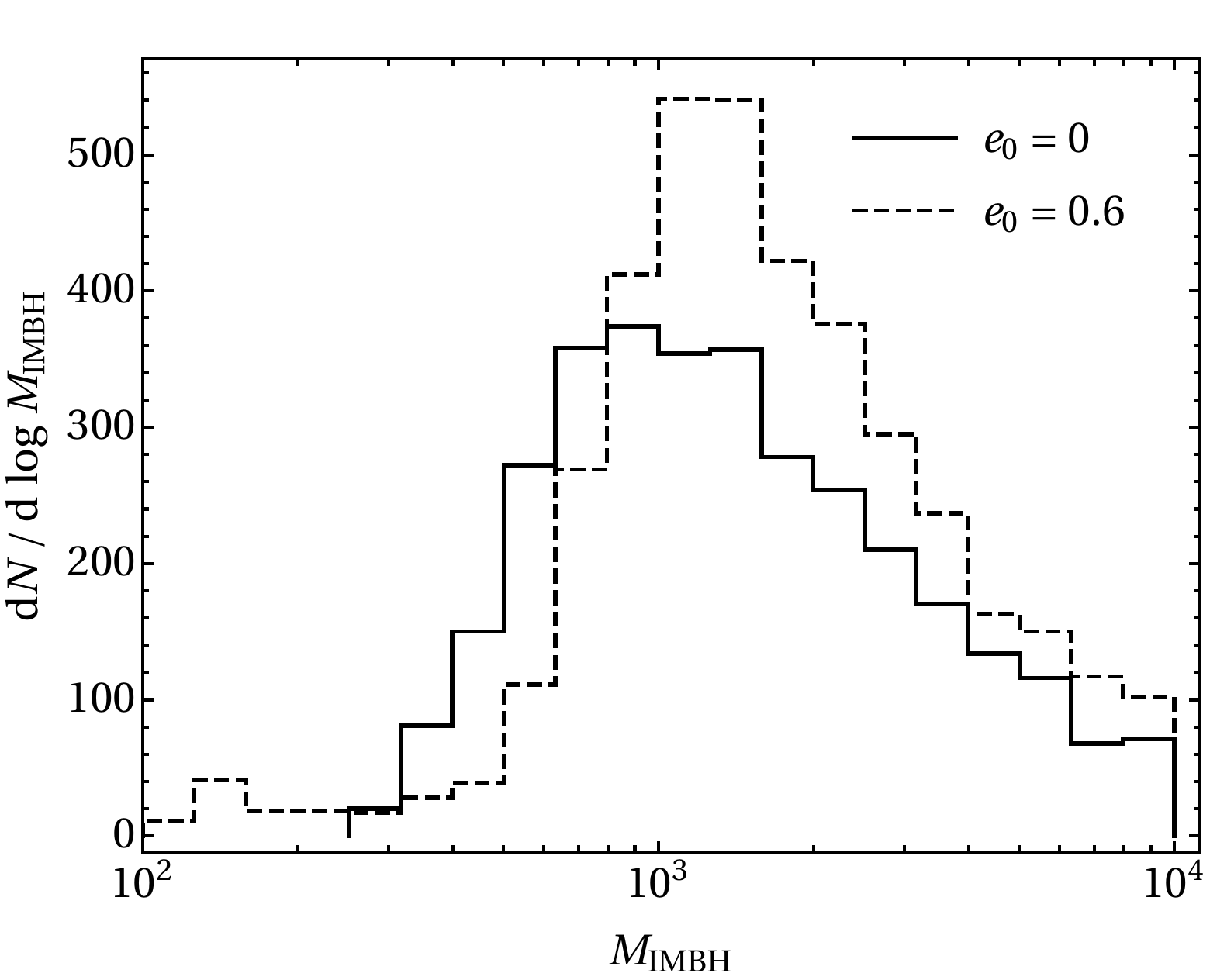}
\caption{The distribution of observed BIMBHs' total masses for $q=1$ and $e_0=0$ (solid) or $e_0=0.6$ (dashed). Top: $\mu=0.01$, bottom: $\mu=0.001$.}
\label{fig:comparison}
\end{figure}

The $\perp$ and $\parallel$ refer to the direction perpendicular and parallel to the orbital angular momentum, respectively, while $\hat{e}_{\perp}$ and $\hat{e}_{\parallel}$ are orthogonal unit vectors in the orbital plane. We have also defined the vector
\begin{equation}
\tilde{\mathbf{S}}=2\frac{\mathbf{\chi}_{2,\perp}+q^2\mathbf{\chi}_{1,\perp}}{(1+q)^2}\ ,
\end{equation}
$\phi_{1}$ as the phase angle of the binary, and $\phi_{\Delta}$ as the angle between the in-plane component of the vector
\begin{equation}
\mathbf{\Delta}=M^2\frac{\mathbf{\chi}_{2}-q\mathbf{\chi}_{1}}{1+q}
\end{equation}
and the infall direction at merger. Finally, we adopt $A=1.2\times 10^4$ km s$^{-1}$, $H=6.9\times 10^3$ km s$^{-1}$, $B=-0.93$, $\xi=145^{\circ}$ \citep{gon07,lou08}, and $V_{1,1}=3678$ km s$^{-1}$, $V_A=2481$ km s$^{-1}$, $V_B=1793$ km s$^{-1}$, $V_C=1507$ km s$^{-1}$ \citep{lou12}. Since GWs radiate not only linear momentum, but also angular momentum and energy during a merger, we adjust the final total spin of the merger product and its mass to account for these losses \citep{lou10}.

% \begin{figure}
% \includegraphics[width=0.48\textwidth]{spin001.pdf}
% \includegraphics[width=0.48\textwidth]{vkick001.pdf}
% \caption{Spins of (top) and kick velocity imparted to (bottom) the remnant from IMBH+IMBH mergers for different values of $q$.}
% \label{fig:spin}
% \end{figure}

\begin{figure}
\includegraphics[width=0.48\textwidth]{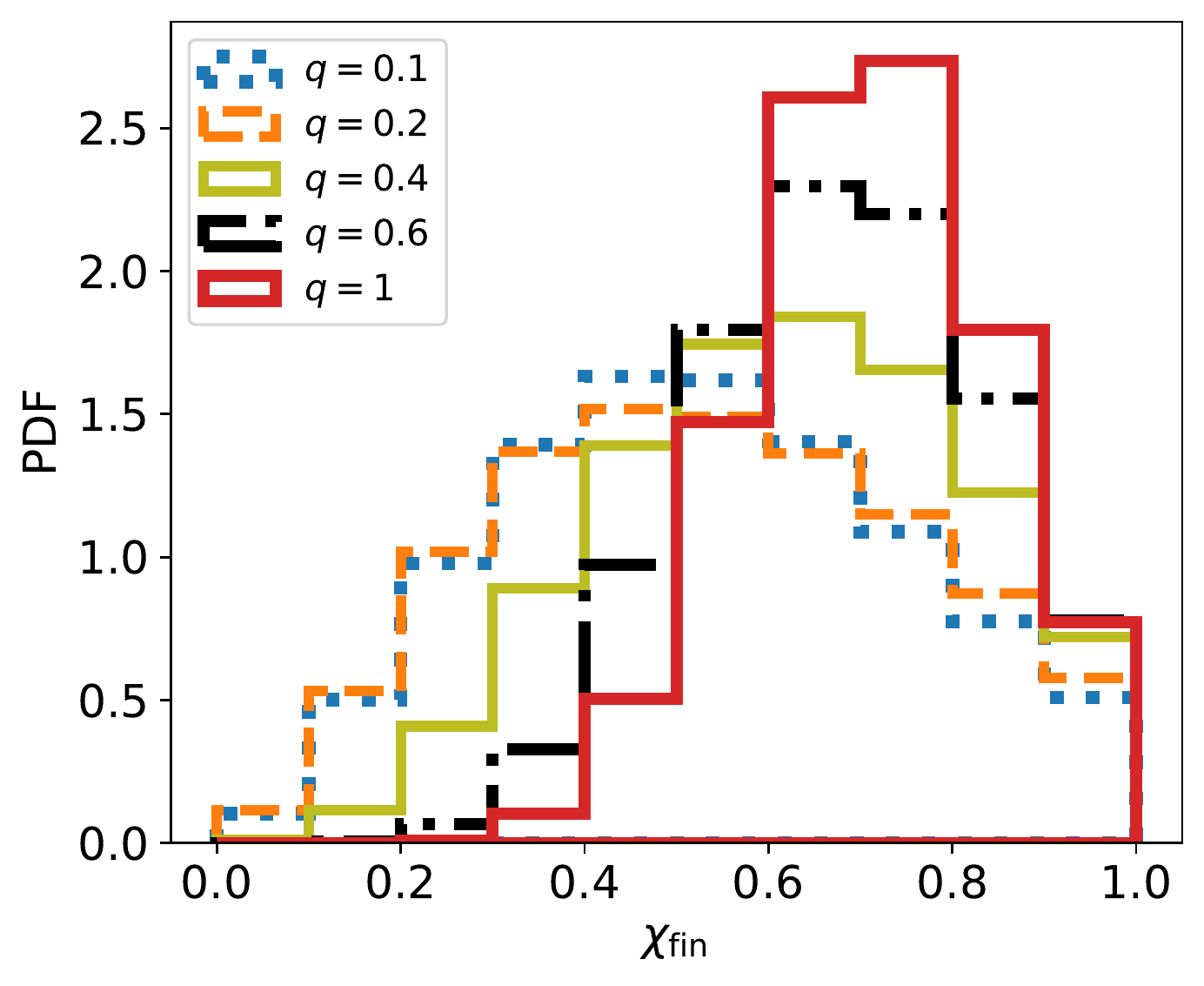}
\includegraphics[width=0.48\textwidth]{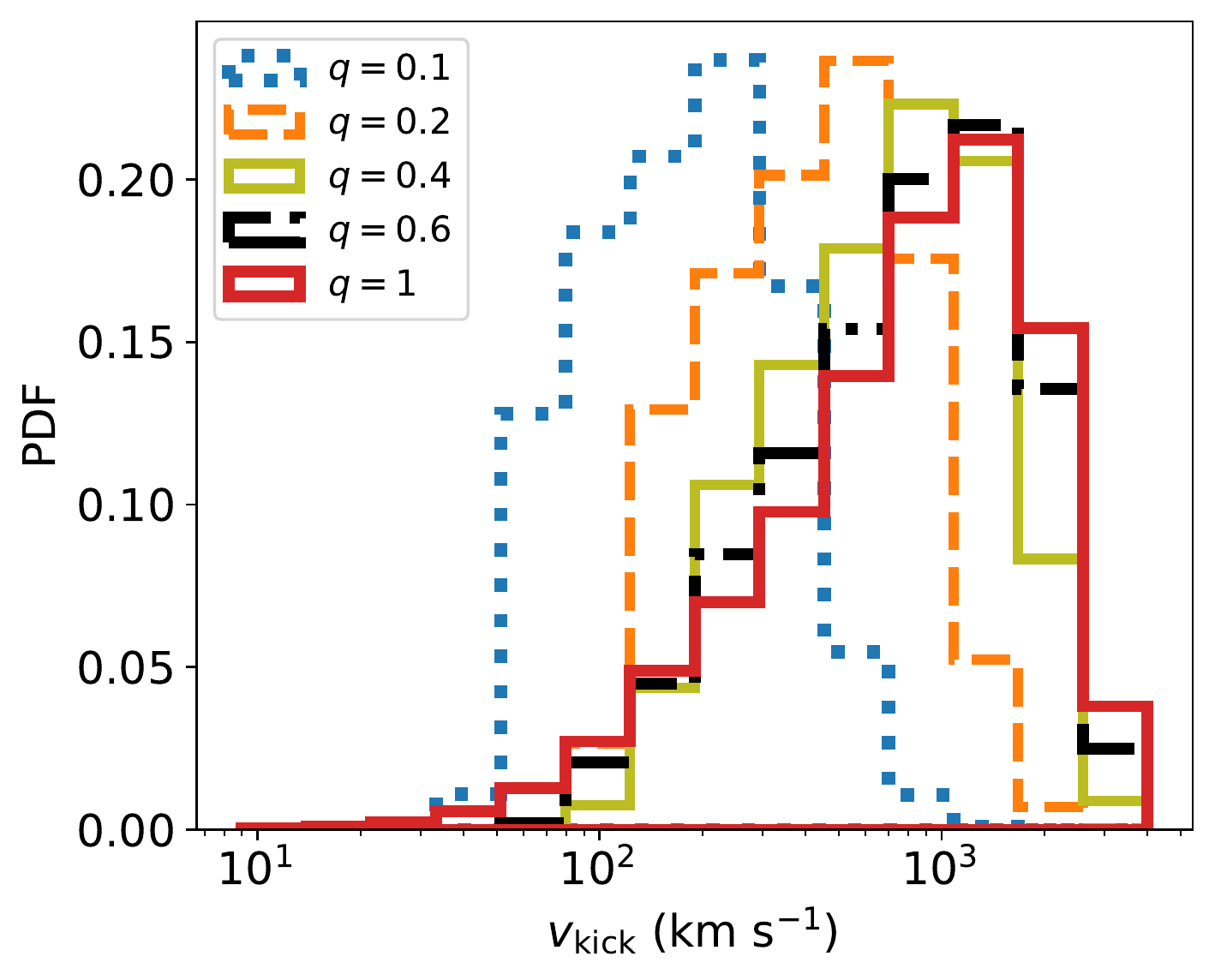}
\caption{Spins of (top) and kick velocity imparted to (bottom) the remnant from IMBH+IMBH mergers for different values of $q$.}
\label{fig:spin}
\end{figure}

In our calculations, we assume that the merging IMBHs have initial reduced spin uniformly distributed. 
We find that the IMBH mass ratio affects both the final spin of the merger remnant and the recoil kick, which is imparted as a result of the asymmetric GW emission. In Fig.~\ref{fig:spin}, we show the distribution of final spins ($\chi_{\rm fin}$) of the remnant from IMBH+IMBH mergers for different values of $q$. The larger the value of $q$, the larger $\chi_{\rm fin}$ and the smaller the width of the distribution. We find that the distribution is peaked at $\sim 0.7$ for $q\gtrsim 0.5$, while for $q\lesssim 0.5$ it is peaked at $\sim 0.5$. Since the spin of the remnant depends only on $\eta$ (hence $q$), the final distribution is not affected by a different value of $\mu$. We also illustrate the recoil kick imparted to the merger remnant in Fig.~\ref{fig:spin}. IMBH of comparable masses receive larger kicks and typically the recoil kick is high enough to eject the remnant from the host GC, and in some cases from the host galaxy, even for $q=0.1$. Therefore, most of the GCs that were born with or hosted a BIMBH are not expected to host any IMBH, if the BIMBH merged through GW emission.

\section{Conclusions}
\label{section:conclusions}

In this paper, we have assumed that binary IMBHs form in a certain fraction of GCs ($\alpha_{\rm IMBH}$) and then harden due to stellar 3-body scatterings. The rate of that hardening depends on the stellar density $\rho_c$ and velocity dispersion $\sigma$ in the GC center. We have take their values from the published catalogue of observed GC properties \citep{harris}; we found both of these parameters correlate with the GC mass, and based on that, have assigned a certain values of $\sigma$ and hardening timescale $t_h$ to every GC in our model (we need $\sigma$ separately to calculate the BIMBH ``hard binary separation''). 

We have also simulated the GC mass $M_{\rm GC}$ loss due to tidal disruption in the galactic field, stellar evolution and ejections of stars via two-body relaxation. Depending on the IMBH mass, we found that the GCs lighter than $\sim\num{1.5e5}$--$\num{4e5}\msun$ evaporate before their BIMBH merge. However, in a small fraction of disrupted GCs the two IMBHs are close enough to merge via GW emission at the moment of disruption.
We found that the coalescence time decreases with $M_{\rm GC}$. As a result, the BIMBH mergers are only produced in GCs more massive than a critical GC mass.

We calculated the rate of BIMBH inspirals and mergers and the prospects for their detection with LISA, LIGO and ET. Assuming $\alpha_{\rm IMBH}=0.1$, { circular BIMBHs} and that every BIMBH mass is $\mu=10^{-2}$ of its GC mass, we found that LISA may detect $\sim100$ BIBMH inspirals over its expected 4-year mission length{, with a weak dependence on $\mu$ ($\propto\mu^{0.39}$)}. 
LIGO and ET would only detect IMBHs with sufficiently low masses; we would expect at least one detection 
if $\mu<0.01$ and $\mu<0.015$, respectively. For lower values of $\mu$, e.g. $10^{-3}$, the detection rate would be $\sim1$ event/yr for LIGO and $\sim5$ events/yr for ET (for the same value of $\alpha_{\rm IMBH}=0.1$). Given the different assumptions about GC number density in the Universe, IMBH masses and sensitivity curves, our results agree with previous papers \citep[][{ see Fig.~\ref{table}}]{fregeau2006,gair2011,santamaria} and also are well within the current aLIGO/Virgo upper limits \citep{ligoIMBH}. { These rate estimates, however, rely on a number of assumptions, such as the absence of loss cone depletion and recoil kicks during 3-body interactions, and should probably be considered upper limits.}
One interesting possibility { that we have not considered} is the possible detection of stochastic GW background produced by numerous unresolved BIMBHs at low frequencies.

{
We have also made some calculations assuming nonzero initial eccentricity of $0.6$. In this case, the merger timescales are decreased by about an order of magnitude. This leads to an increase in detection rate of $\sim 4$--$10$ times for aLIGO and ET and up to $\sim 2.5$ times for LISA (depending on $\mu$). At the same time, shorter merger timescales imply BIMBHs merging at higher redshifts, which makes them harder to detect, which is the reason why the increase is not as high at low $\mu$. We have also found that by the time BIMBHs are detected, they are already deep enough in the GW-dominated regime so that their eccentricity is below $10^{-3}$ at the moment of detection.
}

\begin{table}
\caption{Estimates of BIMBH detection rate from the literature, yr$^{-1}$}
\centering
\renewcommand{\arraystretch}{2}
\renewcommand\cellgape{\Gape[4pt]}
\begin{tabular}{|c|c|c|c|}
\hline
Paper & LISA & aLIGO & ET \\
\hline
\citet{fregeau2006} & 20--50 & $\sim10$ & \\
\hline
\citet{santamaria} & & $\sim1$ & $\sim20$ \\
\hline
\citet{gair2011} & & & $\sim2000$ \\
\hline
This paper, $\alpha=0.1$, $\mu=10^{-2}$ & 22 & 0 & 1.6 \\ 
\hline
This paper, $\alpha=0.1$, $\mu=10^{-3}$ & 8.9 & 1.1 & 4.6 \\
\hline
\end{tabular}
\label{table}
\end{table}

{We calculated the expected distribution of SNR of GW detections with LISA from all BIMBHs in the Universe (Fig.~\ref{fig:snr}). Since these sources originate at cosmological distances and because they are localized in a limited range of frequency, the distribution does not scale like SNR$^{-4}$ as for LIGO sources in the O1 and O2 observing runs \citep{ligoO2}. The LISA SNR distribution of BIMBHs is bimodal, where the two modes respectively correspond to sources which are merging and which inspiral but do not merge during the observation. The distribution is asymptotically proportional to $\dd N/\dd \mathrm{SNR} \propto \mathrm{SNR}^{-2.3}$ for both very small and large SNR. Most of the individually detectable sources represent merging sources.}

We have also calculated the distribution of final spins and kick velocities of BIMBH merger remnants. These are independent of the total BIMBH masses, and may only depend on the mass ratio $q$ {and the initial spin. For a uniform distribution of the initial spin,} the spin distribution peaks at $\sim0.5$ when $q\ll1$ and $\sim0.7$ when $q=1$. Given $q>0.1$, the ejection velocity is in the range $\sim10^2-10^3\,\si{km/s}$ which implies that approximately all IMBH merger remnants will escape their host GC.

In this work, we have only considered the single-cluster BIMBH formation channel. Mergers of two clusters each hosting an IMBH may represent an independent channel. It was recently shown \citep{gcMergers,gcMergers1} that GCs born in the MW disc can undergo close encounters and merge at a rate of $\sim1.8\,\mathrm{Gyr}^{-1}$. Given the redshift dependence of GC megrer rate, the method used in this paper may be generalized to estimate the detection rate in the two-cluster channel.

Apart from an IMBH-IMBH merger considered in this work, IMBHs could also potentially merge with one of stellar-mass BHs in its host GC \citep{frag2018b} or the supermassive BH if the GC gets disrupted close enough to the galactic center \citep{GCinspiral1,GCinspiral2}. Although not as robust as GW emission, there are other ways for a binary IMBH to observationally manifest itself. One of them is tidal disruption of stars and white dwarfs \citep{frag2018b}, which is boosted for a binary IMBH compared to a single one \citep{chen2009,flei2018}. In particular, white dwarfs can only be tidally disrupted by BHs in the IMBH mass range or below. Also, some of the stars a BIMBH interacts with are ejected from the GC with high velocities and have a chance to become detectable as hypervelocity stars in our galaxy \citep{fragual2019MNRAS,subr19}. 

The same dynamical mechanism is known to cause ``scouring'' of galactic cores \citep[Section 8.2.2]{DEGN}, so it is possible that a similar phenomenon happens in GCs as well. To estimate the potential significance of that latter effect, we can compare the total energy transferred from the BIMBH to the ejected stars $E_{\rm bin}\approx G\mimbh/a_{\rm GW}$ (where $a_{\rm GW}$ is the semimajor axis at which hardening due to GW emssion becomes stronger than the one due to stellar ejections) with the GC binding energy $E_{\rm GC}\sim G\mgc/r_h$. Given $\mu=10^{-3}$ and $\mgc=10^6\msun$, $a_{\rm GW}\approx\SI{3}{\mu pc}$. Typical $r_h\sim\SI{1}{pc}$, so that
\eq{
\frac{E_{\rm bin}}{E_{\rm GC}} &= \qty(\frac{\mimbh}{\mgc})^2 \frac{r_h}{a_{\rm GW}} \nonumber\\
&\approx 0.3 \, \qty(\frac{\mu}{10^{-3}})^2 \frac{r_h}{\SI{1}{pc}} \qty(\frac{a_{\rm GW}}{\SI{3}{\mu pc}})^{-1}.
}
These two energies are comparable, and therefore the stellar ejections could significantly affect the GC dynamical evolution, which we leave for a future study.

\acknowledgments
This work received funding from the European Research Council (ERC) under the European Union’s Horizon 2020 Programme for Research and Innovation ERC-2014-STG under grant agreement No. 638435 (GalNUC), and from the Hungarian National Research, Development, and Innovation Office under grant NKFIH KH-125675. GF acknowledges support from a CIERA postdoctoral fellowship at Northwestern University.

\bibliographystyle{yahapj}
\bibliography{bib}

\begin{thebibliography}{}
\providecommand\natexlab[1]{#1}
\providecommand\JournalTitle[1]{#1}

\bibitem[{Abbate {et~al.}(2019)Abbate, Possenti, Colpi, \& Spera}]{Abbate2019}
Abbate, F., Possenti, A., Colpi, M., \& Spera, M. 2019,
  \href{http://dx.doi.org/10.3847/2041-8213/ab46c3}{\JournalTitle{The
  Astrophysical Journal}, 884, L9}

\bibitem[{Abbott {et~al.}(2019{\natexlab{a}})Abbott, Abbott, Abbott, Abraham,
  Acernese, Ackley, Adams, Adhikari, Adya, Affeldt, Agathos, Agatsuma,
  Aggarwal, Aguiar, Aiello, Ain, Ajith, Allen, Allocca, Aloy, Altin, Amato,
  Ananyeva, Anderson, Anderson, Angelova, Antier, Appert, Arai, Araya, Areeda,
  Ar{\`{e}}ne, Arnaud, Arun, Ascenzi, Ashton, Aston, Astone, Aubin, Aufmuth,
  AultONeal, Austin, Avendano, Avila-Alvarez, Babak, Bacon, Badaracco, Bader,
  Bae, Baker, Baldaccini, Ballardin, Ballmer, Banagiri, Barayoga, Barclay,
  Barish, Barker, Barkett, Barnum, Barone, Barr, Barsotti, Barsuglia, Barta,
  Bartlett, Bartos, Bassiri, Basti, Bawaj, Bayley, Bazzan, B{\'{e}}csy, Bejger,
  Belahcene, Bell, Beniwal, Berger, Bergmann, Bernuzzi, Bero, Berry,
  Bersanetti, Bertolini, Betzwieser, Bhandare, Bidler, Bilenko, Bilgili,
  Billingsley, Birch, Birney, Birnholtz, Biscans, Biscoveanu, Bisht, Bitossi,
  Bizouard, Blackburn, Blair, Blair, Blair, Bloemen, Bode, Boer, Boetzel,
  Bogaert, Bondu, Bonilla, Bonnand, Booker, Boom, Booth, Bork, Boschi, Bose,
  Bossie, Bossilkov, Bosveld, Bouffanais, Bozzi, Bradaschia, Brady, Bramley,
  Branchesi, Brau, Briant, Briggs, Brighenti, Brillet, Brinkmann, Brisson,
  Brockill, Brooks, Brown, Brunett, Buikema, Bulik, Bulten, Buonanno,
  Buscicchio, Buskulic, Buy, Byer, Cabero, Cadonati, Cagnoli, Cahillane,
  Bustillo, Callister, Calloni, Camp, Campbell, Canepa, Cannon, Cao, Cao,
  Capocasa, Carbognani, Caride, Carney, Carullo, Diaz, Casentini, Caudill,
  Cavagli{\`{a}}, Cavalier, Cavalieri, Cella, Cerd{\'{a}}-Dur{\'{a}}n,
  Cerretani, Cesarini, Chaibi, Chakravarti, Chamberlin, Chan, Chao, Charlton,
  Chase, Chassande-Mottin, Chatterjee, Chaturvedi, Chatziioannou, Cheeseboro,
  Chen, Chen, Chen, Cheng, Cheong, Chia, Chincarini, Chiummo, Cho, Cho, Cho,
  Christensen, Chu, Chua, Chung, Chung, Ciani, Ciobanu, Ciolfi, Cipriano,
  Cirone, Clara, Clark, Clearwater, Cleva, Cocchieri, Coccia, Cohadon, Cohen,
  Colgan, Colleoni, Collette, Collins, Cominsky, Jr., Conti, Cooper, Corban,
  Corbitt, Cordero-Carri{\'{o}}n, Corley, Cornish, Corsi, Cortese, Costa,
  Cotesta, Coughlin, Coughlin, Coulon, Countryman, Couvares, Covas, Cowan,
  Coward, Cowart, Coyne, Coyne, Creighton, Creighton, Cripe, Croquette,
  Crowder, Cullen, Cumming, Cunningham, Cuoco, Canton, D{\'{a}}lya, Danilishin,
  D'Antonio, Danzmann, Dasgupta, Costa, Datrier, Dattilo, Dave, Davier, Davis,
  Daw, DeBra, Deenadayalan, Degallaix, Laurentis, Del{\'{e}}glise, Pozzo,
  DeMarchi, Demos, Dent, Pietri, Derby, Rosa, Rossi, DeSalvo, de~Varona,
  Dhurandhar, D{\'{\i}}az, Dietrich, Fiore, Giovanni, Girolamo, Lieto, Ding,
  Pace, Palma, Renzo, Dmitriev, Doctor, Donovan, Dooley, Doravari, Dorrington,
  Downes, Drago, Driggers, Du, Ducoin, Dupej, Dwyer, Easter, Edo, Edwards,
  Effler, Ehrens, Eichholz, Eikenberry, Eisenmann, Eisenstein, Essick,
  Estelles, Estevez, Etienne, Etzel, Evans, Evans, Fafone, Fair, Fairhurst,
  Fan, Farinon, Farr, Farr, Fauchon-Jones, Favata, Fays, Fazio, Fee, Feicht,
  Fejer, Feng, Fernandez-Galiana, Ferrante, Ferreira, Ferreira, Ferrini,
  Fidecaro, Fiori, Fiorucci, Fishbach, Fisher, Fishner, Fitz-Axen, Flaminio,
  Fletcher, Flynn, Fong, Font, Forsyth, Fournier, Frasca, Frasconi, Frei,
  Freise, Frey, Frey, Fritschel, Frolov, Fulda, Fyffe, Gabbard, Gadre, Gaebel,
  Gair, Gammaitoni, Ganija, Gaonkar, Garcia, Garc{\'{\i}}a-Quir{\'{o}}s,
  Garufi, Gateley, Gaudio, Gaur, Gayathri, Gemme, Genin, Gennai, George,
  George, Gergely, Germain, Ghonge, Ghosh, Ghosh, Ghosh, Giacomazzo, Giaime,
  Giardina, Giazotto, Gill, Giordano, Glover, Godwin, Goetz, Goetz, Goncharov,
  Gonz{\'{a}}lez, Castro, Gopakumar, Gorodetsky, Gossan, Gosselin, Gouaty,
  Grado, Graef, Granata, Grant, Gras, Grassia, Gray, Gray, Greco, Green, Green,
  Gretarsson, Groot, Grote, Grunewald, Gruning, Guidi, Gulati, Guo, Gupta,
  Gupta, Gustafson, Gustafson, Haegel, Halim, Hall, Hall, Hamilton, Hammond,
  Haney, Hanke, Hanks, Hanna, Hannam, Hannuksela, Hanson, Hardwick, Haris,
  Harms, Harry, Harry, Haster, Haughian, Hayes, Healy, Heidmann, Heintze,
  Heitmann, Hello, Hemming, Hendry, Heng, Hennig, Heptonstall, Vivanco, Heurs,
  Hild, Hinderer, Hoak, Hochheim, Hofman, Holgado, Holland, Holt, Holz,
  Hopkins, Horst, Hough, Howell, Hoy, Hreibi, Huerta, Huet, Hughey, Hulko,
  Husa, Huttner, Huynh-Dinh, Idzkowski, Iess, Ingram, Inta, Intini, Irwin, Isa,
  Isac, Isi, Iyer, Izumi, Jacqmin, Jadhav, Jani, Janthalur, Jaranowski,
  Jenkins, Jiang, Johnson, Jones, Jones, Jones, Jonker, Ju, Junker, Kalaghatgi,
  Kalogera, Kamai, Kandhasamy, Kang, Kanner, Kapadia, Karki, Karvinen, Kashyap,
  Kasprzack, Katsanevas, Katsavounidis, Katzman, Kaufer, Kawabe, Keerthana,
  K{\'{e}}f{\'{e}}lian, Keitel, Kennedy, Key, Khalili, Khan, Khan, Khan, Khan,
  Khazanov, Khursheed, Kijbunchoo, Kim, Kim, Kim, Kim, Kim, Kim, Kimball, King,
  King, Kinley-Hanlon, Kirchhoff, Kissel, Kleybolte, Klika, Klimenko, Knowles,
  Koch, Koehlenbeck, Koekoek, Koley, Kondrashov, Kontos, Koper, Korobko, Korth,
  Kowalska, Kozak, Kringel, Krishnendu, Kr{\'{o}}lak, Kuehn, Kumar, Kumar,
  Kumar, Kumar, Kuo, Kutynia, Kwang, Lackey, Lai, Lam, Landry, Lane, Lang,
  Lange, Lantz, Lanza, Lartaux-Vollard, Lasky, Laxen, Lazzarini, Lazzaro,
  Leaci, Leavey, Lecoeuche, Lee, Lee, Lee, Lee, Lee, Lee, Lehmann, Lenon,
  Leroy, Letendre, Levin, Li, Li, Li, Li, Lin, Linde, Linker, Littenberg, Liu,
  Liu, Lo, Lockerbie, London, Longo, Lorenzini, Loriette, Lormand, Losurdo,
  Lough, Lousto, Lovelace, Lower, Lück, Lumaca, Lundgren, Lynch, Ma, Macas,
  Macfoy, MacInnis, Macleod, Macquet, Maga{\~{n}}a-Sandoval, Zertuche, Magee,
  Majorana, Maksimovic, Malik, Man, Mandic, Mangano, Mansell, Manske,
  Mantovani, Mapelli, Marchesoni, Marion, M{\'{a}}rka, M{\'{a}}rka, Markakis,
  Markosyan, Markowitz, Maros, Marquina, Marsat, Martelli, Martin, Martin,
  Martynov, Mason, Massera, Masserot, Massinger, Masso-Reid, Mastrogiovanni,
  Matas, Matichard, Matone, Mavalvala, Mazumder, McCann, McCarthy, McClelland,
  McCormick, McCuller, McGuire, McIver, McManus, McRae, McWilliams, Meacher,
  Meadors, Mehmet, Mehta, Meidam, Melatos, Mendell, Mercer, Mereni, Merilh,
  Merzougui, Meshkov, Messenger, Messick, Metzdorff, Meyers, Miao, Michel,
  Middleton, Mikhailov, Milano, Miller, Miller, Millhouse, Mills,
  Milovich-Goff, Minazzoli, Minenkov, Mishkin, Mishra, Mistry, Mitra,
  Mitrofanov, Mitselmakher, Mittleman, Mo, Moffa, Mogushi, Mohapatra, Montani,
  Moore, Moraru, Moreno, Morisaki, Mours, Mow-Lowry, Mukherjee, Mukherjee,
  Mukherjee, Mukund, Mullavey, Munch, Mu{\~{n}}iz, Muratore, Murray, Nagar,
  Nardecchia, Naticchioni, Nayak, Neilson, Nelemans, Nelson, Nery, Neunzert,
  Ng, Ng, Nguyen, Nichols, Nissanke, Nocera, North, Nuttall, Obergaulinger,
  Oberling, O'Brien, O'Dea, Ogin, Oh, Oh, Ohme, Ohta, Okada, Oliver, Oppermann,
  Oram, O'Reilly, Ormiston, Ortega, O'Shaughnessy, Ossokine, Ottaway, Overmier,
  Owen, Pace, Pagano, Page, Pai, Pai, Palamos, Palashov, Palomba, Pal-Singh,
  Pan, Pang, Pang, Pankow, Pannarale, Pant, Paoletti, Paoli, Parida, Parker,
  Pascucci, Pasqualetti, Passaquieti, Passuello, Patil, Patricelli, Pearlstone,
  Pedersen, Pedraza, Pedurand, Pele, Penn, Perez, Perreca, Pfeiffer, Phelps,
  Phukon, Piccinni, Pichot, Piergiovanni, Pillant, Pinard, Pirello, Pitkin,
  Poggiani, Pong, Ponrathnam, Popolizio, Porter, Powell, Prajapati, Prasad,
  Prasai, Prasanna, Pratten, Prestegard, Privitera, Prodi, Prokhorov, Puncken,
  Punturo, Puppo, Pürrer, Qi, Quetschke, Quinonez, Quintero, Quitzow-James,
  Raab, Radkins, Radulescu, Raffai, Raja, Rajan, Rajbhandari, Rakhmanov,
  Ramirez, Ramos-Buades, Rana, Rao, Rapagnani, Raymond, Razzano, Read,
  Regimbau, Rei, Reid, Reitze, Ren, Ricci, Richardson, Richardson, Ricker,
  Riles, Rizzo, Robertson, Robie, Robinet, Rocchi, Rolland, Rollins, Roma,
  Romanelli, Romano, Romel, Romie, Rose, Rosi{\'{n}}ska, Rosofsky, Ross, Rowan,
  Rüdiger, Ruggi, Rutins, Ryan, Sachdev, Sadecki, Sakellariadou, Salconi,
  Saleem, Samajdar, Sammut, Sanchez, Sanchez, Sanchis-Gual, Sandberg, Sanders,
  Santiago, Sarin, Sassolas, Sathyaprakash, Saulson, Sauter, Savage, Schale,
  Scheel, Scheuer, Schmidt, Schnabel, Schofield, Schönbeck, Schreiber,
  Schulte, Schutz, Schwalbe, Scott, Scott, Seidel, Sellers, Sengupta, Sennett,
  Sentenac, Sequino, Sergeev, Setyawati, Shaddock, Shaffer, Shahriar, Shaner,
  Shao, Sharma, Shawhan, Shen, Shink, Shoemaker, Shoemaker, ShyamSundar,
  Siellez, Sieniawska, Sigg, Silva, Singer, Singh, Singhal, Sintes,
  Sitmukhambetov, Skliris, Slagmolen, Slaven-Blair, Smith, Smith, Somala, Son,
  Sorazu, Sorrentino, Souradeep, Sowell, Spencer, Spera, Srivastava,
  Srivastava, Staats, Stachie, Standke, Steer, Steinke, Steinlechner,
  Steinlechner, Steinmeyer, Stevenson, Stocks, Stone, Stops, Strain, Stratta,
  Strigin, Strunk, Sturani, Stuver, Sudhir, Summerscales, Sun, Sunil, Suresh,
  Sutton, Swinkels, Szczepa{\'{n}}czyk, Tacca, Tait, Talbot, Talukder, Tanner,
  T{\'{a}}pai, Taracchini, Tasson, Taylor, Thies, Thomas, Thomas, Thondapu,
  Thorne, Thrane, Tiwari, Tiwari, Tiwari, Toland, Tonelli, Tornasi,
  Torres-Forn{\'{e}}, Torrie, Töyrä, Travasso, Traylor, Tringali, Trovato,
  Trozzo, Trudeau, Tsang, Tse, Tso, Tsukada, Tsuna, Tuyenbayev, Ueno, Ugolini,
  Unnikrishnan, Urban, Usman, Vahlbruch, Vajente, Valdes, van Bakel, van
  Beuzekom, van~den Brand, Broeck, Vander-Hyde, van~der Schaaf, van Heijningen,
  van Veggel, Vardaro, Varma, Vass, Vas{\'{u}}th, Vecchio, Vedovato, Veitch,
  Veitch, Venkateswara, Venugopalan, Verkindt, Vetrano, Vicer{\'{e}}, Viets,
  Vine, Vinet, Vitale, Vo, Vocca, Vorvick, Vyatchanin, Wade, Wade, Wade, Walet,
  Walker, Wallace, Walsh, Wang, Wang, Wang, Wang, Wang, Ward, Warden, Warner,
  Was, Watchi, Weaver, Wei, Weinert, Weinstein, Weiss, Wellmann, Wen, Wessel,
  We{\ss}els, Westhouse, Wette, Whelan, Whiting, Whittle, Wilken, Williams,
  Williamson, Willis, Willke, Wimmer, Winkler, Wipf, Wittel, Woan, Woehler,
  Wofford, Worden, Wright, Wu, Wysocki, Xiao, Yamamoto, Yancey, Yang, Yap,
  Yazback, Yeeles, Yu, Yu, Yuen, Yvert, Zadro{\.{z}}ny, Zanolin, Zelenova,
  Zendri, Zevin, Zhang, Zhang, Zhang, Zhao, Zhou, Zhou, Zhu, Zimmerman,
  Zlochower, Zucker, \& and}]{bhmass2}
Abbott, B.~P., Abbott, R., Abbott, T.~D., {et~al.} 2019{\natexlab{a}},
  \href{http://dx.doi.org/10.3847/2041-8213/ab3800}{\JournalTitle{The
  Astrophysical Journal}, 882, L24}

\bibitem[{Abbott {et~al.}(2019{\natexlab{b}})Abbott, Abbott, Abbott, Abraham,
  Acernese, Ackley, Adams, Adhikari, Adya, Affeldt, Agathos, Agatsuma,
  Aggarwal, Aguiar, Aiello, Ain, Ajith, Allen, Allocca, Aloy, Altin, Amato,
  Ananyeva, Anderson, Anderson, Angelova, Antier, Appert, Arai, Araya, Areeda,
  Ar\`ene, Arnaud, Arun, Ascenzi, Ashton, Aston, Astone, Aubin, Aufmuth,
  AultONeal, Austin, Avendano, Avila-Alvarez, Babak, Bacon, Badaracco, Bader,
  Bae, Baker, Baldaccini, Ballardin, Ballmer, Banagiri, Barayoga, Barclay,
  Barish, Barker, Barkett, Barnum, Barone, Barr, Barsotti, Barsuglia, Barta,
  Bartlett, Bartos, Bassiri, Basti, Bawaj, Bayley, Bazzan, B\'ecsy, Bejger,
  Belahcene, Bell, Beniwal, Berger, Bergmann, Bernuzzi, Bero, Berry,
  Bersanetti, Bertolini, Betzwieser, Bhandare, Bidler, Bilenko, Bilgili,
  Billingsley, Birch, Birney, Birnholtz, Biscans, Biscoveanu, Bisht, Bitossi,
  Bizouard, Blackburn, Blackman, Blair, Blair, Blair, Bloemen, Bode, Boer,
  Boetzel, Bogaert, Bondu, Bonilla, Bonnand, Booker, Boom, Booth, Bork, Boschi,
  Bose, Bossie, Bossilkov, Bosveld, Bouffanais, Bozzi, Bradaschia, Brady,
  Bramley, Branchesi, Brau, Briant, Briggs, Brighenti, Brillet, Brinkmann,
  Brisson, Brockill, Brooks, Brown, Brunett, Buikema, Bulik, Bulten, Buonanno,
  Buskulic, Bustamante~Rosell, Buy, Byer, Cabero, Cadonati, Cagnoli, Cahillane,
  Calder\'on~Bustillo, Callister, Calloni, Camp, Campbell, Canepa, Cannon, Cao,
  Cao, Capocasa, Carbognani, Caride, Carney, Carullo, Casanueva~Diaz,
  Casentini, Caudill, Cavagli\`a, Cavalier, Cavalieri, Cella, Cerd\'a-Dur\'an,
  Cerretani, Cesarini, Chaibi, Chakravarti, Chamberlin, Chan, Chao, Charlton,
  Chase, Chassande-Mottin, Chatterjee, Chaturvedi, Chatziioannou, Cheeseboro,
  Chen, Chen, Chen, Cheng, Cheong, Chia, Chincarini, Chiummo, Cho, Cho, Cho,
  Christensen, Chu, Chua, Chung, Chung, Ciani, Ciobanu, Ciolfi, Cipriano,
  Cirone, Clara, Clark, Clearwater, Cleva, Cocchieri, Coccia, Cohadon, Cohen,
  Colgan, Colleoni, Collette, Collins, Cominsky, Constancio, Conti, Cooper,
  Corban, Corbitt, Cordero-Carri\'on, Corley, Cornish, Corsi, Cortese, Costa,
  Cotesta, Coughlin, Coughlin, Coulon, Countryman, Couvares, Covas, Cowan,
  Coward, Cowart, Coyne, Coyne, Creighton, Creighton, Cripe, Croquette,
  Crowder, Cullen, Cumming, Cunningham, Cuoco, Canton, D\'alya, Danilishin,
  D'Antonio, Danzmann, Dasgupta, Da~Silva~Costa, Datrier, Dattilo, Dave,
  Davier, Davis, Daw, DeBra, Deenadayalan, Degallaix, De~Laurentis,
  Del\'eglise, Del~Pozzo, DeMarchi, Demos, Dent, De~Pietri, Derby, De~Rosa,
  De~Rossi, DeSalvo, de~Varona, Dhurandhar, D\'{\i}az, Dietrich, Di~Fiore,
  Di~Giovanni, Di~Girolamo, Di~Lieto, Ding, Di~Pace, Di~Palma, Di~Renzo,
  Dmitriev, Doctor, Donovan, Dooley, Doravari, Dorrington, Downes, Drago,
  Driggers, Du, Ducoin, Dupej, Dwyer, Easter, Edo, Edwards, Effler, Ehrens,
  Eichholz, Eikenberry, Eisenmann, Eisenstein, Essick, Estelles, Estevez,
  Etienne, Etzel, Evans, Evans, Fafone, Fair, Fairhurst, Fan, Farinon, Farr,
  Farr, Fauchon-Jones, Favata, Fays, Fazio, Fee, Feicht, Fejer, Feng,
  Fernandez-Galiana, Ferrante, Ferreira, Ferreira, Ferrini, Fidecaro, Fiori,
  Fiorucci, Fishbach, Fisher, Fishner, Fitz-Axen, Flaminio, Fletcher, Flynn,
  Fong, Font, Forsyth, Fournier, Frasca, Frasconi, Frei, Freise, Frey, Frey,
  Fritschel, Frolov, Fulda, Fyffe, Gabbard, Gadre, Gaebel, Gair, Gammaitoni,
  Ganija, Gaonkar, Garcia, Garc\'{\i}a-Quir\'os, Garufi, Gateley, Gaudio, Gaur,
  Gayathri, Gemme, Genin, Gennai, George, George, Gergely, Germain, Ghonge,
  Ghosh, Ghosh, Ghosh, Giacomazzo, Giaime, Giardina, Giazotto, Gill, Giordano,
  Glover, Godwin, Goetz, Goetz, Goncharov, Gonz\'alez, Gonzalez~Castro,
  Gopakumar, Gorodetsky, Gossan, Gosselin, Gouaty, Grado, Graef, Granata,
  Grant, Gras, Grassia, Gray, Gray, Greco, Green, Green, Gretarsson, Groot,
  Grote, Grunewald, Gruning, Guidi, Gulati, Guo, Gupta, Gupta, Gustafson,
  Gustafson, Haegel, Halim, Hall, Hall, Hamilton, Hammond, Haney, Hanke, Hanks,
  Hanna, Hannam, Hannuksela, Hanson, Hardwick, Haris, Harms, Harry, Harry,
  Haster, Haughian, Hayes, Healy, Heidmann, Heintze, Heitmann, Hello, Hemming,
  Hendry, Heng, Hennig, Heptonstall, Hernandez~Vivanco, Heurs, Hild, Hinderer,
  Hoak, Hochheim, Hofman, Holgado, Holland, Holt, Holz, Hopkins, Horst, Hough,
  Howell, Hoy, Hreibi, Huang, Huerta, Huet, Hughey, Hulko, Husa, Huttner,
  Huynh-Dinh, Idzkowski, Iess, Ingram, Inta, Intini, Irwin, Isa, Isac, Isi,
  Iyer, Izumi, Jacqmin, Jadhav, Jani, Janthalur, Jaranowski, Jenkins, Jiang,
  Johnson, Johnson-McDaniel, Jones, Jones, Jones, Jonker, Ju, Junker,
  Kalaghatgi, Kalogera, Kamai, Kandhasamy, Kang, Kanner, Kapadia, Karki,
  Karvinen, Kashyap, Kasprzack, Katsanevas, Katsavounidis, Katzman, Kaufer,
  Kawabe, Keerthana, K\'ef\'elian, Keitel, Kennedy, Key, Khalili, Khan, Khan,
  Khan, Khan, Khazanov, Khursheed, Kijbunchoo, Kim, Kim, Kim, Kim, Kim, Kim,
  Kimball, King, King, Kinley-Hanlon, Kirchhoff, Kissel, Kleybolte, Klika,
  Klimenko, Knowles, Koch, Koehlenbeck, Koekoek, Koley, Kondrashov, Kontos,
  Koper, Korobko, Korth, Kowalska, Kozak, Kringel, Krishnendu, Kr\'olak, Kuehn,
  Kumar, Kumar, Kumar, Kumar, Kuo, Kutynia, Kwang, Lackey, Lai, Lam, Landry,
  Lane, Lang, Lange, Lantz, Lanza, Lartaux-Vollard, Lasky, Laxen, Lazzarini,
  Lazzaro, Leaci, Leavey, Lecoeuche, Lee, Lee, Lee, Lee, Lee, Lee, Lehmann,
  Lenon, Leroy, Letendre, Levin, Li, Li, Li, Li, Lin, Linde, Linker,
  Littenberg, Liu, Liu, Lo, Lockerbie, London, Longo, Lorenzini, Loriette,
  Lormand, Losurdo, Lough, Lousto, Lovelace, Lower, L\"uck, Lumaca, Lundgren,
  Lynch, Ma, Macas, Macfoy, MacInnis, Macleod, Macquet, Maga\~na Sandoval,
  Maga\~na Zertuche, Magee, Majorana, Maksimovic, Malik, Man, Mandic, Mangano,
  Mansell, Manske, Mantovani, Marchesoni, Marion, M\'arka, M\'arka, Markakis,
  Markosyan, Markowitz, Maros, Marquina, Marsat, Martelli, Martin, Martin,
  Martynov, Mason, Massera, Masserot, Massinger, Masso-Reid, Mastrogiovanni,
  Matas, Matichard, Matone, Mavalvala, Mazumder, McCann, McCarthy, McClelland,
  McCormick, McCuller, McGuire, McIver, McManus, McRae, McWilliams, Meacher,
  Meadors, Mehmet, Mehta, Meidam, Melatos, Mendell, Mercer, Mereni, Merilh,
  Merzougui, Meshkov, Messenger, Messick, Metzdorff, Meyers, Miao, Michel,
  Middleton, Mikhailov, Milano, Miller, Miller, Millhouse, Mills,
  Milovich-Goff, Minazzoli, Minenkov, Mishkin, Mishra, Mistry, Mitra,
  Mitrofanov, Mitselmakher, Mittleman, Mo, Moffa, Mogushi, Mohapatra, Montani,
  Moore, Moraru, Moreno, Morisaki, Mours, Mow-Lowry, Mukherjee, Mukherjee,
  Mukherjee, Mukund, Mullavey, Munch, Mu\~niz, Muratore, Murray, Nagar,
  Nardecchia, Naticchioni, Nayak, Neilson, Nelemans, Nelson, Nery, Neunzert,
  Ng, Ng, Nguyen, Nichols, Nielsen, Nissanke, Nitz, Nocera, North, Nuttall,
  Obergaulinger, Oberling, O'Brien, O'Dea, Ogin, Oh, Oh, Ohme, Ohta, Okada,
  Oliver, Oppermann, Oram, O'Reilly, Ormiston, Ortega, O'Shaughnessy, Ossokine,
  Ottaway, Overmier, Owen, Pace, Pagano, Page, Pai, Pai, Palamos, Palashov,
  Palomba, Pal-Singh, Pan, Pang, Pang, Pankow, Pannarale, Pant, Paoletti,
  Paoli, Papa, Parida, Parker, Pascucci, Pasqualetti, Passaquieti, Passuello,
  Patil, Patricelli, Pearlstone, Pedersen, Pedraza, Pedurand, Pele, Penn,
  Perego, Perez, Perreca, Pfeiffer, Phelps, Phukon, Piccinni, Pichot,
  Piergiovanni, Pillant, Pinard, Pirello, Pitkin, Poggiani, Pong, Ponrathnam,
  Popolizio, Porter, Powell, Prajapati, Prasad, Prasai, Prasanna, Pratten,
  Prestegard, Privitera, Prodi, Prokhorov, Puncken, Punturo, Puppo, P\"urrer,
  Qi, Quetschke, Quinonez, Quintero, Quitzow-James, Raab, Radkins, Radulescu,
  Raffai, Raja, Rajan, Rajbhandari, Rakhmanov, Ramirez, Ramos-Buades, Rana,
  Rao, Rapagnani, Raymond, Razzano, Read, Regimbau, Rei, Reid, Reitze, Ren,
  Ricci, Richardson, Richardson, Ricker, Riemenschneider, Riles, Rizzo,
  Robertson, Robie, Robinet, Rocchi, Rolland, Rollins, Roma, Romanelli, Romano,
  Romel, Romie, Rose, Rosi\ifmmode~\acute{n}\else \'{n}\fi{}ska, Rosofsky,
  Ross, Rowan, R\"udiger, Ruggi, Rutins, Ryan, Sachdev, Sadecki, Sakellariadou,
  Salafia, Salconi, Saleem, Salemi, Samajdar, Sammut, Sanchez, Sanchez,
  Sanchis-Gual, Sandberg, Sanders, Santiago, Sarin, Sassolas, Sathyaprakash,
  Saulson, Sauter, Savage, Schale, Scheel, Scheuer, Schmidt, Schnabel,
  Schofield, Sch\"onbeck, Schreiber, Schulte, Schutz, Schwalbe, Scott, Scott,
  Seidel, Sellers, Sengupta, Sennett, Sentenac, Sequino, Sergeev, Setyawati,
  Shaddock, Shaffer, Shahriar, Shaner, Shao, Sharma, Shawhan, Shen, Shink,
  Shoemaker, Shoemaker, ShyamSundar, Siellez, Sieniawska, Sigg, Silva, Singer,
  Singh, Singhal, Sintes, Sitmukhambetov, Skliris, Slagmolen, Slaven-Blair,
  Smith, Smith, Somala, Son, Sorazu, Sorrentino, Souradeep, Sowell, Spencer,
  Srivastava, Srivastava, Staats, Stachie, Standke, Steer, Steinke,
  Steinlechner, Steinlechner, Steinmeyer, Stevenson, Stocks, Stone, Stops,
  Strain, Stratta, Strigin, Strunk, Sturani, Stuver, Sudhir, Summerscales, Sun,
  Sunil, Suresh, Sutton, Swinkels, Szczepa\ifmmode~\acute{n}\else
  \'{n}\fi{}czyk, Tacca, Tait, Talbot, Talukder, Tanner, T\'apai, Taracchini,
  Tasson, Taylor, Thies, Thomas, Thomas, Thondapu, Thorne, Thrane, Tiwari,
  Tiwari, Tiwari, Toland, Tonelli, Tornasi, Torres-Forn\'e, Torrie, T\"oyr\"a,
  Travasso, Traylor, Tringali, Trovato, Trozzo, Trudeau, Tsang, Tse, Tso,
  Tsukada, Tsuna, Tuyenbayev, Ueno, Ugolini, Unnikrishnan, Urban, Usman,
  Vahlbruch, Vajente, Valdes, van Bakel, van Beuzekom, van~den Brand, Van
  Den~Broeck, Vander-Hyde, van Heijningen, van~der Schaaf, van Veggel, Vardaro,
  Varma, Vass, Vas\'uth, Vecchio, Vedovato, Veitch, Veitch, Venkateswara,
  Venugopalan, Verkindt, Vetrano, Vicer\'e, Viets, Vine, Vinet, Vitale, Vo,
  Vocca, Vorvick, Vyatchanin, Wade, Wade, Wade, Walet, Walker, Wallace, Walsh,
  Wang, Wang, Wang, Wang, Wang, Ward, Warden, Warner, Was, Watchi, Weaver, Wei,
  Weinert, Weinstein, Weiss, Wellmann, Wen, Wessel, We\ss{}els, Westhouse,
  Wette, Whelan, White, Whiting, Whittle, Wilken, Williams, Williamson, Willis,
  Willke, Wimmer, Winkler, Wipf, Wittel, Woan, Woehler, Wofford, Worden,
  Wright, Wu, Wysocki, Xiao, Yamamoto, Yancey, Yang, Yap, Yazback, Yeeles, Yu,
  Yu, Yuen, Yvert, Zadro\ifmmode~\dot{z}\else \.{z}\fi{}ny, Zanolin, Zappa,
  Zelenova, Zendri, Zevin, Zhang, Zhang, Zhang, Zhao, Zhou, Zhou, Zhu,
  Zimmerman, Zlochower, Zucker, \& Zweizig}]{ligoO2}
---. 2019{\natexlab{b}},
  \href{http://dx.doi.org/10.1103/PhysRevX.9.031040}{\JournalTitle{Phys. Rev.
  X}, 9, 031040}

\bibitem[{Abbott {et~al.}(2019{\natexlab{c}})Abbott, Abbott, Abbott, Abraham,
  Acernese, Ackley, Adams, Adams, Adhikari, Adya, Affeldt, Agathos, Agatsuma,
  Aggarwal, Aguiar, Aiello, Ain, Ajith, Allen, Allocca, Aloy, Altin, Amato,
  Anand, Ananyeva, Anderson, Anderson, Angelova, Antier, Appert, Arai, Araya,
  Areeda, Ar\`ene, Arnaud, Aronson, Arun, Ascenzi, Ashton, Aston, Astone,
  Aubin, Aufmuth, AultONeal, Austin, Avendano, Avila-Alvarez, Babak, Bacon,
  Badaracco, Bader, Bae, Baer, Baird, Baker, Baldaccini, Ballardin, Ballmer,
  Bals, Banagiri, Barayoga, Barbieri, Barclay, Barish, Barker, Barkett, Barnum,
  Barone, Barr, Barsotti, Barsuglia, Barta, Bartlett, Bartos, Bassiri, Basti,
  Bawaj, Bayley, Bazzan, B\'ecsy, Bejger, Belahcene, Bell, Beniwal, Benjamin,
  Berger, Bergmann, Bernuzzi, Berry, Bersanetti, Bertolini, Betzwieser,
  Bhandare, Bidler, Biggs, Bilenko, Bilgili, Billingsley, Birney, Birnholtz,
  Biscans, Bischi, Biscoveanu, Bisht, Bitossi, Bizouard, Blackburn, Blackman,
  Blair, Blair, Blair, Bloemen, Bobba, Bode, Boer, Boetzel, Bogaert, Bondu,
  Bonnand, Booker, Boom, Bork, Boschi, Bose, Bossilkov, Bosveld, Bouffanais,
  Bozzi, Bradaschia, Brady, Bramley, Branchesi, Brau, Breschi, Briant, Briggs,
  Brighenti, Brillet, Brinkmann, Brockill, Brooks, Brooks, Brown, Brunett,
  Buikema, Bulik, Bulten, Buonanno, Buskulic, Buy, Byer, Cabero, Cadonati,
  Cagnoli, Cahillane, Calder\'on~Bustillo, Callister, Calloni, Camp, Campbell,
  Canepa, Cannon, Cao, Cao, Carapella, Carbognani, Caride, Carney, Carullo,
  Casanueva~Diaz, Casentini, Caudill, Cavagli\`a, Cavalier, Cavalieri, Cella,
  Cerd\'a-Dur\'an, Cesarini, Chaibi, Chakravarti, Chamberlin, Chan, Chao,
  Charlton, Chase, Chassande-Mottin, Chatterjee, Chaturvedi, Cheeseboro, Chen,
  Chen, Chen, Cheng, Cheong, Chia, Chiadini, Chincarini, Chiummo, Cho, Cho,
  Cho, Christensen, Chu, Chua, Chung, Chung, Ciani, Cie\ifmmode~\acute{s}\else
  \'{s}\fi{}lar, Ciobanu, Ciolfi, Cipriano, Cirone, Clara, Clark, Clearwater,
  Cleva, Coccia, Cohadon, Cohen, Colleoni, Collette, Collins, Colpi, Cominsky,
  Constancio, Conti, Cooper, Corban, Corbitt, Cordero-Carri\'on, Corezzi,
  Corley, Cornish, Corre, Corsi, Cortese, Costa, Cotesta, Coughlin, Coughlin,
  Coulon, Countryman, Couvares, Covas, Cowan, Coward, Cowart, Coyne, Coyne,
  Creighton, Creighton, Cripe, Croquette, Crowder, Cullen, Cumming, Cunningham,
  Cuoco, Canton, D\'alya, D'Angelo, Danilishin, D'Antonio, Danzmann, Dasgupta,
  Da~Silva~Costa, Datrier, Dattilo, Dave, Davier, Davis, Daw, DeBra,
  Deenadayalan, Degallaix, De~Laurentis, Del\'eglise, Del~Pozzo, DeMarchi,
  Demos, Dent, De~Pietri, De~Rosa, De~Rossi, DeSalvo, de~Varona, Dhurandhar,
  D\'{\i}az, Dietrich, Di~Fiore, DiFronzo, Di~Giorgio, Di~Giovanni,
  Di~Giovanni, Di~Girolamo, Di~Lieto, Ding, Di~Pace, Di~Palma, Di~Renzo,
  Divakarla, Dmitriev, Doctor, Donovan, Dooley, Doravari, Dorrington, Downes,
  Drago, Driggers, Du, Ducoin, Dupej, Durante, Dwyer, Easter, Eddolls, Edo,
  Effler, Ehrens, Eichholz, Eikenberry, Eisenmann, Eisenstein, Errico, Essick,
  Estelles, Estevez, Etienne, Etzel, Evans, Evans, Fafone, Fairhurst, Fan,
  Farinon, Farr, Farr, Fauchon-Jones, Favata, Fays, Fazio, Fee, Feicht, Fejer,
  Feng, Ferguson, Fernandez-Galiana, Ferrante, Ferreira, Ferreira, Fidecaro,
  Fiori, Fiorucci, Fishbach, Fisher, Fishner, Fittipaldi, Fitz-Axen, Fiumara,
  Flaminio, Fletcher, Floden, Flynn, Fong, Font, Forsyth, Fournier, Vivanco,
  Frasca, Frasconi, Frei, Freise, Frey, Frey, Fritschel, Frolov, Fronz\`e,
  Fulda, Fyffe, Gabbard, Gadre, Gaebel, Gair, Gammaitoni, Gaonkar,
  Garc\'{\i}a-Quir\'os, Garufi, Gateley, Gaudio, Gaur, Gayathri, Gemme, Genin,
  Gennai, George, George, Gergely, Ghonge, Ghosh, Ghosh, Ghosh, Giacomazzo,
  Giaime, Giardina, Gibson, Gill, Glover, Gniesmer, Godwin, Goetz, Goetz,
  Goncharov, Gonz\'alez, Gonzalez~Castro, Gopakumar, Gossan, Gosselin, Gouaty,
  Grace, Grado, Granata, Grant, Gras, Grassia, Gray, Gray, Greco, Green, Green,
  Gretarsson, Grimaldi, Grimm, Groot, Grote, Grunewald, Gruning, Guidi, Gulati,
  Guo, Gupta, Gupta, Gupta, Gustafson, Gustafson, Haegel, Halim, Hall, Hall,
  Hamilton, Hammond, Haney, Hanke, Hanks, Hanna, Hannam, Hannuksela, Hansen,
  Hanson, Harder, Hardwick, Haris, Harms, Harry, Harry, Hasskew, Haster,
  Haughian, Hayes, Healy, Heidmann, Heintze, Heitmann, Hellman, Hello, Hemming,
  Hendry, Heng, Hennig, Heurs, Hild, Hinderer, Hochheim, Hofman, Holgado,
  Holland, Holt, Holz, Hopkins, Horst, Hough, Howell, Hoy, Huang, H\"ubner,
  Huerta, Huet, Hughey, Hui, Husa, Huttner, Huynh-Dinh, Idzkowski, Iess,
  Inchauspe, Ingram, Inta, Intini, Irwin, Isa, Isac, Isi, Iyer, Jacqmin,
  Jadhav, Jani, Janthalur, Jaranowski, Jariwala, Jenkins, Jiang, Johns,
  Johnson, Jones, Jones, Jones, Jones, Jonker, Ju, Junker, Kalaghatgi,
  Kalogera, Kamai, Kandhasamy, Kang, Kanner, Kapadia, Karki, Kashyap,
  Kasprzack, Katsanevas, Katsavounidis, Katzman, Kaufer, Kawabe, Keerthana,
  K\'ef\'elian, Keitel, Kennedy, Key, Khalili, Khamesra, Khan, Khan, Khazanov,
  Khetan, Khursheed, Kijbunchoo, Kim, Kim, Kim, Kim, Kim, Kim, Kim, Kimball,
  King, Kinley-Hanlon, Kirchhoff, Kissel, Kleybolte, Klika, Klimenko, Knowles,
  Koch, Koehlenbeck, Koekoek, Koley, Kondrashov, Kontos, Koper, Korobko, Korth,
  Kovalam, Kozak, Kr\"amer, Kringel, Krishnendu, Kr\'olak, Krupinski, Kuehn,
  Kumar, Kumar, Kumar, Kumar, Kuo, Kutynia, Kwang, Lackey, Laghi, Laguna, Lai,
  Lam, Landry, Lane, Lang, Lange, Lantz, Lanza, Lartaux-Vollard, Lasky, Laxen,
  Lazzarini, Lazzaro, Leaci, Leavey, Lecoeuche, Lee, Lee, Lee, Lee, Lee, Lee,
  Lehmann, Lenon, Leroy, Letendre, Levin, Li, Li, Li, Li, Li, Lin, Linde,
  Linker, Littenberg, Liu, Liu, Llorens-Monteagudo, Lo, London, Longo,
  Lorenzini, Loriette, Lormand, Losurdo, Lough, Lousto, Lovelace, Lower,
  L\"uck, Lumaca, Lundgren, Lynch, Ma, Macas, Macfoy, MacInnis, Macleod,
  Macquet, Maga\~na Hernandez, Maga\~na Sandoval, Magee, Majorana, Maksimovic,
  Malik, Man, Mandic, Mangano, Mansell, Manske, Mantovani, Mapelli, Marchesoni,
  Marion, M\'arka, M\'arka, Markakis, Markosyan, Markowitz, Maros, Marquina,
  Marsat, Martelli, Martin, Martin, Martinez, Martynov, Masalehdan, Mason,
  Massera, Masserot, Massinger, Masso-Reid, Mastrogiovanni, Matas, Matichard,
  Matone, Mavalvala, McCann, McCarthy, McClelland, McCormick, McCuller,
  McGuire, McIsaac, McIver, McManus, McRae, McWilliams, Meacher, Meadors,
  Mehmet, Mehta, Meidam, Mejuto~Villa, Melatos, Mendell, Mercer, Mereni,
  Merfeld, Merilh, Merzougui, Meshkov, Messenger, Messick, Messina, Metzdorff,
  Meyers, Meylahn, Miani, Miao, Michel, Middleton, Milano, Miller, Millhouse,
  Mills, Milovich-Goff, Minazzoli, Minenkov, Mishkin, Mishra, Mistry, Mitra,
  Mitrofanov, Mitselmakher, Mittleman, Mo, Moffa, Mogushi, Mohapatra,
  Molina-Ruiz, Mondin, Montani, Moore, Moraru, Morawski, Moreno, Morisaki,
  Mours, Mow-Lowry, Muciaccia, Mukherjee, Mukherjee, Mukherjee, Mukherjee,
  Mukund, Mullavey, Munch, Mu\~niz, Muratore, Murray, Nagar, Nardecchia,
  Naticchioni, Nayak, Neil, Neilson, Nelemans, Nelson, Nery, Neunzert, Nevin,
  Ng, Ng, Nguyen, Nguyen, Nichols, Nichols, Nissanke, Nocera, North, Nuttall,
  Obergaulinger, Oberling, O'Brien, Oganesyan, Ogin, Oh, Oh, Ohme, Ohta, Okada,
  Oliver, Oppermann, Oram, O'Reilly, Ormiston, Ortega, O'Shaughnessy, Ossokine,
  Ottaway, Overmier, Owen, Pace, Pagano, Page, Pagliaroli, Pai, Pai, Palamos,
  Palashov, Palomba, Pan, Panda, Pang, Pankow, Pannarale, Pant, Paoletti,
  Paoli, Parida, Parker, Pascucci, Pasqualetti, Passaquieti, Passuello, Patil,
  Patricelli, Payne, Pearlstone, Pechsiri, Pedersen, Pedraza, Pedurand, Pele,
  Penn, Perego, Perez, P\'erigois, Perreca, Petermann, Pfeiffer, Phelps,
  Phukon, Piccinni, Pichot, Piergiovanni, Pierro, Pillant, Pinard, Pinto,
  Pirello, Pitkin, Plastino, Poggiani, Pong, Ponrathnam, Popolizio, Porter,
  Powell, Prajapati, Prasad, Prasai, Prasanna, Pratten, Prestegard, Principe,
  Prodi, Prokhorov, Punturo, Puppo, P\"urrer, Qi, Quetschke, Quinonez, Raab,
  Raaijmakers, Radkins, Radulesco, Raffai, Raja, Rajan, Rajbhandari, Rakhmanov,
  Ramirez, Ramos-Buades, Rana, Rao, Rapagnani, Raymond, Razzano, Read,
  Regimbau, Rei, Reid, Reitze, Rettegno, Ricci, Richardson, Richardson, Ricker,
  Riemenschneider, Riles, Rizzo, Robertson, Robinet, Rocchi, Rolland, Rollins,
  Roma, Romanelli, Romano, Romel, Romie, Rose, Rose, Rose,
  Rosi\ifmmode~\acute{n}\else \'{n}\fi{}ska, Rosofsky, Ross, Rowan, R\"udiger,
  Ruggi, Rutins, Ryan, Sachdev, Sadecki, Sakellariadou, Salafia, Salconi,
  Saleem, Samajdar, Sammut, Sanchez, Sanchez, Sanchis-Gual, Sanders, Santiago,
  Santos, Sarin, Sassolas, Sathyaprakash, Sauter, Savage, Schale, Scheel,
  Scheuer, Schmidt, Schnabel, Schofield, Sch\"onbeck, Schreiber, Schulte,
  Schutz, Scott, Scott, Seidel, Sellers, Sengupta, Sennett, Sentenac, Sequino,
  Sergeev, Setyawati, Shaddock, Shaffer, Shahriar, Shaner, Sharma, Sharma,
  Shawhan, Shen, Shink, Shoemaker, Shoemaker, Shukla, ShyamSundar, Siellez,
  Sieniawska, Sigg, Singer, Singh, Singh, Singhal, Sintes, Sitmukhambetov,
  Skliris, Slagmolen, Slaven-Blair, Smith, Smith, Somala, Son, Soni, Sorazu,
  Sorrentino, Souradeep, Sowell, Spencer, Spera, Srivastava, Srivastava,
  Staats, Stachie, Standke, Steer, Steinke, Steinlechner, Steinlechner,
  Steinmeyer, Stevenson, Stocks, Stolle-McAllister, Stone, Stops, Strain,
  Stratta, Strigin, Strunk, Sturani, Stuver, Sudhir, Summerscales, Sun, Sunil,
  Sur, Suresh, Sutton, Swinkels, Szczepa\ifmmode~\acute{n}\else \'{n}\fi{}czyk,
  Tacca, Tait, Talbot, Tanner, Tao, T\'apai, Tapia, Tasson, Taylor, Tenorio,
  Terkowski, Thomas, Thomas, Thondapu, Thorne, Thrane, Tiwari, Tiwari, Tiwari,
  Toland, Tonelli, Tornasi, Torres-Forn\'e, Torrie, T\"oyr\"a, Travasso,
  Traylor, Tringali, Tripathee, Trovato, Trozzo, Tsang, Tse, Tso, Tsukada,
  Tsuna, Tsutsui, Tuyenbayev, Ueno, Ugolini, Unnikrishnan, Urban, Usman,
  Vahlbruch, Vajente, Valdes, Valentini, van Bakel, van Beuzekom, van~den
  Brand, Van Den~Broeck, Vander-Hyde, van~der Schaaf, VanHeijningen, van
  Veggel, Vardaro, Varma, Vass, Vas\'uth, Vecchio, Vedovato, Veitch, Veitch,
  Venkateswara, Venugopalan, Verkindt, Vetrano, Vicer\'e, Viets, Vinciguerra,
  Vine, Vinet, Vitale, Vo, Vocca, Vorvick, Vyatchanin, Wade, Wade, Wade, Walet,
  Walker, Wallace, Walsh, Wang, Wang, Wang, Wang, Wang, Ward, Warden, Warner,
  Was, Watchi, Weaver, Wei, Weinert, Weinstein, Weiss, Wellmann, Wen, Wessel,
  We\ss{}els, Westhouse, Wette, Whelan, Whiting, Whittle, Wilken, Williams,
  Williamson, Willis, Willke, Winkler, Wipf, Wittel, Woan, Woehler, Wofford,
  Wright, Wu, Wysocki, Xiao, Xu, Yamamoto, Yancey, Yang, Yang, Yang, Yap,
  Yazback, Yeeles, Yoon, Yu, Yu, Yuen, Zadro\ifmmode~\dot{z}\else \.{z}\fi{}ny,
  Zadro\ifmmode~\dot{z}\else \.{z}\fi{}ny, Zanolin, Zelenova, Zendri, Zevin,
  Zhang, Zhang, Zhang, Zhao, Zhao, Zhou, Zhou, Zhu, Zucker, Zweizig, Salemi, \&
  Papa}]{ligoIMBH}
---. 2019{\natexlab{c}},
  \href{http://dx.doi.org/10.1103/PhysRevD.100.064064}{\JournalTitle{Phys. Rev.
  D}, 100, 064064}

\bibitem[{{Amaro-Seoane} \& {Freitag}(2006)}]{asf2006}
{Amaro-Seoane}, P., \& {Freitag}, M. 2006,
  \href{http://dx.doi.org/10.1086/510405}{\JournalTitle{\apjl}, 653, L53}

\bibitem[{Amaro-Seoane \& Santamar{\'{\i}}a(2010)}]{santamaria}
Amaro-Seoane, P., \& Santamar{\'{\i}}a, L. 2010,
  \href{http://dx.doi.org/10.1088/0004-637x/722/2/1197}{\JournalTitle{The
  Astrophysical Journal}, 722, 1197}

\bibitem[{{Amaro-Seoane} {et~al.}(2017){Amaro-Seoane}, {Audley}, {Babak},
  {Baker}, {Barausse}, {Bender}, {Berti}, {Binetruy}, {Born}, {Bortoluzzi},
  {Camp}, {Caprini}, {Cardoso}, {Colpi}, {Conklin}, {Cornish}, {Cutler},
  {Danzmann}, {Dolesi}, {Ferraioli}, {Ferroni}, {Fitzsimons}, {Gair}, {Gesa
  Bote}, {Giardini}, {Gibert}, {Grimani}, {Halloin}, {Heinzel}, {Hertog},
  {Hewitson}, {Holley-Bockelmann}, {Hollington}, {Hueller}, {Inchauspe},
  {Jetzer}, {Karnesis}, {Killow}, {Klein}, {Klipstein}, {Korsakova}, {Larson},
  {Livas}, {Lloro}, {Man}, {Mance}, {Martino}, {Mateos}, {McKenzie},
  {McWilliams}, {Miller}, {Mueller}, {Nardini}, {Nelemans}, {Nofrarias},
  {Petiteau}, {Pivato}, {Plagnol}, {Porter}, {Reiche}, {Robertson},
  {Robertson}, {Rossi}, {Russano}, {Schutz}, {Sesana}, {Shoemaker}, {Slutsky},
  {Sopuerta}, {Sumner}, {Tamanini}, {Thorpe}, {Troebs}, {Vallisneri},
  {Vecchio}, {Vetrugno}, {Vitale}, {Volonteri}, {Wanner}, {Ward}, {Wass},
  {Weber}, {Ziemer}, \& {Zweifel}}]{lisa}
{Amaro-Seoane}, P., {Audley}, H., {Babak}, S., {et~al.} 2017,
  \JournalTitle{arXiv e-prints}, arXiv:1702.00786

\bibitem[{{Antonini} \& {Rasio}(2016)}]{ANTORASIO2016}
{Antonini}, F., \& {Rasio}, F.~A. 2016,
  \href{http://dx.doi.org/10.3847/0004-637X/831/2/187}{\JournalTitle{\apj},
  831, 187}

\bibitem[{{Arca Sedda} {et~al.}(2019){Arca Sedda}, {Askar}, \&
  {Giersz}}]{mocca}
{Arca Sedda}, M., {Askar}, A., \& {Giersz}, M. 2019, \JournalTitle{arXiv
  e-prints}, arXiv:1905.00902

\bibitem[{{Barack} \& {Cutler}(2004)}]{BarackCutler}
{Barack}, L., \& {Cutler}, C. 2004,
  \href{http://dx.doi.org/10.1103/PhysRevD.69.082005}{\JournalTitle{\prd}, 69,
  082005}

\bibitem[{Baumgardt(2016)}]{imbhBaumgardt}
Baumgardt, H. 2016,
  \href{http://dx.doi.org/10.1093/mnras/stw2488}{\JournalTitle{Monthly Notices
  of the Royal Astronomical Society}, 464, 2174}

\bibitem[{{Baumgardt} {et~al.}(2019){Baumgardt}, {He}, {Sweet}, {Drinkwater},
  {Sollima}, {Hurley}, {Usher}, {Kamann}, {Dalgleish}, {Dreizler}, \&
  {Husser}}]{baum2019}
{Baumgardt}, H., {He}, C., {Sweet}, S.~M., {et~al.} 2019,
  \href{http://dx.doi.org/10.1093/mnras/stz2060}{\JournalTitle{\mnras}, 488,
  5340}

\bibitem[{{Begelman} {et~al.}(1980){Begelman}, {Blandford}, \&
  {Rees}}]{Begelman1980}
{Begelman}, M.~C., {Blandford}, R.~D., \& {Rees}, M.~J. 1980,
  \href{http://dx.doi.org/10.1038/287307a0}{\JournalTitle{\nat}, 287, 307}

\bibitem[{{Bennett} {et~al.}(2014){Bennett}, {Larson}, {Weiland}, \&
  {Hinshaw}}]{cosmology}
{Bennett}, C.~L., {Larson}, D., {Weiland}, J.~L., \& {Hinshaw}, G. 2014,
  \href{http://dx.doi.org/10.1088/0004-637X/794/2/135}{\JournalTitle{\apj},
  794, 135}

\bibitem[{{Binney} \& {Tremaine}(2008)}]{BinneyTremaine}
{Binney}, J., \& {Tremaine}, S. 2008, {Galactic Dynamics: Second Edition}
  (Princeton University Press)

\bibitem[{{Bonetti} {et~al.}(2020){Bonetti}, {Rasskazov}, {Sesana}, {Dotti},
  {Haardt}, {Leigh}, {Arca Sedda}, {Fragione}, \& {Rossi}}]{bonetti}
{Bonetti}, M., {Rasskazov}, A., {Sesana}, A., {et~al.} 2020,
  \href{http://dx.doi.org/10.1093/mnrasl/slaa018}{\JournalTitle{\mnras}, 493,
  L114}

\bibitem[{{Bortolas} {et~al.}(2016){Bortolas}, {Gualandris}, {Dotti}, {Spera},
  \& {Mapelli}}]{Bortolas2016}
{Bortolas}, E., {Gualandris}, A., {Dotti}, M., {Spera}, M., \& {Mapelli}, M.
  2016, \href{http://dx.doi.org/10.1093/mnras/stw1372}{\JournalTitle{\mnras},
  461, 1023}

\bibitem[{Chandar {et~al.}(2010{\natexlab{a}})Chandar, Fall, \&
  Whitmore}]{gcmass4}
Chandar, R., Fall, S.~M., \& Whitmore, B.~C. 2010{\natexlab{a}},
  \href{http://dx.doi.org/10.1088/0004-637x/711/2/1263}{\JournalTitle{The
  Astrophysical Journal}, 711, 1263}

\bibitem[{Chandar {et~al.}(2011)Chandar, Whitmore, Calzetti, Nino, Kennicutt,
  Regan, \& Schinnerer}]{gcmass6}
Chandar, R., Whitmore, B.~C., Calzetti, D., {et~al.} 2011,
  \href{http://dx.doi.org/10.1088/0004-637x/727/2/88}{\JournalTitle{The
  Astrophysical Journal}, 727, 88}

\bibitem[{Chandar {et~al.}(2010{\natexlab{b}})Chandar, Whitmore, Kim, Kaleida,
  Mutchler, Calzetti, Saha, O'Connell, Balick, Bond, Carollo, Disney, Dopita,
  Frogel, Hall, Holtzman, Kimble, McCarthy, Paresce, Silk, Trauger, Walker,
  Windhorst, \& Young}]{gcmass5}
Chandar, R., Whitmore, B.~C., Kim, H., {et~al.} 2010{\natexlab{b}},
  \href{http://dx.doi.org/10.1088/0004-637x/719/1/966}{\JournalTitle{The
  Astrophysical Journal}, 719, 966}

\bibitem[{{Chatterjee} {et~al.}(2003){Chatterjee}, {Hernquist}, \&
  {Loeb}}]{Chatterjee2003}
{Chatterjee}, P., {Hernquist}, L., \& {Loeb}, A. 2003,
  \href{http://dx.doi.org/10.1086/375552}{\JournalTitle{\apj}, 592, 32}

\bibitem[{{Chen} {et~al.}(2009){Chen}, {Madau}, {Sesana}, \& {Liu}}]{chen2009}
{Chen}, X., {Madau}, P., {Sesana}, A., \& {Liu}, F.~K. 2009,
  \href{http://dx.doi.org/10.1088/0004-637X/697/2/L149}{\JournalTitle{\apjl},
  697, L149}

\bibitem[{{Chilingarian} {et~al.}(2018){Chilingarian}, {Katkov}, {Zolotukhin},
  {Grishin}, {Beletsky}, {Boutsia}, \& {Osip}}]{Zolotukhin2018}
{Chilingarian}, I.~V., {Katkov}, I.~Y., {Zolotukhin}, I.~Y., {et~al.} 2018,
  \href{http://dx.doi.org/10.3847/1538-4357/aad184}{\JournalTitle{\apj}, 863,
  1}

\bibitem[{Coleman~Miller \& Hamilton(2002)}]{miller2002}
Coleman~Miller, M., \& Hamilton, D.~P. 2002,
  \href{http://dx.doi.org/10.1046/j.1365-8711.2002.05112.x}{\JournalTitle{Monthly
  Notices of the Royal Astronomical Society}, 330, 232}

\bibitem[{Davis {et~al.}(2011)Davis, Narayan, Zhu, Barret, Farrell, Godet,
  Servillat, \& Webb}]{xray2}
Davis, S.~W., Narayan, R., Zhu, Y., {et~al.} 2011,
  \href{http://dx.doi.org/10.1088/0004-637x/734/2/111}{\JournalTitle{The
  Astrophysical Journal}, 734, 111}

\bibitem[{{Deme} {et~al.}(2020){Deme}, {Meiron}, \& {Kocsis}}]{Deme2019}
{Deme}, B., {Meiron}, Y., \& {Kocsis}, B. 2020,
  \href{http://dx.doi.org/10.3847/1538-4357/ab7921}{\JournalTitle{\apj}, 892,
  130}

\bibitem[{El-Badry {et~al.}(2018)El-Badry, Quataert, Weisz, Choksi, \&
  Boylan-Kolchin}]{ElBadry}
El-Badry, K., Quataert, E., Weisz, D.~R., Choksi, N., \& Boylan-Kolchin, M.
  2018, \href{http://dx.doi.org/10.1093/mnras/sty3007}{\JournalTitle{Monthly
  Notices of the Royal Astronomical Society}, 482, 4528–4552}

\bibitem[{{Enoki} \& {Nagashima}(2007)}]{EnokiNagashima}
{Enoki}, M., \& {Nagashima}, M. 2007,
  \href{http://dx.doi.org/10.1143/PTP.117.241}{\JournalTitle{Progress of
  Theoretical Physics}, 117, 241}

\bibitem[{Fabbiano(2006)}]{xray1}
Fabbiano, G. 2006,
  \href{http://dx.doi.org/10.1146/annurev.astro.44.051905.092519}{\JournalTitle{Annual
  Review of Astronomy and Astrophysics}, 44, 323}

\bibitem[{Fragione {et~al.}(2019)Fragione, Antonini, \&
  Gnedin}]{millisecondPulsars}
Fragione, G., Antonini, F., \& Gnedin, O.~Y. 2019,
  \href{http://dx.doi.org/10.3847/2041-8213/aafc62}{\JournalTitle{The
  Astrophysical Journal}, 871, L8}

\bibitem[{{Fragione} {et~al.}(2018{\natexlab{a}}){Fragione}, {Ginsburg}, \&
  {Kocsis}}]{GCinspiral2}
{Fragione}, G., {Ginsburg}, I., \& {Kocsis}, B. 2018{\natexlab{a}},
  \href{http://dx.doi.org/10.3847/1538-4357/aab368}{\JournalTitle{\apj}, 856,
  92}

\bibitem[{{Fragione} \& {Gualandris}(2019)}]{fragual2019MNRAS}
{Fragione}, G., \& {Gualandris}, A. 2019,
  \href{http://dx.doi.org/10.1093/mnras/stz2451}{\JournalTitle{\mnras}, 489,
  4543}

\bibitem[{{Fragione} \& {Leigh}(2018)}]{flei2018}
{Fragione}, G., \& {Leigh}, N. 2018,
  \href{http://dx.doi.org/10.1093/mnras/sty1600}{\JournalTitle{\mnras}, 479,
  3181}

\bibitem[{{Fragione} {et~al.}(2018{\natexlab{b}}){Fragione}, {Leigh},
  {Ginsburg}, \& {Kocsis}}]{frag2018b}
{Fragione}, G., {Leigh}, N. W.~C., {Ginsburg}, I., \& {Kocsis}, B.
  2018{\natexlab{b}},
  \href{http://dx.doi.org/10.3847/1538-4357/aae486}{\JournalTitle{\apj}, 867,
  119}

\bibitem[{{Fregeau} {et~al.}(2006){Fregeau}, {Larson}, {Miller},
  {O'Shaughnessy}, \& {Rasio}}]{fregeau2006}
{Fregeau}, J.~M., {Larson}, S.~L., {Miller}, M.~C., {O'Shaughnessy}, R., \&
  {Rasio}, F.~A. 2006,
  \href{http://dx.doi.org/10.1086/507106}{\JournalTitle{\apjl}, 646, L135}

\bibitem[{Freitag {et~al.}(2006)Freitag, Gürkan, \& Rasio}]{freitag2006}
Freitag, M., Gürkan, M.~A., \& Rasio, F.~A. 2006,
  \href{http://dx.doi.org/10.1111/j.1365-2966.2006.10096.x}{\JournalTitle{Monthly
  Notices of the Royal Astronomical Society}, 368, 141}

\bibitem[{{Gair} {et~al.}(2011){Gair}, {Mandel}, {Miller}, \&
  {Volonteri}}]{gair2011}
{Gair}, J.~R., {Mandel}, I., {Miller}, M.~C., \& {Volonteri}, M. 2011,
  \href{http://dx.doi.org/10.1007/s10714-010-1104-3}{\JournalTitle{General
  Relativity and Gravitation}, 43, 485}

\bibitem[{Gieles(2009)}]{gcmass2}
Gieles, M. 2009,
  \href{http://dx.doi.org/10.1111/j.1365-2966.2009.14473.x}{\JournalTitle{Monthly
  Notices of the Royal Astronomical Society}, 394, 2113}

\bibitem[{{Giersz} {et~al.}(2015){Giersz}, {Leigh}, {Hypki}, {L{\"u}tzgendorf},
  \& {Askar}}]{giers2015}
{Giersz}, M., {Leigh}, N., {Hypki}, A., {L{\"u}tzgendorf}, N., \& {Askar}, A.
  2015, \href{http://dx.doi.org/10.1093/mnras/stv2162}{\JournalTitle{\mnras},
  454, 3150}

\bibitem[{Gnedin {et~al.}(2014)Gnedin, Ostriker, \& Tremaine}]{GOT}
Gnedin, O.~Y., Ostriker, J.~P., \& Tremaine, S. 2014,
  \href{http://dx.doi.org/10.1088/0004-637x/785/1/71}{\JournalTitle{The
  Astrophysical Journal}, 785, 71}

\bibitem[{{Gonz{\'a}lez} {et~al.}(2007){Gonz{\'a}lez}, {Sperhake},
  {Br{\"u}gmann}, {Hannam}, \& {Husa}}]{gon07}
{Gonz{\'a}lez}, J.~A., {Sperhake}, U., {Br{\"u}gmann}, B., {Hannam}, M., \&
  {Husa}, S. 2007,
  \href{http://dx.doi.org/10.1103/PhysRevLett.98.091101}{\JournalTitle{Physical
  Review Letters}, 98, 091101}

\bibitem[{{Gratton} {et~al.}(2012){Gratton}, {Carretta}, \&
  {Bragaglia}}]{gratton2012}
{Gratton}, R.~G., {Carretta}, E., \& {Bragaglia}, A. 2012,
  \href{http://dx.doi.org/10.1007/s00159-012-0050-3}{\JournalTitle{\aapr}, 20,
  50}

\bibitem[{{Greene} {et~al.}(2019){Greene}, {Strader}, \& {Ho}}]{Greene2019}
{Greene}, J.~E., {Strader}, J., \& {Ho}, L.~C. 2019, \JournalTitle{arXiv
  e-prints}, arXiv:1911.09678

\bibitem[{{Gualandris} {et~al.}(2010){Gualandris}, {Gillessen}, \&
  {Merritt}}]{gual2010}
{Gualandris}, A., {Gillessen}, S., \& {Merritt}, D. 2010,
  \href{http://dx.doi.org/10.1111/j.1365-2966.2010.17373.x}{\JournalTitle{\mnras},
  409, 1146}

\bibitem[{{G{\"u}rkan} {et~al.}(2006){G{\"u}rkan}, {Fregeau}, \&
  {Rasio}}]{gurkan2006}
{G{\"u}rkan}, M.~A., {Fregeau}, J.~M., \& {Rasio}, F.~A. 2006,
  \href{http://dx.doi.org/10.1086/503295}{\JournalTitle{\apjl}, 640, L39}

\bibitem[{{Harris}(2010)}]{harris}
{Harris}, W.~E. 2010, \JournalTitle{arXiv e-prints},
  \href{http://arxiv.org/abs/1012.3224}{{\sffamily arXiv:1012.3224
  [astro-ph.GA]}}

\bibitem[{{Harris} {et~al.}(2017){Harris}, {Blakeslee}, \&
  {Harris}}]{harris2017}
{Harris}, W.~E., {Blakeslee}, J.~P., \& {Harris}, G. L.~H. 2017,
  \href{http://dx.doi.org/10.3847/1538-4357/836/1/67}{\JournalTitle{\apj}, 836,
  67}

\bibitem[{{Hild} {et~al.}(2011){Hild}, {Abernathy}, {Acernese}, {Amaro-Seoane},
  {Andersson}, {Arun}, {Barone}, {Barr}, {Barsuglia}, {Beker}, {Beveridge},
  {Birindelli}, {Bose}, {Bosi}, {Braccini}, {Bradaschia}, {Bulik}, {Calloni},
  {Cella}, {Chassande Mottin}, {Chelkowski}, {Chincarini}, {Clark}, {Coccia},
  {Colacino}, {Colas}, {Cumming}, {Cunningham}, {Cuoco}, {Danilishin},
  {Danzmann}, {De Salvo}, {Dent}, {De Rosa}, {Di Fiore}, {Di Virgilio},
  {Doets}, {Fafone}, {Falferi}, {Flaminio}, {Franc}, {Frasconi}, {Freise},
  {Friedrich}, {Fulda}, {Gair}, {Gemme}, {Genin}, {Gennai}, {Giazotto},
  {Glampedakis}, {Gr{\"a}f}, {Granata}, {Grote}, {Guidi}, {Gurkovsky},
  {Hammond}, {Hannam}, {Harms}, {Heinert}, {Hendry}, {Heng}, {Hennes}, {Hough},
  {Husa}, {Huttner}, {Jones}, {Khalili}, {Kokeyama}, {Kokkotas}, {Krishnan},
  {Li}, {Lorenzini}, {L{\"u}ck}, {Majorana}, {Mandel}, {Mandic}, {Mantovani},
  {Martin}, {Michel}, {Minenkov}, {Morgado}, {Mosca}, {Mours},
  {M{\"u}ller─Ebhardt}, {Murray}, {Nawrodt}, {Nelson}, {Oshaughnessy}, {Ott},
  {Palomba}, {Paoli}, {Parguez}, {Pasqualetti}, {Passaquieti}, {Passuello},
  {Pinard}, {Plastino}, {Poggiani}, {Popolizio}, {Prato}, {Punturo}, {Puppo},
  {Rabeling}, {Rapagnani}, {Read}, {Regimbau}, {Rehbein}, {Reid}, {Ricci},
  {Richard}, {Rocchi}, {Rowan}, {R{\"u}diger}, {Santamar{\'\i}a}, {Sassolas},
  {Sathyaprakash}, {Schnabel}, {Schwarz}, {Seidel}, {Sintes}, {Somiya},
  {Speirits}, {Strain}, {Strigin}, {Sutton}, {Tarabrin}, {Th{\"u}ring}, {van
  den Brand}, {van Veggel}, {van den Broeck}, {Vecchio}, {Veitch}, {Vetrano},
  {Vicere}, {Vyatchanin}, {Willke}, {Woan}, \& {Yamamoto}}]{ETSensitivity}
{Hild}, S., {Abernathy}, M., {Acernese}, F., {et~al.} 2011,
  \href{http://dx.doi.org/10.1088/0264-9381/28/9/094013}{\JournalTitle{Classical
  and Quantum Gravity}, 28, 094013}

\bibitem[{{Hurley} {et~al.}(2000){Hurley}, {Pols}, \& {Tout}}]{hur00}
{Hurley}, J.~R., {Pols}, O.~R., \& {Tout}, C.~A. 2000,
  \href{http://dx.doi.org/10.1046/j.1365-8711.2000.03426.x}{\JournalTitle{\mnras},
  315, 543}

\bibitem[{{Ideta} \& {Makino}(2004)}]{omegacen}
{Ideta}, M., \& {Makino}, J. 2004,
  \href{http://dx.doi.org/10.1086/426505}{\JournalTitle{\apjl}, 616, L107}

\bibitem[{Khoperskov {et~al.}(2018)Khoperskov, Mastrobuono-Battisti, Di~Matteo,
  \& Haywood}]{gcMergers}
Khoperskov, S., Mastrobuono-Battisti, A., Di~Matteo, P., \& Haywood, M. 2018,
  \href{http://dx.doi.org/10.1051/0004-6361/201833534}{\JournalTitle{A\&A},
  620, A154}

\bibitem[{{Kroupa}(2001)}]{KroupaIMF}
{Kroupa}, P. 2001,
  \href{http://dx.doi.org/10.1046/j.1365-8711.2001.04022.x}{\JournalTitle{\mnras},
  322, 231}

\bibitem[{{Larsen}(2009)}]{gcmass3}
{Larsen}, S.~S. 2009,
  \href{http://dx.doi.org/10.1051/0004-6361:200811212}{\JournalTitle{\aap},
  494, 539}

\bibitem[{{Lin} {et~al.}(2018){Lin}, {Strader}, {Carrasco}, {Page},
  {Romanowsky}, {Homan}, {Irwin}, {Remillard}, {Godet}, {Webb}, {Baumgardt},
  {Wijnands}, {Barret}, {Duc}, {Brodie}, \& {Gwyn}}]{imbhTDE}
{Lin}, D., {Strader}, J., {Carrasco}, E.~R., {et~al.} 2018,
  \href{http://dx.doi.org/10.1038/s41550-018-0493-1}{\JournalTitle{Nature
  Astronomy}, 2, 656}

\bibitem[{{Lousto} {et~al.}(2010){Lousto}, {Campanelli}, {Zlochower}, \&
  {Nakano}}]{lou10}
{Lousto}, C.~O., {Campanelli}, M., {Zlochower}, Y., \& {Nakano}, H. 2010,
  \href{http://dx.doi.org/10.1088/0264-9381/27/11/114006}{\JournalTitle{Classical
  and Quantum Gravity}, 27, 114006}

\bibitem[{{Lousto} \& {Zlochower}(2008)}]{lou08}
{Lousto}, C.~O., \& {Zlochower}, Y. 2008,
  \href{http://dx.doi.org/10.1103/PhysRevD.77.044028}{\JournalTitle{\prd}, 77,
  044028}

\bibitem[{{Lousto} {et~al.}(2012){Lousto}, {Zlochower}, {Dotti}, \&
  {Volonteri}}]{lou12}
{Lousto}, C.~O., {Zlochower}, Y., {Dotti}, M., \& {Volonteri}, M. 2012,
  \href{http://dx.doi.org/10.1103/PhysRevD.85.084015}{\JournalTitle{\prd}, 85,
  084015}

\bibitem[{{MacLeod} {et~al.}(2016){MacLeod}, {Guillochon}, {Ramirez-Ruiz},
  {Kasen}, \& {Rosswog}}]{imbhWD2016}
{MacLeod}, M., {Guillochon}, J., {Ramirez-Ruiz}, E., {Kasen}, D., \& {Rosswog},
  S. 2016,
  \href{http://dx.doi.org/10.3847/0004-637X/819/1/3}{\JournalTitle{\apj}, 819,
  3}

\bibitem[{Madau \& Rees(2001)}]{Madau_2001}
Madau, P., \& Rees, M.~J. 2001,
  \href{http://dx.doi.org/10.1086/319848}{\JournalTitle{The Astrophysical
  Journal}, 551, L27}

\bibitem[{{Mapelli}(2016)}]{Mapelli2016}
{Mapelli}, M. 2016,
  \href{http://dx.doi.org/10.1093/mnras/stw869}{\JournalTitle{\mnras}, 459,
  3432}

\bibitem[{{Martynov} {et~al.}(2016){Martynov}, {Hall}, {Abbott}, {Abbott},
  {Abbott}, {Adams}, {Adhikari}, {Anderson}, {Anderson}, {Arai}, {Arain},
  {Aston}, {Austin}, {Ballmer}, {Barbet}, {Barker}, {Barr}, {Barsotti},
  {Bartlett}, {Barton}, {Bartos}, {Batch}, {Bell}, {Belopolski}, {Bergman},
  {Betzwieser}, {Billingsley}, {Birch}, {Biscans}, {Biwer}, {Black}, {Blair},
  {Bogan}, {Bork}, {Bridges}, {Brooks}, {Celerier}, {Ciani}, {Clara}, {Cook},
  {Countryman}, {Cowart}, {Coyne}, {Cumming}, {Cunningham}, {Damjanic},
  {Dannenberg}, {Danzmann}, {Costa}, {Daw}, {DeBra}, {DeRosa}, {DeSalvo},
  {Dooley}, {Doravari}, {Driggers}, {Dwyer}, {Effler}, {Etzel}, {Evans},
  {Evans}, {Factourovich}, {Fair}, {Feldbaum}, {Fisher}, {Foley}, {Frede},
  {Fritschel}, {Frolov}, {Fulda}, {Fyffe}, {Galdi}, {Giaime}, {Giardina},
  {Gleason}, {Goetz}, {Gras}, {Gray}, {Greenhalgh}, {Grote}, {Guido}, {Gushwa},
  {Gustafson}, {Gustafson}, {Hammond}, {Hanks}, {Hanson}, {Hardwick}, {Harry},
  {Heefner}, {Heintze}, {Heptonstall}, {Hoak}, {Hough}, {Ivanov}, {Izumi},
  {Jacobson}, {James}, {Jones}, {Kandhasamy}, {Karki}, {Kasprzack}, {Kaufer},
  {Kawabe}, {Kells}, {Kijbunchoo}, {King}, {King}, {Kinzel}, {Kissel},
  {Kokeyama}, {Korth}, {Kuehn}, {Kwee}, {Landry}, {Lantz}, {Le Roux}, {Levine},
  {Lewis}, {Lhuillier}, {Lockerbie}, {Lormand}, {Lubinski}, {Lundgren},
  {MacDonald}, {MacInnis}, {Macleod}, {Mageswaran}, {Mailand }, {M{\'a}rka},
  {M{\'a}rka}, {Markosyan}, {Maros}, {Martin}, {Martin}, {Marx}, {Mason},
  {Massinger}, {Matichard}, {Mavalvala}, {McCarthy}, {McClelland}, {McCormick},
  {McIntyre}, {McIver}, {Merilh}, {Meyer}, {Meyers}, {Miller}, {Mittleman},
  {Moreno}, {Mueller}, {Mueller}, {Mullavey}, {Munch}, {Nuttall}, {Oberling},
  {O'Dell}, {Oppermann}, {Oram}, {O'Reilly}, {Osthelder}, {Ottaway},
  {Overmier}, {Palamos}, {Paris}, {Parker}, {Patrick}, {Pele}, {Penn},
  {Phelps}, {Pickenpack}, {Pierro}, {Pinto}, {Poeld}, {Principe}, {Prokhorov},
  {Puncken}, {Quetschke}, {Quintero}, {Raab}, {Radkins}, {Raffai}, {Ramet},
  {Reed}, {Reid}, {Reitze}, {Robertson}, {Rollins}, {Roma}, {Romie}, {Rowan},
  {Ryan}, {Sadecki}, {Sanchez}, {Sandberg}, {Sannibale}, {Savage}, {Schofield},
  {Schultz}, {Schwinberg}, {Sellers}, {Sevigny}, {Shaddock}, {Shao}, {Shapiro},
  {Shawhan}, {Shoemaker}, {Sigg}, {Slagmolen}, {Smith}, {Smith},
  {Smith-Lefebvre}, {Sorazu}, {Staley}, {Stein}, {Stochino}, {Strain},
  {Taylor}, {Thomas}, {Thomas}, {Thorne}, {Thrane}, {Torrie}, {Traylor},
  {Vajente}, {Valdes}, {van Veggel}, {Vargas}, {Vecchio}, {Veitch},
  {Venkateswara}, {Vo}, {Vorvick}, {Waldman}, {Walker}, {Ward}, {Warner},
  {Weaver}, {Weiss}, {Welborn}, {We{\ss}els}, {Wilkinson}, {Willems},
  {Williams}, {Willke}, {Winkelmann}, {Wipf}, {Worden}, {Wu}, {Yamamoto},
  {Yancey}, {Yu}, {Zhang}, {Zucker}, \& {Zweizig}}]{ligoSensitivity}
{Martynov}, D.~V., {Hall}, E.~D., {Abbott}, B.~P., {et~al.} 2016,
  \href{http://dx.doi.org/10.1103/PhysRevD.93.112004}{\JournalTitle{\prd}, 93,
  112004}

\bibitem[{Mastrobuono-Battisti {et~al.}(2019)Mastrobuono-Battisti, Khoperskov,
  Di~Matteo, \& Haywood}]{gcMergers1}
Mastrobuono-Battisti, A., Khoperskov, S., Di~Matteo, P., \& Haywood, M. 2019,
  \href{http://dx.doi.org/10.1051/0004-6361/201834087}{\JournalTitle{A\&A},
  622, A86}

\bibitem[{McKernan {et~al.}(2014)McKernan, Ford, Kocsis, Lyra, \&
  Winter}]{mckernan2014}
McKernan, B., Ford, K. E.~S., Kocsis, B., Lyra, W., \& Winter, L.~M. 2014,
  \href{http://dx.doi.org/10.1093/mnras/stu553}{\JournalTitle{Monthly Notices
  of the Royal Astronomical Society}, 441, 900}

\bibitem[{McKernan {et~al.}(2012)McKernan, Ford, Lyra, \&
  Perets}]{mckernan2012}
McKernan, B., Ford, K. E.~S., Lyra, W., \& Perets, H.~B. 2012,
  \href{http://dx.doi.org/10.1111/j.1365-2966.2012.21486.x}{\JournalTitle{Monthly
  Notices of the Royal Astronomical Society}, 425, 460}

\bibitem[{{Merritt}(2013)}]{DEGN}
{Merritt}, D. 2013, {Dynamics and Evolution of Galactic Nuclei} (Princeton:
  Princeton University Pres)

\bibitem[{Merritt \& Ferrarese(2001)}]{m-sigma}
Merritt, D., \& Ferrarese, L. 2001,
  \href{http://dx.doi.org/10.1086/318372}{\JournalTitle{The Astrophysical
  Journal}, 547, 140}

\bibitem[{{Mezcua} {et~al.}(2018){Mezcua}, {Civano}, {Marchesi}, {Suh},
  {Fabbiano}, \& {Volonteri}}]{Mezcua2018}
{Mezcua}, M., {Civano}, F., {Marchesi}, S., {et~al.} 2018,
  \href{http://dx.doi.org/10.1093/mnras/sty1163}{\JournalTitle{\mnras}, 478,
  2576}

\bibitem[{{Navarro} {et~al.}(1997){Navarro}, {Frenk}, \& {White}}]{NFW}
{Navarro}, J.~F., {Frenk}, C.~S., \& {White}, S. D.~M. 1997,
  \href{http://dx.doi.org/10.1086/304888}{\JournalTitle{\apj}, 490, 493}

\bibitem[{{O'Leary} {et~al.}(2009){O'Leary}, {Kocsis}, \& {Loeb}}]{OLeary2009}
{O'Leary}, R.~M., {Kocsis}, B., \& {Loeb}, A. 2009,
  \href{http://dx.doi.org/10.1111/j.1365-2966.2009.14653.x}{\JournalTitle{\mnras},
  395, 2127}

\bibitem[{\"{O}zel {et~al.}(2010)\"{O}zel, Psaltis, Narayan, \&
  McClintock}]{bhmass1}
\"{O}zel, F., Psaltis, D., Narayan, R., \& McClintock, J.~E. 2010,
  \href{http://dx.doi.org/10.1088/0004-637x/725/2/1918}{\JournalTitle{The
  Astrophysical Journal}, 725, 1918}

\bibitem[{{Peng} {et~al.}(2019){Peng}, {Yang}, {Shen}, {Wang}, {Zou}, \&
  {Zhang}}]{peng2019}
{Peng}, Z.-K., {Yang}, Y.-S., {Shen}, R.-F., {et~al.} 2019,
  \href{http://dx.doi.org/10.3847/2041-8213/ab481b}{\JournalTitle{\apjl}, 884,
  L34}

\bibitem[{{Peters}(1964)}]{Peters64}
{Peters}, P.~C. 1964,
  \href{http://dx.doi.org/10.1103/PhysRev.136.B1224}{\JournalTitle{Physical
  Review}, 136, 1224}

\bibitem[{{Peters} \& {Mathews}(1963)}]{PetersMathews}
{Peters}, P.~C., \& {Mathews}, J. 1963,
  \href{http://dx.doi.org/10.1103/PhysRev.131.435}{\JournalTitle{Physical
  Review}, 131, 435}

\bibitem[{Petts \& Gualandris(2017)}]{GCinspiral1}
Petts, J.~A., \& Gualandris, A. 2017,
  \href{http://dx.doi.org/10.1093/mnras/stx296}{\JournalTitle{Monthly Notices
  of the Royal Astronomical Society}, 467, 3775}

\bibitem[{Portegies-Zwart \& McMillan(2002)}]{Portegies_Zwart_2002}
Portegies-Zwart, S.~F., \& McMillan, S. L.~W. 2002,
  \href{http://dx.doi.org/10.1086/341798}{\JournalTitle{The Astrophysical
  Journal}, 576, 899}

\bibitem[{Punturo {et~al.}(2010)Punturo, Abernathy, Acernese, Allen, Andersson,
  Arun, Barone, Barr, Barsuglia, Beker, Beveridge, Birindelli, Bose, Bosi,
  Braccini, Bradaschia, Bulik, Calloni, Cella, Mottin, Chelkowski, Chincarini,
  Clark, Coccia, Colacino, Colas, Cumming, Cunningham, Cuoco, Danilishin,
  Danzmann, Luca, Salvo, Dent, Derosa, Fiore, Virgilio, Doets, Fafone, Falferi,
  Flaminio, Franc, Frasconi, Freise, Fulda, Gair, Gemme, Gennai, Giazotto,
  Glampedakis, Granata, Grote, Guidi, Hammond, Hannam, Harms, Heinert, Hendry,
  Heng, Hennes, Hild, Hough, Husa, Huttner, Jones, Khalili, Kokeyama, Kokkotas,
  Krishnan, Lorenzini, Lück, Majorana, Mandel, Mandic, Martin, Michel,
  Minenkov, Morgado, Mosca, Mours, Müller-Ebhardt, Murray, Nawrodt, Nelson,
  Oshaughnessy, Ott, Palomba, Paoli, Parguez, Pasqualetti, Passaquieti,
  Passuello, Pinard, Poggiani, Popolizio, Prato, Puppo, Rabeling, Rapagnani,
  Read, Regimbau, Rehbein, Reid, Rezzolla, Ricci, Richard, Rocchi, Rowan,
  Rüdiger, Sassolas, Sathyaprakash, Schnabel, Schwarz, Seidel, Sintes, Somiya,
  Speirits, Strain, Strigin, Sutton, Tarabrin, van~den Brand, van Leewen, van
  Veggel, van~den Broeck, Vecchio, Veitch, Vetrano, Vicere, Vyatchanin, Willke,
  Woan, Wolfango, \& Yamamoto}]{ET}
Punturo, M., Abernathy, M., Acernese, F., {et~al.} 2010,
  \href{http://dx.doi.org/10.1088/0264-9381/27/8/084007}{\JournalTitle{Classical
  and Quantum Gravity}, 27, 084007}

\bibitem[{Quinlan(1996)}]{Quinlan1996}
Quinlan, G.~D. 1996,
  \href{http://dx.doi.org/10.1016/S1384-1076(96)00003-6}{\JournalTitle{New
  Astronomy}, 1, 35}

\bibitem[{{Rasskazov} {et~al.}(2019){Rasskazov}, {Fragione}, {Leigh}, {Tagawa},
  {Sesana}, {Price-Whelan}, \& {Rossi}}]{Rasskazov2019}
{Rasskazov}, A., {Fragione}, G., {Leigh}, N. W.~C., {et~al.} 2019,
  \href{http://dx.doi.org/10.3847/1538-4357/ab1c5d}{\JournalTitle{\apj}, 878,
  17}

\bibitem[{Rasskazov \& Merritt(2017{\natexlab{a}})}]{Rasskazov2017}
Rasskazov, A., \& Merritt, D. 2017{\natexlab{a}},
  \href{http://dx.doi.org/10.3847/1538-4357/aa6188}{\JournalTitle{\apj}, 837,
  135}

\bibitem[{Rasskazov \& Merritt(2017{\natexlab{b}})}]{RasskazovMerritt}
---. 2017{\natexlab{b}},
  \href{http://dx.doi.org/10.3847/1538-4357/aa6188}{\JournalTitle{\apj}, 837,
  135}

\bibitem[{{Robson} {et~al.}(2019){Robson}, {Cornish}, \& {Liu}}]{robson}
{Robson}, T., {Cornish}, N.~J., \& {Liu}, C. 2019,
  \href{http://dx.doi.org/10.1088/1361-6382/ab1101}{\JournalTitle{Classical and
  Quantum Gravity}, 36, 105011}

\bibitem[{Rodriguez {et~al.}(2018)Rodriguez, Amaro-Seoane, Chatterjee, Kremer,
  Rasio, Samsing, Ye, \& Zevin}]{Rodriguez}
Rodriguez, C.~L., Amaro-Seoane, P., Chatterjee, S., {et~al.} 2018,
  \href{http://dx.doi.org/10.1103/PhysRevD.98.123005}{\JournalTitle{Phys. Rev.
  D}, 98, 123005}

\bibitem[{{Rosswog} {et~al.}(2009){Rosswog}, {Ramirez-Ruiz}, \&
  {Hix}}]{imbhWD2009}
{Rosswog}, S., {Ramirez-Ruiz}, E., \& {Hix}, W.~R. 2009,
  \href{http://dx.doi.org/10.1088/0004-637X/695/1/404}{\JournalTitle{\apj},
  695, 404}

\bibitem[{{Sch{\"o}del} {et~al.}(2018){Sch{\"o}del}, {Gallego-Cano}, {Dong},
  {Nogueras-Lara}, {Gallego-Calvente}, {Amaro-Seoane}, \&
  {Baumgardt}}]{sch2018}
{Sch{\"o}del}, R., {Gallego-Cano}, E., {Dong}, H., {et~al.} 2018,
  \href{http://dx.doi.org/10.1051/0004-6361/201730452}{\JournalTitle{\aap},
  609, A27}

\bibitem[{Sesana {et~al.}(2006)Sesana, Haardt, \& Madau}]{Sesana2006}
Sesana, A., Haardt, F., \& Madau, P. 2006,
  \href{http://dx.doi.org/10.1086/507596}{\JournalTitle{\apj}, 651, 392}

\bibitem[{{Sesana} \& {Khan}(2015)}]{sesana2015}
{Sesana}, A., \& {Khan}, F.~M. 2015,
  \href{http://dx.doi.org/10.1093/mnrasl/slv131}{\JournalTitle{\mnras}, 454,
  L66}

\bibitem[{Stone \& Ostriker(2015)}]{GCdensity}
Stone, N.~C., \& Ostriker, J.~P. 2015,
  \href{http://dx.doi.org/10.1088/2041-8205/806/2/l28}{\JournalTitle{The
  Astrophysical Journal}, 806, L28}

\bibitem[{{Tagawa} {et~al.}(2020){Tagawa}, {Haiman}, \& {Kocsis}}]{Tagawa2019}
{Tagawa}, H., {Haiman}, Z., \& {Kocsis}, B. 2020,
  \href{http://dx.doi.org/10.3847/1538-4357/ab7922}{\JournalTitle{\apj}, 892,
  36}

\bibitem[{{The LIGO Scientific Collaboration} {et~al.}(2019){The LIGO
  Scientific Collaboration}, {the Virgo Collaboration}, {Salemi}, \&
  {Papa}}]{ligo}
{The LIGO Scientific Collaboration}, {the Virgo Collaboration}, {Salemi}, F.,
  \& {Papa}, M.~A. 2019, \JournalTitle{arXiv e-prints}, arXiv:1906.08000

\bibitem[{{Vasiliev} {et~al.}(2015){Vasiliev}, {Antonini}, \&
  {Merritt}}]{Vasiliev2015}
{Vasiliev}, E., {Antonini}, F., \& {Merritt}, D. 2015,
  \href{http://dx.doi.org/10.1088/0004-637X/810/1/49}{\JournalTitle{\apj}, 810,
  49}

\bibitem[{{{\v{S}}ubr} {et~al.}(2019){{\v{S}}ubr}, {Fragione}, \&
  {Dabringhausen}}]{subr19}
{{\v{S}}ubr}, L., {Fragione}, G., \& {Dabringhausen}, J. 2019,
  \href{http://dx.doi.org/10.1093/mnras/stz162}{\JournalTitle{\mnras}, 484,
  2974}

\bibitem[{Wen(2003)}]{Wen2003}
Wen, L. 2003, \href{http://dx.doi.org/10.1086/378794}{\JournalTitle{The
  Astrophysical Journal}, 598, 419}

\bibitem[{Whalen \& Fryer(2012)}]{Whalen_2012}
Whalen, D.~J., \& Fryer, C.~L. 2012,
  \href{http://dx.doi.org/10.1088/2041-8205/756/1/l19}{\JournalTitle{The
  Astrophysical Journal}, 756, L19}

\bibitem[{Woods {et~al.}(2017)Woods, Heger, Whalen, Haemmerl{\'{e}}, \&
  Klessen}]{Woods_2017}
Woods, T.~E., Heger, A., Whalen, D.~J., Haemmerl{\'{e}}, L., \& Klessen, R.~S.
  2017, \href{http://dx.doi.org/10.3847/2041-8213/aa7412}{\JournalTitle{The
  Astrophysical Journal}, 842, L6}

\bibitem[{{Wu} {et~al.}(2016){Wu}, {Czerny}, {Grzedzielski}, {Janiuk}, {Gu},
  {Dong}, {Cao}, {You}, {Yan}, \& {Sun}}]{wu2016}
{Wu}, Q., {Czerny}, B., {Grzedzielski}, M., {et~al.} 2016,
  \href{http://dx.doi.org/10.3847/1538-4357/833/1/79}{\JournalTitle{\apj}, 833,
  79}

\bibitem[{{Yu} \& {Tremaine}(2003)}]{yut03}
{Yu}, Q., \& {Tremaine}, S. 2003,
  \href{http://dx.doi.org/10.1086/379546}{\JournalTitle{\apj}, 599, 1129}

\bibitem[{{Zhang} \& {Fall}(1999)}]{gcmass1}
{Zhang}, Q., \& {Fall}, S.~M. 1999,
  \href{http://dx.doi.org/10.1086/312412}{\JournalTitle{\apjl}, 527, L81}

\end{thebibliography}

\end{document}